%% file: flexibledecays.tex
\documentclass[preprint]{elsarticle}

\usepackage[utf8]{inputenc}
\usepackage[T1]{fontenc}
\usepackage{lmodern}
\usepackage{amsmath,amssymb,mathtools}
\usepackage{booktabs,tabularx}
\usepackage{geometry}
\usepackage{listings}
\usepackage{multirow}
\usepackage{scalefnt}
\usepackage{xspace}
\usepackage[usenames]{xcolor}\definecolor{fscolor}{RGB}{44,118,255}
\usepackage[absolute]{textpos}
\usepackage[normalem]{ulem}
\usepackage{listings}
\usepackage{relsize}
\usepackage{bbold}
\usepackage[shortcuts]{extdash} 

\usepackage[
  pdftitle={FlexibleDecay: An automated calculator of scalar decay widths},
  pdfauthor={Peter Athron, Adam B\"{u}chner, Dylan Harries,
    Wojciech Kotlarski, Dominik St\"{o}ckinger, Alexander Voigt},
  pdfkeywords={FlexibleSUSY,decays},
  bookmarks=true, linktocpage, colorlinks=true, allbordercolors=white,
  allcolors=fscolor]{hyperref}
\usepackage{orcidlink} 

\allowdisplaybreaks

\newcommand{\HD}{\texttt{HDECAY}\@\xspace}
\newcommand{\THDECAY}{\texttt{2HDECAY}\@\xspace}
\newcommand{\sHD}{\texttt{sHDECAY}\@\xspace}
\newcommand{\FH}{\texttt{FeynHiggs}\@\xspace}
\newcommand{\FS}{\texttt{Flex\-ib\-le\-SUSY}\@\xspace}
\newcommand{\FEFTH}{\texttt{Flex\-ib\-le\-EFT\-Higgs}\@\xspace}
\newcommand{\HB}{\texttt{HiggsBounds}\@\xspace}
\newcommand{\HS}{\texttt{HiggsSignals}\@\xspace}
\newcommand{\FD}{\texttt{Flex\-ib\-le\-De\-cay}\@\xspace}
\newcommand{\SP}{\texttt{SPheno}\@\xspace}
\newcommand{\eHD}{\texttt{eHDECAY}\@\xspace}
\newcommand{\SSUSY}{\texttt{SOFTSUSY}\@\xspace}
\newcommand{\hfold}{\texttt{HFOLD}\@\xspace}
\newcommand{\shit}{\texttt{SUSY-HIT}\@\xspace}
\newcommand{\THDMC}{\texttt{2HDMC}\xspace}
\newcommand{\Collier}{\texttt{Collier}\xspace}
\newcommand{\LT}{\texttt{LoopTools}\xspace}
\newcommand{\FeynArts}{\texttt{FeynArts}\xspace}
\newcommand{\FormCalc}{\texttt{FormCalc}\xspace}
\newcommand{\gmtwocalc}{\texttt{GM2Calc}\xspace}
\newcommand{\Himalaya}{\texttt{Himalaya}\xspace}
\newcommand{\Tsil}{\texttt{TSIL}\xspace}
\newcommand{\modelname}[1]{\texttt{#1}\@\xspace}

\newcommand{\hcoup}{\texttt{H-COUP}\@\xspace}
\newcommand{\mathematica}{\texttt{Ma\-the\-ma\-ti\-ca}\xspace}
\newcommand{\sarah}{\texttt{SARAH}\@\xspace}
\newcommand{\spheno}{\SP}

\newcommand{\abbrev}[1]{{\scalefont{.9}\text{#1}}} 
\newcommand{\DiLi}{\text{Li}_2}
\newcommand{\code}[1]{\lstinline|#1|}  
\newcommand{\as}{\alpha_3} 
\newcommand{\alphah}{\alpha_{h}} 
\newcommand{\ol}[1]{\overline{#1}}
\newcommand{\MSbar}{\ensuremath{\ol{\abbrev{MS}}}\xspace}
\newcommand{\DRbar}{\ensuremath{\ol{\abbrev{DR}}}\xspace}
\newcommand{\MSUSY}{\ensuremath{m_{\abbrev{SUSY}}}\xspace}
\newcommand{\unit}[1]{\,\text{#1}}      
\newcommand{\pole}{\text{pole}}
\newcommand{\full}{\text{full}}

\newcommand{\order}[1]{\mathcal{O}\left(#1\right)}
\newcommand{\amp}{\mathcal{A}}
\newcommand{\SM}{\ensuremath{\abbrev{SM}}\xspace}
\newcommand{\BSM}{\ensuremath{\abbrev{BSM}}\xspace}
\newcommand{\QCD}{\ensuremath{\abbrev{QCD}}}
\newcommand{\QED}{\ensuremath{\abbrev{QED}}}
\newcommand{\LO}{\ensuremath{\abbrev{LO}}}
\newcommand{\NLO}{\ensuremath{\abbrev{NLO}}}
\newcommand{\NNLO}{\ensuremath{\abbrev{NNLO}}}
\newcommand{\NNNLO}{\ensuremath{\abbrev{NNNLO}}}
\newcommand{\NNNNLO}{\ensuremath{\abbrev{N$^4$LO}}}
\newcommand{\smc}[1]{\hat{#1}}   
\renewcommand{\imath}{\text{i}}  
\newcommand{\figref}[1]{\figurename~\ref{#1}}
\newcommand{\secref}[1]{Section~\ref{#1}}
\newcommand{\subsecref}[1]{Subsection~\ref{#1}}
\newcommand{\appref}[1]{\ref{#1}}
\newcommand{\tabref}[1]{\tablename~\ref{#1}}
\newcommand{\Rplus}{\protect\hspace{-.1em}\protect\raisebox{.35ex}{\smaller{\smaller\textbf{+}}}}
\newcommand{\CPP}{\mbox{C\Rplus\Rplus}\xspace}

\DeclareMathOperator{\sign}{sign}

\journal{Computer Physics Communications}

\begin{document}
\begin{frontmatter}

  \title{\Large \bf FlexibleDecay: An automated calculator of scalar
    decay widths}
  \author[nanjing,monash]{Peter Athron\orcidlink{0000-0003-2966-5914}}
  \author[dresden]{Adam B\"uchner}
  \author[prague]{Dylan Harries\orcidlink{0000-0002-2476-6989}}
  \author[dresden]{Wojciech Kotlarski\orcidlink{0000-0002-1191-6343}\corref{cor1}}
  \ead{wojciech.kotlarski@tu-dresden.de}
  \cortext[cor1]{Corresponding author}
  \author[dresden]{Dominik St\"ockinger}
  \author[flensburg,rwth]{Alexander Voigt\orcidlink{0000-0001-8963-6512}}
  \address[nanjing]{Department  of  Physics  and  Institute  of  Theoretical  Physics,  Nanjing  Normal  University, Nanjing, Jiangsu 210023, China}
  \address[monash]{ARC Centre of Excellence for Particle Physics at
  the Terascale, School of Physics, Monash University, Melbourne,
  Victoria 3800, Australia}
  \address[dresden]{Institut f\"ur Kern- und Teilchenphysik,
    TU Dresden, Zellescher Weg 19, 01069 Dresden, Germany}
  \address[prague]{Institute of Particle and Nuclear Physics, Faculty of
    Mathematics and Physics, Charles University in Prague, V
    Hole\v{s}ovi\v{c}k\'{a}ch 2, 180 00 Praha 8, Czech Republic}
  \address[flensburg]{Fachbereich Energie und Biotechnologie, Hochschule Flensburg, Kanzleistraße 91--93, 24943 Flensburg, Germany}
  \address[rwth]{Institute for Theoretical Particle Physics and Cosmology, RWTH Aachen University, Sommerfeldstraße 16, 52074 Aachen, Germany}

\input{tex/abstract}

\begin{keyword}
    decays, Higgs boson, BSM
\end{keyword}

\end{frontmatter}

\begin{textblock*}{10em}(\textwidth,1.5cm)
\raggedleft\noindent\footnotesize
P3H--21--019 \\
TTK--21--11
\end{textblock*}

\clearpage

\noindent
{\bf PROGRAM SUMMARY} 

\bigskip

\begin{small}
\noindent
{\em Program Title: \FD}\\[0.5em]
{\em CPC Library link to program files:} (to be added by Technical Editor)\\[0.5em]
{\em Developer's repository link:} \url{https://github.com/FlexibleSUSY/FlexibleSUSY}\\[0.5em]
{\em Code Ocean capsule:} (to be added by Technical Editor)\\[0.5em]
{\em Licensing provisions:} GPLv3\\[0.5em]
{\em Programming language:} \CPP, Wolfram Language, Fortran, Bourne shell\\[0.5em]
{\em Supplementary material:} none\\[0.5em]
{\em Nature of problem:} 
Calculation of decay widths of scalar bosons in extensions of the
Standard Model of particle physics.  For the calculation of Higgs
boson decay widths known higher-order contributions are taken into
account to achieve a high precision.
\\[0.5em]
{\em Solution method:} 
The decay widths of the scalar bosons are expressed in terms of
Feynman diagrams in a general renormalizable quantum field theory and
specialized to the considered model.  The resulting expressions are
numerically evaluated using special functions and numerical
integration.
\\[0.5em]
{\em Additional comments including restrictions and unusual features:} 
\FD\ is an addon to \FS\ [1, 2].  The decay widths can only be
calculated in models that can be treated with \FS.  Decay widths of
fermions or vector bosons are not calculated.  For non-Higgs scalar
bosons the calculation is restricted to the leading order in the
perturbation series.

\end{small}

\numberwithin{table}{section}
\clearpage
\newgeometry{top=2.2cm,left=3cm,right=3cm,bottom=4cm,footskip=3em}
\tableofcontents
\clearpage

\input{tex/introduction}
\input{tex/quick}
\input{tex/renormalization_scheme}
\input{tex/implementation}
\input{tex/comparison}
\input{tex/limitations}
\input{tex/conclusions}
\input{tex/acknowledgements}

\clearpage
\appendix

\input{tex/configuration}
\input{tex/matrix_elements}
\input{tex/spheno_setup}

\clearpage

\addcontentsline{toc}{section}{References}

\bibliography{bibliography}
\bibliographystyle{elsarticle-num}
\biboptions{sort&compress}
\end{document}

%% file: tex/abstract.tex
\begin{abstract}
We present \FD, a tool to calculate decays of scalars in a broad class of
BSM models.
The tool aims for high precision particularly in the case of Higgs boson
decays.
In the
case of scalar and pseudoscalar Higgs boson decays the known higher
order SM QED, QCD and EW effects are taken into account where
possible.  The program works in a modified $\MSbar$ scheme
that exhibits a decoupling property with respect to heavy BSM
physics, with BSM parameters themselves treated in the
$\MSbar/\DRbar$-scheme allowing for an
easy connection to high scale tests for, e.g., perturbativity and vacuum
stability, and the many observable calculations readily available in
$\MSbar/\DRbar$ programs. Pure BSM effects are taken into account at the leading order,
including all one-loop contributions to loop-induced processes. The program is
implemented as an extension to \FS, which determines the mass spectrum
for arbitrary BSM models, and does not require any extra configuration
from the user. We compare our predictions for
Higgs decays in the SM, singlet extended SM, type II THDM, CMSSM and
MRSSM, as well as for squark decays in the CMSSM against a selection
of publicly available tools.  The numerical differences between our
and other programs are explained.
The release of \FD officially deprecates the old effective couplings routines in \FS.
\end{abstract}

%% file: tex/introduction.tex
\section{Introduction}

Higgs boson properties are rapidly transforming into high-precision
observables, less than a decade after the discovery of the Higgs boson
at the LHC \cite{Aad:2012tfa,Chatrchyan:2012xdj}. The Higgs decay
branching fractions are being determined with increasing accuracy
\cite{Sirunyan:2018koj,ATLAS-CONF-2018-031} and combined in global
fits (see e.g. Ref.\ \cite{Bernon:2014vta}).  Potential deviations from
predictions of the Standard Model (SM) can provide important
fingerprints of effects from physics beyond the SM (BSM).  The latest
results show that the Higgs boson is mostly SM\-/like---without any
visible BSM deviations.  This provides a major constraint on any BSM
physics that modifies the Higgs sector or Higgs decay modes, making it
essential to rigorously test these models against this data.

This requires precise predictions of Higgs boson decay widths and
branching ratios.  Many software packages that calculate Higgs decays
have been developed over the years, including
\HD\ \cite{Djouadi:1997yw,Djouadi:2018xqq},
\THDECAY\ \cite{Krause:2018wmo,Krause:2016oke,Krause:2016xku,Denner:2018opp},
\FH\ \cite{Hahn:2010te,Bahl:2018qog}, \THDMC\ \cite{Eriksson:2009ws},
\hcoup\ \cite{Kanemura:2017gbi,Kanemura:2019slf},
\eHD\ \cite{Contino:2014aaa}, \sHD\ \cite{Costa:2015llh}, \hfold
\cite{Frisch:2010gw}, \SP\ \cite{Porod:2003um,Porod:2011nf},
\SSUSY\ \cite{Allanach:2017hcf}, \texttt{NMSSMCALC}
\cite{Baglio:2013iia}, and \texttt{N2HDECAY} \cite{Engeln:2018mbg}.
Most of these codes work only for specific models such as the Minimal
Supersymmetric Standard Model (MSSM) or the two-Higgs doublet model
(THDM), their slight extensions or arbitrary models through higher
dimensional operator parametrization. However this represents just a
tiny fraction of the models that are viable candidates for physics
beyond the standard model and have been proposed in the literature.

In this paper we present an alternative \FD, which is an extension of
\FS~\cite{Athron:2014yba,Athron:2016fuq,Athron:2017fvs}, that can work
for a very broad class of models. \FS is a spectrum generator generator: once a BSM
model is specified via model files, it generates a spectrum generator
for the desired model, which computes the pole masses after
integrating renormalization-group equations and solving the
electroweak symmetry breaking conditions.\footnote{\FS uses \sarah
  \cite{Staub:2009bi,Staub:2010jh,Staub:2012pb,Staub:2013tta} to get
  analytical expressions for masses, vertices and
  renormalization-group equations, and \FS also uses some numerical
  routines originally from \SSUSY
  \cite{Allanach:2001kg,Allanach:2013kza}.} This makes it much easier
to study the phenomenology of BSM scenarios for which no
model-specific codes exist and also allows comparisons of well-known
models on an equal footing.  The only other code with this capability is
\sarah/\SP
\cite{Staub:2009bi,Staub:2010jh,Staub:2012pb,Staub:2013tta,Goodsell:2014bna,Goodsell:2015ira,Gabelmann:2018axh,Porod:2003um,Porod:2011nf}.

\FS also implements many model-specific precision corrections, so that
it provides the same---or higher---precision as the model specific
codes.  For example in the MSSM \FS has state-of-the-art Higgs mass
calculations with up to three-loop fixed-order, effective field theory and hybrid
calculations
\cite{Athron:2017fvs,Harlander:2017kuc,Harlander:2018yhj,Harlander:2019dge,Kwasnitza:2020wli},
see also Ref.\ \cite{Slavich:2020zjv} for a review.  For arbitrary
models \FS implements the \FEFTH hybrid calculation
\cite{Athron:2016fuq}, providing the most precise predictions for the
Higgs mass for BSM models without model-specific Higgs mass
calculators.  \FS has also been extended to compute other important
observables, such as the anomalous magnetic moment of the muon, lepton
electric dipole moments and $\mu \to e \gamma$.

For decays, generic calculations of loop-induced effective vertices for
the important $h\to\gamma\gamma$ and $h\to gg$ processes were
implemented in both \FS and \sarah/\SP in Ref.\ \cite{Staub:2016dxq}.
\sarah/\SP is also able to compute decay widths in two different ways:
one takes into account full BSM one-loop effects
\cite{Goodsell:2017pdq}, and the other takes into account SM-like
higher-order effects partially beyond one-loop order
\cite{Porod:2011nf,Staub:2013tta}.

\FD extends \FS with an automated calculation of decays of scalars (and Higgs
bosons in particular) in a broad class of BSM models.
\FD specifically  aims  for the highest level
of precision and reliability possible in Higgs decays.
This is important since no BSM
effects in the Higgs sector have been observed so far, and any
potential BSM effects amount to small corrections to the
SM predictions. An important observation is therefore that taking
into account higher-order SM corrections even at the multi-loop level
is indispensable, while purely BSM loop effects are typically very
small in appropriate renormalization schemes.

The precision goal of \FD is achieved in two ways.
First, for Higgs decays into Standard Model states we implement all known
higher order corrections that are applicable to general models,
ensuring state-of-the-art precision.
Second, the BSM effects are treated at leading order, but in a
special, decoupling renormalization scheme. This scheme minimizes
potential pure BSM loop effects from heavy states and thus maximizes
the accuracy of the leading order treatment.
 This implementation supersedes the previous effective
coupling routines available in \FS.  While the emphasis of the current
release is on the Higgs boson decays, the code is capable of
automatically evaluating leading-order contributions to the partial width
of any scalar.

Among the differences between \FD and other codes are not only the
selection of included higher-order corrections and the choice of
renormalization scheme but also the treatment of subtleties related to
potential numerical violations of Ward identities. The present paper
describes all choices made in \FD in detail and presents an extensive
comparison to numerical results of other codes, including those that are
model-specific and to the decay calculations of \sarah/\SP.

The paper is structured as follows.  In \secref{sec:quick} we give a
quick start guide to \FD.  \secref{sec:renormalization_scheme}
describes the special renormalization scheme used in the code.  In
\secref{sec:implementation} we describe the implementation in detail,
particularly focusing on specifying the higher order corrections in
Higgs decays. Here we also explain the treatment of potential
violations of the Ward identities in loop-induced decays.
\secref{sec:comparison} compares our results with existing codes for a
wide selection of models. The comparison includes supersymmetric and
non-supersymmetric models.  The limitations of the current version of
\FD are discussed in \secref{sec:limitations}, and our conclusions are
presented in \secref{sec:conclusions}.

%% file: tex/quick.tex
\section{Quick start guide}
\label{sec:quick}

\FD\ is an extension to the \FS\ package and is distributed with \FS\
from version 2.6.0.  In the following we briefly describe the initial
steps to download, setup and run the program to calculate decay widths.

\subsection{Downloading}
\label{sec:dl}

\FS\ 2.6.2 (current version) can be downloaded from
[\url{http://flexiblesusy.hepforge.org}] by running
\begin{lstlisting}[language=bash]
$ wget https://www.hepforge.org/archive/flexiblesusy/FlexibleSUSY-2.6.2.tar.gz
$ tar -xf FlexibleSUSY-2.6.2.tar.gz
$ cd FlexibleSUSY-2.6.2
\end{lstlisting}
Alternatively, \FS\ 2.6.2 can be downloaded via the version control system
\texttt{git} from [\url{https://github.com/FlexibleSUSY/FlexibleSUSY}]
by running
\begin{lstlisting}
$ git clone https://github.com/FlexibleSUSY/FlexibleSUSY
$ cd FlexibleSUSY
\end{lstlisting}

\subsection{Prerequisites}

\FS\ requires the following programs/libraries to be available:%
\footnote{There are also optional dependencies that are not required
  for \FD: \Tsil \cite{Martin:2005qm}, \gmtwocalc
  \cite{Athron:2015rva} and \Himalaya
  \cite{Harlander:2017kuc,Harlander:2018yhj,Harlander:2019dge} (using
  the three-loop ${\cal O}(\alpha_t \as^2)$ corrections from Refs.\
  \cite{Harlander:2008ju,Kant:2010tf}), which are only needed for
  specific models.}
\begin{itemize}
\item \mathematica/\texttt{Wolfram Engine}, version 11.0 or higher \cite{Mathematica}
\item \sarah, version 4.11.0 or higher
  \cite{Staub:2009bi,Staub:2010jh,Staub:2012pb,Staub:2013tta,Goodsell:2014bna,Goodsell:2015ira,Gabelmann:2018axh}
  [\url{http://sarah.hepforge.org}]
\item \CPP{}14 compatible compiler (\texttt{g++} 5.0.0 or higher,
  \texttt{clang++} 3.8.1 or higher, \texttt{icpc} 17.0.0 or higher)
\item FORTRAN compiler (\texttt{gfortran}, \texttt{ifort}, etc.)
\item GNU Scientific Library \cite{gsl} [\url{http://www.gnu.org/software/gsl}]
\item Eigen library, version 3.1 or higher \cite{eigenweb}
  [\url{http://eigen.tuxfamily.org}]
\item Boost library, version 1.37.0 or higher \cite{BoostLibrary}
  [\url{http://www.boost.org}]
\item \LT\ version 2.8 or higher \cite{Hahn:1998yk}
  [\url{http://www.feynarts.de/looptools}] and/or \Collier\
  \cite{Denner:2002ii,Denner:2005nn,Denner:2010tr,Denner:2016kdg}
  [\url{http://collier.hepforge.org}]
\end{itemize}
\sarah\ can be installed automatically by running
\begin{lstlisting}
./install-sarah
\end{lstlisting}

\subsection{Running \FD}

To generate a spectrum generator that calculates decays for a given
model, the \code{FSCalculateDecays} variable must be set to
\code{True} in the corresponding model file
\path{model_files/<model>/FlexibleSUSY.m.in}:
\begin{lstlisting}
FSCalculateDecays = True;
\end{lstlisting}
If the \code{FSCalculateDecays} variable is not set in the model file,
it is assumed to be \code{False}, so decays are not calculated by
default.  However, for many commonly studied models that are
distributed with \FS, such as the \modelname{SM}, \modelname{MSSM},
\modelname{THDMII}, \modelname{UMSSM}, \modelname{NMSSM},
\modelname{MRSSM}, \modelname{munuSSM}, \modelname{E6SSM}, and others,
the \code{FSCalculateDecays} variable is set to \code{True} in the
model file.  Note that by default only decays of neutral scalar,
pseudoscalar and charged Higgs bosons are calculated.  See
\appref{sec:configuration} for further configuration options.

To generate a spectrum generator for the singlet-extended Standard
Model (\modelname{SSM}) the following commands must be run:
\begin{lstlisting}[language=bash]
$ ./createmodel --name=SSM
$ ./configure --with-models=SSM --with-loop-libraries=collier,looptools
$ make
\end{lstlisting}
Note that either \LT\ \cite{Hahn:1998yk} or \Collier\
\cite{Denner:2002ii,Denner:2005nn,Denner:2010tr,Denner:2016kdg} are
required for the decays.
Once the spectrum generator has been compiled, the code can then be
run using an SLHA\-/like input file
\cite{Skands:2003cj,Allanach:2008qq}, for example:
\begin{lstlisting}[language=bash]
$ cd models/SSM
$ ./run_SSM.x --slha-input-file=LesHouches.in.SSM
\end{lstlisting}
For a detailed description of the build- and run-time configuration
options for \FS\ and \FD\ see \appref{sec:configuration} and
Ref.~\cite{FlexibleSUSY}.

\FS\ also generates a standalone program that accepts the SLHA
spectrum file as input and calculates the decays.  Using the
\code{SSM} model from above, for example, one can calculate the decays
by running:
\begin{lstlisting}[language=bash]
$ cd models/SSM
$ ./run_SSM.x --slha-input-file=LesHouches.in.SSM --slha-output-file=LesHouches.out.SSM
$ ./run_decays_SSM.x LesHouches.out.SSM
\end{lstlisting}

For the variant of the \code{SSM} model the output of \FD may read
\begin{lstlisting}
Block DCINFO
     1   FlexibleSUSY
     2   2.6.2
     5   SSMMhInput2
     9   4.14.3
DECAY        25     3.20822225E-03   # hh(1) decays
     5.82132951E-01   2          -5         5  # BR(hh(1) -> barFd(3) Fd(3))
     2.10420263E-01   2         -24        24  # BR(hh(1) -> conjVWp VWp)
     8.56749173E-02   2          21        21  # BR(hh(1) -> VG VG)
     6.19478919E-02   2         -15        15  # BR(hh(1) -> barFe(3) Fe(3))
     2.87695050E-02   2          -4         4  # BR(hh(1) -> barFu(2) Fu(2))
     2.67970867E-02   2          23        23  # BR(hh(1) -> VZ VZ)
     2.29077094E-03   2          22        22  # BR(hh(1) -> VP VP)
     1.48184338E-03   2          22        23  # BR(hh(1) -> VP VZ)
     2.64746094E-04   2          -3         3  # BR(hh(1) -> barFd(2) Fd(2))
     2.19309212E-04   2         -13        13  # BR(hh(1) -> barFe(2) Fe(2))
DECAY        35     8.56617667E-01   # hh(2) decays
     6.81973285E-01   2         -24        24  # BR(hh(2) -> conjVWp VWp)
     3.04006836E-01   2          23        23  # BR(hh(2) -> VZ VZ)
     1.21750789E-02   2          25        25  # BR(hh(2) -> hh(1) hh(1))
     8.72302442E-04   2          -5         5  # BR(hh(2) -> barFd(3) Fd(3))
     7.25578847E-04   2          21        21  # BR(hh(2) -> VG VG)
     1.06892644E-04   2         -15        15  # BR(hh(2) -> barFe(3) Fe(3))
     7.66833204E-05   2          22        23  # BR(hh(2) -> VP VZ)
     4.35647959E-05   2          -4         4  # BR(hh(2) -> barFu(2) Fu(2))
     1.89983116E-05   2          22        22  # BR(hh(2) -> VP VP)
\end{lstlisting}
The \code{DCINFO} block contains information about the calculation.
The decay output conforms to the SLHA standard \cite{Skands:2003cj}.

%% file: tex/renormalization_scheme.tex
\section{Renormalization scheme}
\label{sec:renormalization_scheme}

\FS\ is a tool that allows one to create spectrum generators for large
classes of BSM models of an a priori unknown phenomenology.  Due to
the large degree of automation required, the \MSbar\ and \DRbar\
renormalization schemes are \FS's default choice to perform the
necessary loop calculations.  While these schemes are well suited for
the calculation of properties of pure BSM observables, these schemes
are not always the optimal choice when it comes to studying properties
of some low-energy SM observables, such as the decays of the SM\-/like
Higgs boson.  That is because in the presence of a heavy BSM state
those schemes tend to introduce large higher-order corrections,
resulting in a large theoretical uncertainty in the calculated
low-energy observables.  This can for instance lead to a
non\-/decoupling behavior of the low-energy observables when the BSM
states become heavy.

To circumvent this issue, the decay calculation in \FD\ is performed
in a specific ``decoupling'' renormalization scheme.  This scheme is
designed in a way that can be automatically applied to all BSM models
that can be studied with \FS.  The scheme is similar to the familiar
decoupling scheme for the strong coupling $\as$
\cite{Collins:1978wz}, which has been applied to a non\-/minimal
SUSY model e.g.\ in Ref.\ \cite{Diessner:2017ske}, and recovers SM results in case when all BSM particles are heavy.

In the decoupling scheme of \FD\ all SM\-/like \MSbar/\DRbar\
couplings of the BSM
model (the electroweak gauge couplings $g_1$ and $g_2$,
the strong gauge coupling $g_3$, the Yukawa couplings
$Y_u$, $Y_d$ and $Y_e$ and the SM\-/like vacuum
expectation value $v$) are set equal to the corresponding SM
\MSbar\ couplings.  Thus, formally, the decoupling scheme of \FD\ is
defined by the following set of renormalization conditions:
\begin{subequations}
\begin{align}
  g_i(m_X) &= \smc{g}_i(m_X), \\
  Y_f(m_X) &= \smc{Y}_f(m_X), \\
  v(m_X)   &= \smc{v}(m_X),
\end{align}\label{eqs:decoupling_conditions}%
\end{subequations}
where $i=1,2,3$ and $f=u,d,e$.  On the r.h.s.\ of
Eqs.~\eqref{eqs:decoupling_conditions} the $\smc{g}_i$, $\smc{Y}_f$
and $\smc{v}$ denote the SM gauge couplings, Yukawa couplings and
vacuum expectation value, defined in the \MSbar\ scheme with $6$ quark
flavors.
The conditions \eqref{eqs:decoupling_conditions} are imposed
individually for every decay calculation of a particle $X$ with mass
$m_X$ at the respective renormalization scale $m_X$.
In the decoupling scheme, the non\-/SM\-/like BSM parameters remain
defined in the \MSbar/\DRbar\ scheme by the boundary conditions
specified in the \FS\ model file. Note that the renormalization conditions
\eqref{eqs:decoupling_conditions} are imposed as relations on the
finite parameters; they could equivalently be imposed via relations on
renormalization constants, however the values of renormalization
constants are not needed for this work.

A nontrivial point is that relations of the form
\eqref{eqs:decoupling_conditions} can always be identified by \FS\ for
any supported BSM model, even in models that have a more complicated structure
of Yukawa couplings or vacuum expectation values.  For a given model
the relations \eqref{eqs:decoupling_conditions} are extracted from the
mandatory\footnote{Previously the \code{LowScaleInput} boundary condition was not required in models with \code{FlexibleEFTHiggs = True} because the \FEFTH algorithm for the spectrum generation does not require this.  However this block \textit{is} required if decays are enabled in the model file via \code{FSCalculateDecays = True}.} \code{LowScaleInput} variable from the corresponding \FS\
model file.  Consider the MSSM as an example: If the model file
contains the lines
\begin{lstlisting}
LowScaleInput = {
   {g1, Sqrt[5/3] EDRbar / Cos[ThetaWDRbar]},
   {g2, EDRbar / Sin[ThetaWDRbar]},
   {g3, Sqrt[4 Pi AlphaS]},
   {Yu, Sqrt[2] Tp[upQuarksDRbar] / vu},
   {Yd, Sqrt[2] Tp[downQuarksDRbar] / vd},
   {Ye, Sqrt[2] Tp[downLeptonsDRbar] / vd},
   {vd, 2 MZDRbar / Sqrt[3/5 g1^2 + g2^2] Cos[ArcTan[TanBeta]]},
   {vu, 2 MZDRbar / Sqrt[3/5 g1^2 + g2^2] Sin[ArcTan[TanBeta]]}
};
\end{lstlisting}
then the corresponding relations \eqref{eqs:decoupling_conditions}
take the specific form
\begin{subequations}
\begin{align}
  g_1(m_X) &= \sqrt{\frac{5}{3}} \frac{\smc{e}(m_X)}{\cos[\smc{\theta}_W(m_X)]}, \\
  g_2(m_X) &= \frac{\smc{e}(m_X)}{\sin[\smc{\theta}_W(m_X)]}, \\
  g_3(m_X) &= \sqrt{4 \pi \smc{\alpha}_3(m_X)}, \\
  Y_u(m_X) &= \frac{\sqrt{2} [\smc{m}_u(m_X)]^T}{v_u(m_X)}, \\
  Y_d(m_X) &= \frac{\sqrt{2} [\smc{m}_d(m_X)]^T}{v_d(m_X)}, \\
  Y_e(m_X) &= \frac{\sqrt{2} [\smc{m}_e(m_X)]^T}{v_d(m_X)}, \\
  v_d(m_X) &= \frac{2 \smc{m}_Z(m_X) \cos(\beta)}{\sqrt{\frac{3}{5} [g_1(m_X)]^2 + [g_2(m_X)]^2}}, \label{eq:decoupling_conditions_MSSM_vd} \\
  v_u(m_X) &= \frac{2 \smc{m}_Z(m_X) \sin(\beta)}{\sqrt{\frac{3}{5} [g_1(m_X)]^2 + [g_2(m_X)]^2}}. \label{eq:decoupling_conditions_MSSM_vu}
\end{align}\label{eqs:decoupling_conditions_MSSM}%
\end{subequations}
Note that in many BSM models \FS\ can determine the SM\-/like gauge and
Yukawa couplings automatically from corresponding SM parameters.  In
this case the definitions in the \code{LowScaleInput} variable can be
simplified as follows:
\begin{lstlisting}
LowScaleInput = {
   {Yu, Automatic},
   {Yd, Automatic},
   {Ye, Automatic},
   {vd, 2 MZDRbar / Sqrt[3/5 g1^2 + g2^2] Cos[ArcTan[TanBeta]]},
   {vu, 2 MZDRbar / Sqrt[3/5 g1^2 + g2^2] Sin[ArcTan[TanBeta]]}
};
\end{lstlisting}
Furthermore, \FS\ provides the symbols \code{EDRbar},
\code{ThetaWDRbar}, \code{MZDRbar} and many more that can be used to
define relations between couplings and masses in \code{LowScaleInput},
see \tabref{tab:symbols}.
\begin{table}[b]
  \centering
  \begin{tabular}{ll}
    \toprule
    Symbol & Description \\
    \midrule
    \code{EDRbar}           & running electromagnetic gauge coupling \\
    \code{MWDRbar}          & running $W$ boson mass \\
    \code{MZDRbar}          & running $Z$ boson mass \\
    \code{ThetaWDRbar}      & running weak mixing angle $\theta_W$ \\
    \code{upQuarksDRbar}    & running diagonal up-quark $3\times 3$ mass matrix \\
    \code{downQuarksDRbar}  & running diagonal down-quark $3\times 3$ mass matrix \\
    \code{downLeptonsDRbar} & running diagonal down-lepton $3\times 3$ mass matrix \\
    \code{neutrinoDRbar}    & running diagonal up-lepton $3\times 3$ mass matrix \\
    \code{VEV}              & running SM-like vacuum expectation value \\
    \code{CKM}              & running CKM matrix \\
    \code{PMNS}             & running PMNS matrix \\
    \bottomrule
  \end{tabular}
  \caption{\FS\ model file symbols to define relations between BSM and SM quantities.}
  \label{tab:symbols}
\end{table}
Note further, that in Eqs.~\eqref{eqs:decoupling_conditions} the expression
\begin{align}
  \frac{2 \smc{m}_Z(m_X)}{\sqrt{\frac{3}{5} [g_1(m_X)]^2 + [g_2(m_X)]^2}} =: \smc{v}(m_X)
\end{align}
corresponds to the Higgs vacuum expectation value $\smc{v}(m_X)$
in the SM in the \MSbar\ scheme at the scale $m_X$.  Thus,
Eqs.~\eqref{eq:decoupling_conditions_MSSM_vd}--\eqref{eq:decoupling_conditions_MSSM_vu}
correspond to
\begin{subequations}
\begin{align}
  v_d(m_X) &= \smc{v}(m_X) \cos(\beta), \\
  v_u(m_X) &= \smc{v}(m_X) \sin(\beta),
\end{align}
\end{subequations}
which implicitly fixes the combination
\begin{align}
  v(m_X) := \sqrt{[v_u(m_X)]^2 + [v_d(m_X)]^2} = \smc{v}(m_X),
\end{align}
but leaves the ratio of the up- and down-type vacuum expectation
values as
\begin{align}
  \frac{v_u(m_X)}{v_d(m_X)} = \tan\beta.
\end{align}

After imposing the decoupling conditions
\eqref{eqs:decoupling_conditions}, the dependent parameters that are
fixed by the electroweak symmetry breaking conditions are calculated
using the tree-level tadpole equations with the SM\-/like decoupling
scheme parameters defined in \eqref{eqs:decoupling_conditions}.  For
example, in the MSSM these dependent parameters may be soft-breaking
Higgs doublet mass parameters. Following this, all tree-level masses
are calculated in the decoupling scheme.
The BSM pole masses are not calculated in the decoupling scheme, but
in the traditional (non\-/decoupling) \MSbar/\DRbar\ scheme as in
standard \FS\ \cite{Athron:2014yba,Athron:2016fuq,Athron:2017fvs}.

The advantage of this decoupling scheme is that it does not suffer
from artificial large quantum corrections in the presence of only heavy BSM
particles.  Instead, in the limit of infinitely heavy BSM physics one
recovers SM results.  Furthermore, equating SM\-/like parameters between
the SM and the BSM, as done in \eqref{eqs:decoupling_conditions}, also
makes it possible to take over known higher-order SM corrections to light Higgs
decays from the literature, which are necessary for a meaningful
comparison between theory and experimental results.\footnote{%
We note that despite this desirable decoupling property, the
tree-level matching does mean 
that the computation can miss important genuine \BSM\ higher-order
effects.
For example it is
well known in the MSSM that SUSY corrections to the bottom quark
Yukawa coupling, the so\-/called $\Delta_b$ corrections, can be
sizeable at large $\tan\beta$. These contributions are particularly
important in the case where the additional non-SM-like Higgs states
remain light, i.e.\ where the set of low energy states is that of the two-Higgs
Doublet Model (2HDM), see e.g.\ Ref.\ \cite{Haber:2000kq}.
}

\begin{figure}[tb]
  \centering
  \includegraphics[width=\textwidth]{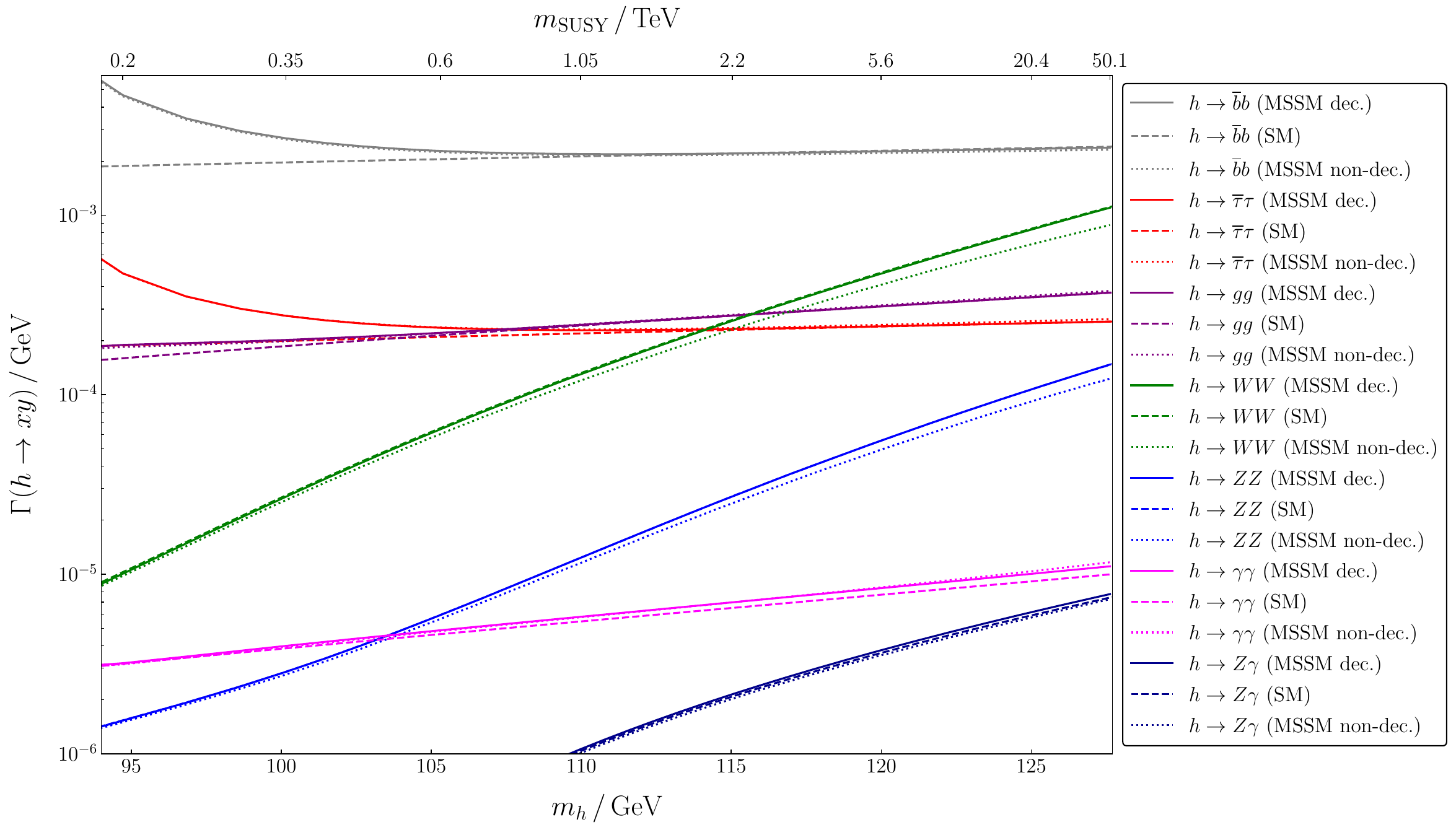}
  \caption{Decoupling properties in Higgs decay partial widths for
    \FS's \modelname{MSSMEFTHiggs} model as a function of the
    SM\-/like Higgs boson pole mass $m_h$ and the common SUSY scale
    $\MSUSY$ for a fixed $\tan\beta(\MSUSY) =10$. We color code the partial widths into different states
    and use a solid line for all partial widths calculated in the
    \modelname{MSSMEFTHiggs} decoupling scheme, a dotted line for the
    non\-/decoupling \DRbar\ scheme and a dashed line for SM.}
  \label{fig:MSSMEFTHiggs_decoupling}
\end{figure}

In \figref{fig:MSSMEFTHiggs_decoupling} we show this decoupling
property for the MSSM, using \FS's \modelname{MSSMEFTHiggs} model
file, with all the new BSM masses set to a common \MSUSY\ mass (for
details see Eq.~(10) of Ref.\ \cite{Athron:2016fuq}) and $\tan\beta(\MSUSY) =10$. This allows us
to test the decoupling property by increasing \MSUSY\ so that the new
MSSM particles get heavier and seeing if their effects really decouple
by plotting them against the Higgs decay partial widths in the SM.
Therefore we show color coded partial widths for our decoupling scheme
(solid lines), the regular \DRbar\ scheme (dotted lines), and the SM
(dashed lines). Our focus is on the differences between the SM and
MSSM results and the scheme differences. Since increasing \MSUSY\ also
increases the Higgs mass, we show $m_h$ on the lower horizontal and
\MSUSY\ on the upper horizontal axis. Please note that it is the
changes in \MSUSY\ which are driving the differences between the
results and not the change in the Higgs mass, which is not essential
for the differences between the results and is simply a feature of the
MSSM where the Higgs mass is a prediction and increases
logarithmically with \MSUSY.  We have chosen the
\modelname{MSSMEFTHiggs} model here because it also calculates the
Higgs mass precisely at large \MSUSY, however we have also checked
that we get the same behavior for the fixed\-/order Higgs mass
calculation in the MSSM.

As can be seen in the figure, in most of the partial widths the
decoupling scheme result interpolates between the \DRbar\ result at
low \MSUSY\ to the SM result at high \MSUSY.  The agreement with the SM result at high \MSUSY\ indicates the decoupling property of
this scheme. Only for $h \to \gamma \gamma$ we see a very mild
deviation from the SM at high \MSUSY. We have traced this to a
violation of the QED Ward identity at higher orders, which is an
interesting issue we discuss in \secref{sec:WardIdentity}.  The
deviation increases with the splitting between the Higgs pole mass
and the \DRbar mass, so that it is maximized at the largest \MSUSY,
growing to about $10\%$ on the right hand side of the plot.  This
deviation comes from higher order corrections and is therefore part of
the theory uncertainty of the calculation and due to the small size
of this partial width it does not appreciably affect the BRs for other
channels.  This also shows up in the $h \to \gamma Z$ decay, and the
same remarks apply there, but in that case the effect is barely
visible on the plot.  At low \MSUSY the two different
renormalization schemes give very similar results in the MSSM. The
reason is that at low scales the two schemes differ only by
non-enhanced higher-order corrections. The largest scheme differences
occur in $h\to gg $ and $h\to b\bar b$, which are affected by strong
interactions.  In total, the results confirm that the \FD predictions
in the decoupling scheme show the expected properties and the
associated theory uncertainty is under control. The significant
deviations between the MSSM and SM results at low \MSUSY are therefore
an example physics prediction which allows one to test and constrain a BSM
scenario against experimental data.

%% file: tex/implementation.tex
\section{Implementation}
\label{sec:implementation}

In this section we describe the quantum field theoretical results that
are used in the implementation of the decay processes in \FD. Since
SM\-/like higher\-/order corrections, even at the multi-loop level,
are very important for Higgs boson decays into SM particles we take
special care of the treatment of these decay modes and almost always
include appropriate higher-order corrections.  In particular, we
include a large set of higher-order corrections that can be
incorporated in a model-independent way, making our calculation of
Higgs decay modes close to state-of-the-art already in the standard
model and actually state-of-the-art in most standard model extensions.
In the case of a decay into a gauge boson pair we moreover include
single and double off-shell decays to SM fermion pairs.  Regarding
decays of (i) Higgs bosons into one or more BSM final states, (ii)
decays of non\-/Higgs BSM scalars, however, we restrict ourselves to a
pure leading\-/order calculation in the decoupling scheme, which
already gives a good approximation of the true result.

In \subsecref{sec:generic_decays} we describe the implementation of
generic, tree-level or loop-induced, 2-body decays of a scalar $S$
used in the latter cases. In the subsequent
\subsecref{sec:treeleveldecays} we then go through the important SM
decay modes of Higgs bosons which exist already at the tree level,
while in \subsecref{sec:loopinduceddecays} we discuss relevant
loop-induced decay modes. In the case of CP-conservation we use the symbols $H$ and $A$
for generic CP-even and CP-odd Higgs bosons, respectively, also
collectively denoted as $\Phi$.  We reserve the symbol $S$ for generic
scalars which may or may not be Higgs bosons.  Some corrections are
applicable irrespective of the CP properties of Higgs bosons and are
applied in all cases, as stated in the text.  A special
\subsecref{sec:WardIdentity} is devoted to a discussion of the QED
Ward identity that is relevant for the processes of $H\to\gamma\gamma$
and $H\to \gamma Z$, and which can be numerically violated in a
generic spectrum generator.

\subsection{Generic result for $S \to AB$}
\label{sec:generic_decays}

The generic formula for the leading order decay of a scalar particle
$S$ into particles $A$ and $B$ (where $A$ or $B$ can be of type
scalar, fermion or vector) reads
\begin{equation}
  \Gamma(S \to A B) = \frac{1}{16\pi m_S} \lambda^{1/2}\left(1, \frac{m_A^2}{m_S^2}, \frac{m_B^2}{m_S^2}\right) \left|\amp_{S \to A B}\right|^2,
  \label{eq:generic_decay_width}
\end{equation}
where $\lambda$ is the K\"all\'en function
\begin{equation}
  \lambda(x, y, z) \equiv (x-y-z)^2 - 4 y z
\end{equation}
and $\amp_{S \to A B}$ are the matrix elements, computed at the
leading order, be it at tree or a one-loop level.  The
$\left|\amp_{S \to A B}\right|^2$ are the squared matrix elements (if
applicable summed over spins or polarizations) listed in
\appref{app:ME} with potential extra symmetry and/or color factors.
The automatized process of obtaining one-loop amplitudes is described
in more detail in \subsecref{sec:genericloopinduceddecays}.  The
formulas are applied in all those cases that are not listed in the
remainder of this section, in particular in the case of Higgs decay if
any of the BSM particles appears in the final state or the decay of
$H \to \gamma Z$.  For such decays we use pole masses in the flux,
phase space factors and factors arising when squaring the matrix
elements.  Running parameters are used in the calculation of the
vertices and, if present, inside any loops.

All processes for which we don't describe a special treatment below are treated in the manner described in this subsection.
\subsection{Decays allowed at tree-level}
\label{sec:treeleveldecays}

\subsubsection{Special treatment of $\Phi \to q \bar{q}$}

In the case of a Higgs decay into a quark--anti-quark pair of the same
flavor\footnote{Flavor violating decays, if present in a given
  model, are computed according to \secref{sec:generic_decays},
  i.e., without any higher order corrections.} in the final
state we include several higher-order effects as this mode is
particularly important since the SM\-/like Higgs and many BSM Higgs bosons
have dominant decay modes into $b\bar{b}$ or $t\bar{t}$ pairs.  It is
therefore of the utmost importance to calculate such decay widths as
precisely as possible.  Due to the similarities in the calculation of
the decay widths of scalar and pseudoscalar Higgs bosons we present
them here together.  Following \HD, we implement a scheme which allows
us to use Higgs boson masses calculated in an arbitrary
renormalization scheme.  In this way Higgs boson masses calculated
for instance in an \MSbar\ or an on-shell (OS) scheme can be used,
whichever is appropriate in the given BSM model.

We start with the description of the implementation of the NLO QCD
corrections to the $\Phi \to q \bar{q}$ decays in the OS scheme.
Similarly to Eq.~\eqref{eq:generic_decay_width} the loop corrected
decay width can be written as
\begin{equation}
  \label{eq:hbbbar_os_only_qcd}
  \Gamma(\Phi \to q \bar{q}) = \frac{1}{16 \pi m_\Phi} \lambda^{1/2}\left(1, \frac{m_q^2}{m_\Phi^2}, \frac{m_q^2}{m_\Phi^2}\right) \left|\amp_{\Phi \to q\bar{q}}^{\text{OS-Yukawa}}\right|^2 \left(1 + C_F\frac{\as}{\pi}\Delta_\Phi(\beta)\right) ,
\end{equation}
where $\as \equiv g_3^2/(4\pi)$, $C_F = 4/3$, and
$\beta = (1 - 4 m_q^2/m_\Phi^2)^{1/2}$.
The squared matrix element $\left|\amp_{\Phi \to q\bar{q}}^{\text{OS-Yukawa}}\right|^2$ is defined similarly as in Eq.~\eqref{eq:hbbbar_os_only_qcd} but with an on-shell Yukawa coupling.
In the OS scheme for quark
masses and Yukawa couplings the NLO correction coefficients
$\Delta_\Phi$ are given by \cite{Drees:1989du}
\begin{align}
\label{eq:DeltaH}
  \Delta_H(\beta) &= \frac{1}{\beta} A(\beta) + \frac{1}{16 \beta^3} \left(3 + 34\beta^2 - 13 \beta^4\right) \ln\left(\frac{1+\beta}{1-\beta}\right) + \frac{3}{8\beta^2}\left(7\beta^2-1\right), \\
  \Delta_A(\beta) &= \frac{1}{\beta} A(\beta) + \frac{1}{16 \beta} \left(19 + 2\beta^2 + 3 \beta^4\right) \ln\left(\frac{1+\beta}{1-\beta}\right) + \frac{3}{8}\left(7-\beta^2\right),
  \label{eq:DeltaA}
\end{align}
where
\begin{align}
\begin{split}
  A(\beta) = \left(1+\beta^2\right) \bigg[& 4\,\DiLi \left(\frac{1-\beta}{1+\beta}\right) + 2\,\DiLi \left(-\frac{1-\beta}{1+\beta}\right)
  -3 \ln \left(\frac{1+\beta}{1-\beta}\right) \ln\left(\frac{2}{1+\beta}\right) \\
  & -2 \ln\left(\frac{1+\beta}{1-\beta}\right) \ln \beta \bigg]
  -3 \beta \ln\left(\frac{4}{1-\beta^2}\right) - 4\beta \ln \beta .
\end{split}
\end{align}
Eq.~\eqref{eq:hbbbar_os_only_qcd} is applicable as long as $m_q$ is of
the order of $m_\Phi$.  Otherwise, in the limit
$m_q^2/m_\Phi^2 \to 0$, the term $\Delta_\Phi$ contains a large
logarithmic contribution
\begin{equation}
  \Delta_H \approx \Delta_A \approx \frac{9}{4} - \frac{3}{2} \ln \frac{m_\Phi^2}{m_q^2} ,
\end{equation}
which spoils the perturbative expansion. Hence, for light quarks a
different approximation is required.  Consider the one-loop relation
between the pole mass $m_q$ and the corresponding running mass
$\smc{m}_q$ in pure QCD,
\begin{equation}
  m_q = \smc{m}_q \left \{1 + C_F \frac{\as}{\pi} \left(1 + \frac{3}{4}\ln \frac{\mu^2}{m_q^2} \right ) \right \}.
  \label{eq:mq_pole_to_mq}
\end{equation}
Since at the order $\as$ there are no contributions to the vacuum
expectation value, the relation \eqref{eq:mq_pole_to_mq} also holds
for the quark Yukawa couplings.  The mass\-/singular logarithms in
Eqs.~\eqref{eq:DeltaH}--\eqref{eq:DeltaA} can therefore be resumed by
using the running Yukawa coupling at the scale of the Higgs mass in
$\amp_{\Phi \to q\bar q}$ \cite{PhysRevD.22.715}.  A similar
argument holds for the $\order{\alpha}$ QED
correction.\footnote{Note that while there are no
  $\order{\alpha}$ QED corrections to the vacuum expectation
  value, there \emph{are} electroweak corrections, which cannot be
  resummed in this way.}  For small ratios of $m_q^2/m_H^2$ the more
appropriate expression for the partial width is therefore
\begin{align}
  \label{eq:Hbbbar_in_MSbar}
  \Gamma^\text{light,massless}(\Phi \to q \bar{q}) ={}& \frac{1}{16 \pi m_H} \lambda^{1/2}\left(1, \frac{m_q^2}{m_\Phi^2}, \frac{m_q^2}{m_\Phi^2}\right) \left|\amp^{\MSbar \text{ Yukawa}}_{\Phi \to q\bar q}\right|^2 \\
  & \times \left(1 + \Delta_{qq}^\QCD  + \Delta_{qq}^\QED + \Delta_{qq}^{\QED \times \QCD} + \Delta^2_\Phi\right) , \nonumber
\end{align}
where
\begin{align}
  \begin{split}
  \Delta_{qq}^{\QCD} ={}&
  \frac{17}{3} \frac{\as}{\pi} \\
  & + \left(\frac{10801}{144} - \frac{19}{12}\pi^2 - \frac{39}{2}\zeta_3 - N_f \left[ -\frac{65}{24} + \frac{1}{18}\pi^2 + \frac{2}{3}\zeta_3\right]\right) \frac{\as^2}{\pi^2} \\
  & + \left(\frac{6163613}{5184} - \frac{3535}{72}\pi^2 - \frac{109735}{216}\zeta_3 + \frac{815}{12}\zeta_5\right. \\
  & + N_f \left [ -\frac{46147}{486} + \frac{277}{72}\pi^2 + \frac{262}{9}\zeta_3 - \frac{5}{6}\zeta_4 - \frac{25}{9}\zeta_5\right] \\
  & \left. + N_f^2 \left[ \frac{15511}{11664} - \frac{11}{162}\pi^2 - \frac{1}{3}\zeta_3\right] \right) \frac{\as^3}{\pi^3}
  \\
  & + (39.34 - 220.9 N_f + 9.685 N_f^2 - 0.0205 N_f^3) \frac{\as^4}{\pi^4} ,
  \end{split} \label{eq:deltaqqQCD} \\
  \Delta_{qq}^{\QED} ={}& \frac{17}{4} Q_q^2 \frac{\alpha}{\pi} , \\
  \Delta_{qq}^{\QED\times \QCD} ={}& \left( \frac{691}{24} - 6 \zeta_3 - \pi^2 \right) Q_q^2 \frac{\alpha\as}{\pi^2}, \\
  \Delta_{H}^2 ={}& \frac{g_{t}^H}{g_{q}^H} \left( \frac{8}{3} - \frac{\pi^2}{9} - \frac{2}{3} \ln \frac{m_H^2}{m_t^2} + \frac{1}{9} \ln^2 \frac{\smc{m}_q^2}{m_H^2}\right) \frac{\as^2}{\pi^2},   \label{eq:chiral_break1}\\
  \Delta_{A}^2 ={}& \frac{g_{t}^A}{g_{q}^A} \left( \frac{23}{6} - \ln \frac{m_A^2}{m_t^2} + \frac{1}{6} \ln^2 \frac{\smc{m}_q^2}{m_A^2} \right) \frac{\as^2}{\pi^2},
  \label{eq:chiral_break2}
\end{align}
where $\zeta_n \equiv \zeta(n)$ is the Riemann Zeta function with
$\zeta_2 = \pi^2/6$, $\zeta_3 \approx 1.202$ etc.\ and $N_f$ is the
number of quark flavors lighter than $\Phi$.  The Yukawa couplings in Eq.~\eqref{eq:Hbbbar_in_MSbar}, the couplings $\alpha$ and $\as$ as well as light quark
masses $\smc{m}_f$ in
Eqs.~\eqref{eq:chiral_break1}--\eqref{eq:chiral_break2} are \MSbar\
parameters defined in the SM with $N_f$ active flavors.\footnote{In
  \FD\ we assume that $\Phi$ is always heavier than the $b$-quark,
  such that $N_f \geq 5$.}  The term $\Delta_{qq}^{\QCD}$ contains the
higher-order corrections in massless QCD up to the four-loop level
\cite{Chetyrkin:1996sr,Baikov:2005rw}. Similarly, $\Delta_{qq}^{\QED}$
contains the one-loop QED correction for quarks with electric charge
$Q_q$, while $\Delta_{qq}^{\QED \times \QCD}$ contains two-loop mixed
QED--QCD correction \cite{Kataev:1997cq}.  The contributions $\Delta_{\Phi}^2$ of
$\order{\as^2}$ are only added in the case of a CP-conserving Higgs
sector.  They originate from a flavor singlet diagram with both
bottom and top quark loops (see Fig.~1a of Ref.~\cite{Chetyrkin:1995pd}).
This explains the occurrence of the two weight factors $g_t^\Phi$ and
$g_q^\Phi$, defined as
\begin{align}
  g_q^\Phi = \frac{v \, C_{\Phi q \bar{q}}}{\smc{m}_q},
\end{align}
where $C_{\Phi q \bar{q}}$ is the $\Phi q \bar{q}$ coupling.  Note
that the weight factors $g_t^\Phi$ and $g_q^\Phi$ reflect the BSM
nature of the model, whereas $g_t^\Phi = g_q^\Phi = 1$ in the SM.
Furthermore, since the $\Delta_{\Phi}^2$ contain corrections from a
top loop, they are added only for decays into quarks lighter than the
top quark.  The loop contributions from lighter quarks to
$\Phi \to q\bar{q}$ are suppressed by the small masses of those quarks
and are therefore neglected.

To calculate decays of particles with an arbitrary ratio $m_q/m_\Phi$
in a uniform fashion we follow the approach of \HD\ and implement a
linear interpolation in $m_q^2/m_\Phi^2$ between the two regimes:
\begin{equation}
  \Gamma(\Phi \to q \bar q) =
  \left(1 - \frac{4 m_q^2}{m_\Phi^2}\right) \Gamma^\text{light}(\Phi \to q q)
  + \frac{4 m_q^2}{m_\Phi^2} \Gamma^\text{heavy}(\Phi \to q q) .
  \label{eq:Gamma_Phi_qq_interpolation}
\end{equation}
Moreover, we replace $\Gamma^{\text{light,massless}}$ by
$\Gamma^{\text{light}}$ to include the full mass effects at NLO
through
\begin{align}
\begin{split}
  \Gamma^{\text{light}}(\Phi \to q \bar q) ={}&
  \frac{1}{16 \pi m_\Phi} \lambda^{1/2}\left(1, \frac{\smc{m}_q^2}{m_\Phi^2}, \frac{\smc{m}_q^2}{m_\Phi^2}\right) \left|\amp^{\MSbar \text{Yukawa}}_{\Phi \to q\bar{q}}\right|^2 \\
  & \times \left(
    1
    + C_F \frac{\as}{\pi} \left[\Delta_\Phi(\hat\beta)
    + \Delta_\Phi^{\text{mass}} \right]
    + \Delta_{qq}^{\QED}
    +\Delta_{qq}^{\QCD \geq \text{2-loop}}
    + \Delta_{qq}^{\QED \times \QCD}
    + \Delta_\Phi^2
    \right) ,
\end{split}\label{eq:Gammaqq_light}
\end{align}
where $\hat \beta = (1- 4 \smc{m}^2_q/m_\Phi^2)^{1/2}$ and
\begin{align}
\Delta_H^{\text{mass}} &=
    2 \left(1 + \frac{3}{4} \ln \frac{m_H^2}{\smc{m}_q^2}\right)
    \frac{1- 10 \frac{\smc{m}_q^2}{m_\Phi^2}}{1 - 4 \frac{\smc{m}_q^2}{m_\Phi^2}} , \\
\Delta_A^{\text{mass}} &=
    2 \left(1 + \frac{3}{4} \ln \frac{m_H^2}{\smc{m}_q^2}\right)
    \frac{1- 6 \frac{\smc{m}_q^2}{m_\Phi^2}}{1 - 4 \frac{\smc{m}_q^2}{m_\Phi^2}} .
\end{align}
The contribution $\Delta_\Phi^{\text{mass}}$ originates from
expressing the tree-level decay in terms of the running quark mass up
to $\order{\as}$.  One also has
$\lim_{\smc{m}_q^2/m_H^2 \to 0} \left( \Delta_\Phi +
  \Delta_\Phi^{\text{mass}}\right) = 17/4$, as expected from
Eq.~\eqref{eq:deltaqqQCD}.  The term
$\Delta_{qq}^{\QCD \geq \text{2-loop}}$ is obtained from
Eq.~\eqref{eq:deltaqqQCD} by dropping only the $\order{\as}$
term.\footnote{Since \HD\ does not include QED effects in the running
  of the quark Yukawa coupling, it uses $\Delta_\Phi$ instead of
  $\Delta_{qq}^{\QED}$ in the analog of Eq.~\eqref{eq:Gammaqq_light}.}
Furthermore, Eq.~\eqref{eq:Gammaqq_light} differs from
Eq.~\eqref{eq:Hbbbar_in_MSbar} by terms of
$\order{\as\smc{m}_q^2/m_\Phi^2}$ and higher.  Apart from the
corrections from Eq.~\eqref{eq:hbbbar_os_only_qcd},
$\Gamma^\text{heavy}$ also contains one-loop QED corrections in the OS
scheme
\begin{align}
  \label{eq:Gammaqq_heavy}
  \Gamma^{\text{heavy}}(\Phi \to q \bar q) =  \frac{1}{16 \pi m_\Phi} \lambda^{1/2}\left(1, \frac{m_q^2}{m_\Phi^2}, \frac{m_q^2}{m_\Phi^2}\right) \left|\amp_{\Phi \to q\bar{q}}^{\text{OS-Yukawa}}\right|^2 \left[1 + (C_F \as + \alpha) \frac{1}{\pi} \Delta_\Phi (\beta)
  \right] .
\end{align}
Since all couplings in \FS\ are running couplings, we obtain the OS
Yukawa coupling needed in the above equation from the relation
\begin{align}
  \left|\amp_{\Phi \to q\bar{q}}^{\text{OS-Yukawa}}\right|^2
  = \frac{m_q^2}{\smc{m}_q^2} \left|\amp_{\Phi \to q\bar{q}}\right|^2 .
\end{align}

As there is no interference between scalar and pseudoscalar parts, in models with CP-violation we add both contributions in the final result.

Finally, we emphasise, that no higher-order weak corrections are applied in Eqs.~\eqref{eq:Hbbbar_in_MSbar} and \eqref{eq:Gammaqq_light}.

\subsubsection{Special treatment of $\Phi \to l^+ l^-$}
\label{sec:Phitollbar}

For the case of a pair of charged leptons in the final state we follow the
generic approach from Eq.~\eqref{eq:generic_decay_width} but include
universal one-loop QED corrections in massless limit as
\begin{equation}
  \Gamma(\Phi \to l^+ l^-) = \frac{1}{16 \pi m_\Phi} \lambda^{1/2}\left(1, \frac{m^2_{l}}{m_\Phi^2}, \frac{m_{l}^2}{m_\Phi^2}\right) \left|\amp_{\Phi \to l^+l^-}\right|^2 \left(1 + \frac{17}{4} \frac{\alpha}{\pi}\right),
  \label{eq:Gamma_Phi_ll}
\end{equation}
where $\alpha$ denotes the SM \MSbar\ electromagnetic coupling.
In contrast to the $\Phi \to q\bar{q}$ case we always choose an $\alpha$ with $6$ active quark flavors and ignore 2-loop contributions.

Eq.~\eqref{eq:Gamma_Phi_ll} can be obtained immediately by adapting
Eq.~\eqref{eq:Hbbbar_in_MSbar} to the color singlet case at the
one-loop level. This is sufficient, because QED corrections to
$\Phi \to l^+ l^-$ are small and the leptonic decay rates of the Higgs are
much smaller than decay rates to quarks.

\subsubsection{Special treatment of $H \to Z Z$ and $H \to W^+ W^-$}

For the SM\-/like Higgs boson, 3-body decays through off-shell $W$ and
$Z$ bosons are phenomenologically very relevant.  We therefore include
both single and double off-shell diagrams as well as two body on-shell
decays in this case.

The single off-shell decay width, assuming the SM decay pattern and
neglecting fermion masses, is given by \cite{Pocsik:1980ta,Rizzo:1980gz,Keung:1984hn,Djouadi:2005gi}
\begin{align}
  \label{eq:single_offshell}
  \Gamma(H \to V V^*) = \frac{3 |C_{H V V}|^2}{256 \pi^3 m_H} \frac{R_T(m_V^2/m_H^2)}{m_V^2/m_H^2} \frac{g_2^2}{2} \delta_V ,
\end{align}
where
\begin{align}
  \label{eq:RT}
  R_T(x) \equiv \frac{3(1-8x + 20x^2)}{(4x-1)^{1/2}}\arccos\left( \frac{3x-1}{2 x^{3/2}}\right) - \frac{1-x}{2x} \left(2-13x+47x^2\right) - \frac{3}{2}\left(1-6x+4x^2\right)\ln(x) ,
\end{align}
$C_{HVV}$ is the coupling coefficient in front
of the metric tensor, and
\begin{align}
 \delta_W &\equiv 1 , \\
 \delta_Z &\equiv \frac{1}{\hat c_w^2} \left(\frac{7}{12} - \frac{10}{9} \hat s_w^2 + \frac{40}{27} \hat s_w^4\right) ,
\end{align}
with $\smc{c}_w$ and $\smc{s}_w$ being the cosine and sine of the
running weak mixing angle, respectively.

The double off-shell decay width is given by \cite{Cahn:1988ru,Grau:1990uu,Djouadi:2005gi}
\begin{align}
  \label{eq:double_offshell}
  \Gamma(H \to V^* V^*) = \frac{1}{\pi^2} \int_0^{m_H^2} ds_1 \frac{m_V \Gamma_V}{(s_1 - m_V^2)^2 + m_V^2 \Gamma_V^2} \int_0^{(m_H - s_1)^2} ds_2 \frac{m_V \Gamma_V}{(s_2 - m_V^2)^2 + m_V^2 \Gamma_V^2} \Gamma_0 ,
\end{align}
where
\begin{align}
  \label{eq:Gamma_0}
  \Gamma_0 =  \frac{|C_{HVV}|^2 m_H^3}{128 \pi m_V^4} S_V \lambda^{1/2} \left(1, \frac{s_1}{m_H^2}, \frac{s_2}{m_H^2}\right)
  \left[\lambda \left(1, \frac{s_1}{m_H^2}, \frac{s_2}{m_H^2}\right) + \frac{12 s_1 s_2}{m_H^4} \right] .
\end{align}
while $\Gamma_V$ is the width of the final state vector boson $V$.
In Eq.~\eqref{eq:Gamma_0} the coupling $C_{HVV}$ is the same as in the
case of the 3-body decay and
\begin{align}
  S_V =
  \begin{cases}
    2 & \text{for } V = W,\\
    1 & \text{for } V = Z.
  \end{cases}
\end{align}

We implicitly assume, that an off-shell boson can only ``decay'' to SM
leptons or quarks other than the top quark.  This assumption enters
either by directly considering only SM decays in
Eq.~\eqref{eq:single_offshell} or by using the SM $W$ and $Z$ widths
in Eq.~\eqref{eq:double_offshell}.  This assumption is no longer valid
if $V$ couples to two BSM electroweak particles or to one BSM and one
SM particle, whose sum of masses is smaller than $m_H - m_V$.
\FD automatically detects such cases and disables the calculation of
those off-shell decay widths, while still allowing all other decay
calculations to proceed as normal.
In that case the off-shell partial
width is effectively set to zero.
As a result, all the branching ratios printed
by \FD\ in such a case will be incorrect, though the partial widths
themselves are correct.

By default, \FD uses double off-shell decays both for $m_H < m_V$ as
well as for $m_V < m_H < 2 m_V$ and two-body decays, according to
Eq.~\eqref{eq:generic_decay_width}, if $m_H > 2 m_V$.  The user is
free to select which off-shell decays are used for which mass
ordering, though.
This is controlled by the flag 4 in block \texttt{FlexibleDecay} as
described in \tabref{tab:FS_SLHA_configuration_FD-block}.
Technically, Eq.~\eqref{eq:double_offshell} is integrated numerically
in \FD\ using an algorithm from the GNU Scientific Library \cite{gsl}
with a relative error goal of $< 0.1\%$.

\subsection{Loop-induced decays}\label{sec:loopinduceddecays}

Several important decay channels of Higgs and BSM scalars do not arise
at tree level but instead are loop\-/induced. These loop\-/induced
decays are computed by \FD\ to at least the full one-loop level. As in
the case of tree-level decays, loop-induced decays of high
phenomenological importance are augmented by higher order SM\-/like
corrections.  This concerns the decays of $\Phi \to \gamma \gamma$ and
$\Phi \to g g$.  Regarding the loop-induced decay of $H \to \gamma Z$,
the SM QCD corrections are small
\cite{Spira:1991tj,Gehrmann:2015dua,Bonciani:2015eua}, so the calculation is
restricted to the leading order, which is sufficient for current and
near future experiments \cite{deBlas:2019rxi}.
In the following we first describe the general one-loop calculation
and afterwards focus on the details of the implementation of
$\Phi \to \gamma \gamma$ and $\Phi \to g g$.

\subsubsection{General one-loop computation of loop-induced decays}
\label{sec:genericloopinduceddecays}

\begin{figure}[tb]
  \centering
  \includegraphics[width=0.6\textwidth]{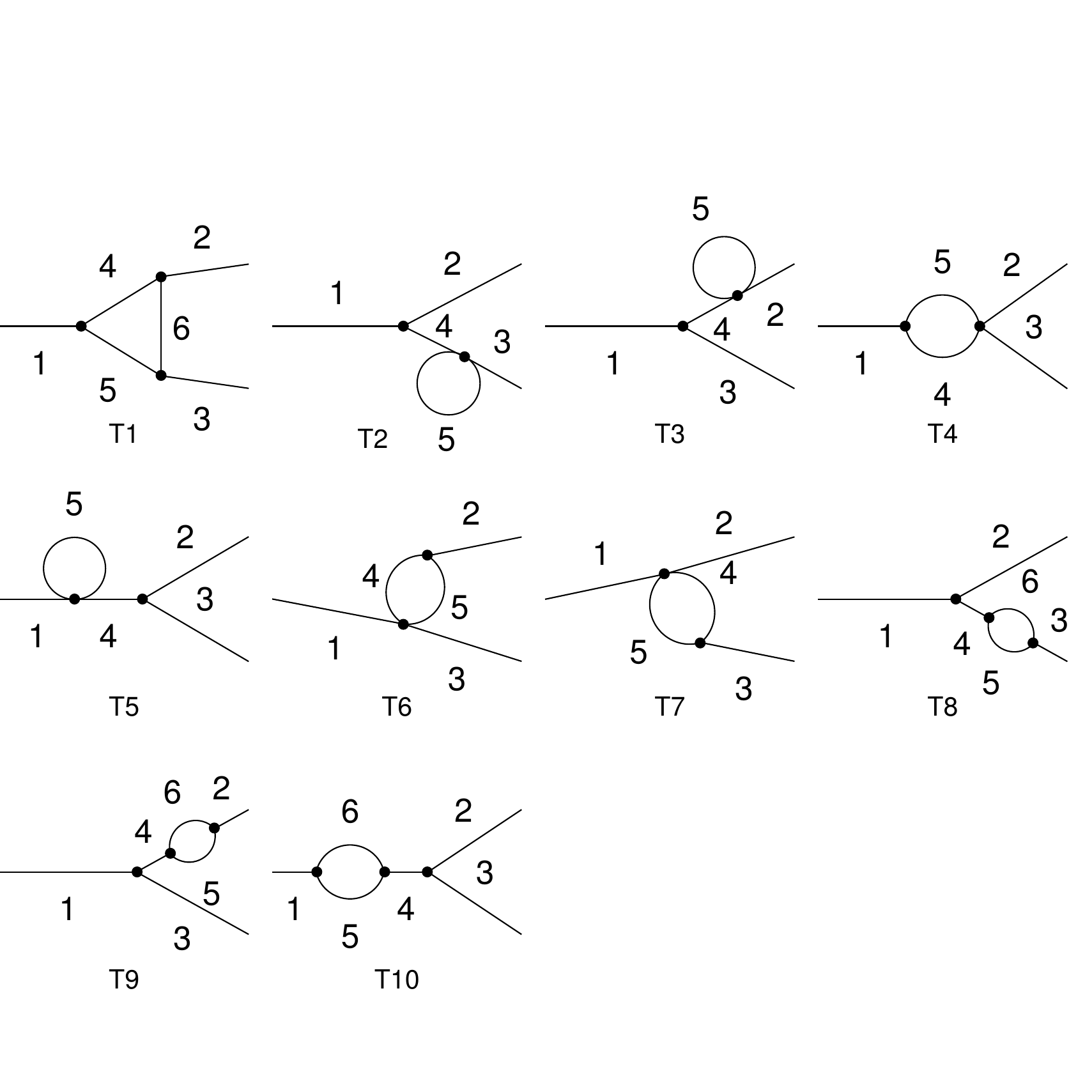}
  \caption{Diagrams contributing to the one-loop amplitude for decay
    of particle 1 to particles 2 and 3.  We take particle 1 to be
    always a scalar.  Each of the particles 2--6 can be a scalar,
    fermion or (massive or massless) vector boson, as long as the
    couplings are compatible with Lorentz and gauge invariance.}
  \label{fig:1l_decay_topologies}
\end{figure}

Whenever some decay amplitude with two final-state particles vanishes
analytically at tree level, \FD\ switches to computing the amplitude
at full one-loop level.%
\footnote{\label{footnote:pm_check} The check is performed analytically at the level of particle
  multiplets in the mass eigenstate basis, as defined in the \sarah\
  model file.
  This means that currently no loop-induced flavor-violating decays are generated.
}
In general, the ten different diagram topologies shown in
\figref{fig:1l_decay_topologies} can contribute.  \FD\ has implemented
all these topologies as well as the analytic expressions for the
corresponding amplitudes after inserting all possible concrete
particle types (scalars, fermions, vectors).  These generic analytic
expressions were pre-generated using \FeynArts\ and \FormCalc\
\cite{Hahn:2000kx,Hahn:1998yk} and are distributed with \FS.  For a
given concrete model the diagram topologies are filled with concrete
particle insertions during the \CPP\ code generation of \FS.

\subsubsection{Special treatment of $\Phi \to \gamma \gamma$}
\label{sec:Phigamgam}

Following the outlined procedure, we first evaluate all one-loop
diagrams for $\Phi\to\gamma\gamma$ in a BSM model.  The amplitude can
be split into loops with fermions, scalars and vector bosons as
\begin{equation}
  \amp = \sum_f \amp_f + \sum_{S,V} \amp_{S, V} .
\end{equation}
Some of those individual terms are then augmented by SM\-/like
higher-order corrections, which is described in the following.

First, fermion triangle loops (topology \texttt{T1} from
\figref{fig:1l_decay_topologies}) for $\Phi\to\gamma\gamma$ consisting
of a closed loop of quarks of the same flavor are multiplied by a
correction factor that implements two-loop QCD corrections.
Specifically, every triangle loop with a color triplet fermion $f$,
which can be written as
\begin{equation}
  A_f^{\texttt{T1}} = A_f^{\texttt{T1},S} \left[k^2 (\epsilon_1^* \epsilon_2^*) - (k_1 \epsilon_2^*) \, (k_2 \epsilon_1^*)\right] +  A_f^{\texttt{T1},P} \det(\epsilon_1^*, \epsilon_2^*, k_1, k_2) ,
\end{equation}
where $k_i$ and $\epsilon_i$ are the momentum and polarization vectors
of photon $i$, respectively, is replaced by
\begin{equation}
  A_f^{\texttt{T1},X} \to A_f^{\texttt{T1},X} \left[1 + \frac{\as}{\pi} \left(C_1^X(\tau_q) + C_2^X(\tau_q) \ln \frac{\mu^2}{\smc{m}_f^2} \right)\right] ,
\end{equation}
where $\smc{m}_f$ is the \MSbar\ fermion mass, $\as$ is the
\MSbar\ strong coupling in the SM,
$X\in\{S,P\}$, and
$\tau_f \equiv m_\Phi^2/(4 \smc{m}_f^2)$.
Similarly to Sec.~\ref{sec:Phitollbar} we do not apply NNLO corrections and always choose the 6-flavour scheme.
The correction coefficients
$C_i^S$ were first evaluated numerically for intermediate $\tau_f$s in \cite{Djouadi:1990aj}, and $C_i^X$ were given analytically in the limit of $\tau_f \gg 1$ and $\tau_f \ll 1$ in \cite{Djouadi:1993ji}.
Here we implement fully analytic result without any approximations
as given in Eqs.~(2.7)--(2.8) ($C_i^S$) and
in Eqs.~(2.19)--(2.20) ($C_i^P$) of Ref.\ \cite{Harlander:2005rq}.
In the case of the SM top\-/quark loop it would be sufficiently accurate
to implement only the first few terms of the Taylor expansion around
$\tau=0$. However we implement the full expressions in order to
improve the precision for decays of arbitrary Higgs bosons and BSM
quark\-/like fermion loops with arbitrary Higgs/fermion mass ratios.

In the limit of $\tau_f \to 1$, $C^P_i$ exhibits a Coulomb singularity  \cite{Harlander:2005rq} which is regulated by a finite fermion $f$ width and should be resummed \cite{Melnikov:1994jb}.
Here we resort to a solution where we simply do not apply higher order corrections when $|1-\tau_f| \leq 10^{-2}$.
This case is also flagged, with an appropriate warning printed when it occurs.

The two-loop corrections to closed scalar loops with \texttt{T1} and
\texttt{T4} topologies are implemented according to Ref.\
\cite{Aglietti:2006tp}.  Similar as before, closed scalar loops are
multiplied by the correction factor
\begin{equation}
  \amp_S^{\text{\texttt{T1+T4}}} \to \amp_S^{\text{\texttt{T1+T4}}} \left[1 + \frac{\as}{\pi} C(R)\frac{\mathcal{F}_0^{(2l,a)}(r) + \left[\mathcal{F}_0^{(2l,b)}(r)+ \mathcal{F}_0^{(2l,c)}(r)\right] \ln \frac{\smc{m}_0^2}{\mu^2}}{\mathcal{F}_0^{(1l)}(r)}\right],
\end{equation}
where $\smc{m}_0$ is a running scalar mass in the loop,
$\as$ is the \MSbar\ strong coupling in the SM with $6$ active quark flavors,
$r \equiv m_H^2/\smc{m}^2_0$ and $C(R)$ is the Casimir of the color
representation of the scalar.  Currently, \FD\ handles color triplets
with $C_F = (N_c^2 - 1)/(2 N_c)$ and octets with $C_A = N_c$.
The functions $\mathcal{F}_{0}$ are available analytically as a series
expansion.  For $r \ll 1$ one has
\begin{align}
\mathcal{F}_{0}^{(1l)}(r) ={}&   - \frac{1}{3}
                               - \frac{2}{45} r
			       - \frac{1}{140} r^2
			       - \frac{2}{1575} r^3
                               - \frac{1}{4158} r^4
			       + \order{r^{5}} ,\\
\mathcal{F}_{0}^{(2l,a)}(r)   ={}&
        		   - \frac{3}{4}
        		   - \frac{29}{216} r
        		   - \frac{4973}{226800} r^2
        		   - \frac{3137}{882000} r^3
        		   - \frac{1180367}{2095632000} r^4
			   + \order{r^{5}} , \\
\mathcal{F}_{0}^{(2l,b)}(r)   ={}&
        		   - \frac{1}{4}
        		   - \frac{1}{15} r
        		   - \frac{9}{560} r^2
        		   - \frac{2}{525} r^3
        		   - \frac{5}{5544} r^4
			   + \order{r^{5}} ,
\end{align}
while for $r \gg 1$
\begin{align}
\mathcal{F}_{0}^{(1l)}(r) ={}&
                          \frac{4}{r}
			  + 4 \ln^2(-r) \frac{1}{r^2}
			  + \order{\frac{1}{r^3}} ,\\
\begin{split}
\mathcal{F}_{0}^{(2l,a)}(r)  ={}&  \left[
            14
	  - 3 \, \ln(-r)
	     \right] \frac{1}{r}
       + \Bigl[
            6
          - \frac{72}{5} \zeta_2^2
          - 24 \zeta_3
          - 8 ( 1 - \zeta_2 - 4 \zeta_3 ) \, \ln(-r) \\
&
          + ( 17 - 8 \zeta_2 ) \, \ln^2(-r)
          - \frac{5}{3} \, \ln^3(-r)
          - \frac{1}{6} \, \ln^4(-r)
	     \Bigr] \frac{1}{r^2}
          + \order{\frac1{r^{3}}} ,
\end{split} \\
\mathcal{F}_{0}^{(2l,b)}(r)  ={}& \left[
	    6 \, \ln(-r)
	  - 3 \, \ln^2(-r)
	     \right] \frac{1}{r^2}
          + \order{\frac1{r^{3}}} .
\end{align}

Throughout the calculation we choose a renormalization scale $\mu = m_\Phi$, as opposed, for example, to a $\mu = m_\Phi/2$ choice of Ref.~\cite{Djouadi:1993ji}.
Finally, the total amplitude for $\Phi \to \gamma \gamma$ is multiplied by $\alpha(0)/\alpha(m_\Phi)$.
This effectively incorporates certain two-loop QED corrections and minimizes the remaining theoretical uncertainty.\footnote{\label{footnote:flag3} This is controlled by flag 3 in block \texttt{FlexibleDecay} (see \tabref{tab:FS_SLHA_configuration_FD-block}).}

For the SM Higgs boson it is known that these implemented two-loop QCD
corrections increase the partial width by around $2\%$. In addition to
these corrections, also the complete SM three-loop QCD
\cite{Maierhofer:2012vv} and the complete SM two-loop electroweak
corrections \cite{Degrassi:2005mc,Actis:2008ts} are known.  For the SM Higgs, the
three-loop QCD part increases the partial with by $\lesssim 0.1\%$;
therefore we opt to not include them in the current version of \FD.
The two-loop order $\smc{y}_t^2$ corrections are known in an analytic
form in the large $m_t$ limit and for $m_W < m_H < 2 m_W$
\cite{Fugel:2004ug}.  These corrections turn out to not be dominant,
being on par with the one from the pure Yang--Mills sector.  For the
SM\-/like Higgs the total electroweak correction is negative and
accounts for around $-2.5\%$.  Since those corrections cannot be
easily incorporated into our calculation we neglect them. The precision of
$\Gamma(h \to \gamma \gamma)/\Gamma(h \to \gamma \gamma)_\SM$ ratio at
the high luminosity LHC is estimated to be
$3.8\%$ \cite{deBlas:2019rxi}, so neglecting them will not be the biggest
source of uncertainty when confronting our calculation with
experimental data.

This implementation supersedes the previously implemented calculation of effective $\Phi \to \gamma \gamma$ coupling \cite{Staub:2016dxq}.
Compared to the old case we now compute branching ratios rather than effective couplings, include two-loop QCD corrections to fermion and scalar loops and improve handling of CP-violating models.

\subsubsection{Special treatment of $\Phi \to \gamma Z$}

In the double limit of a heavy color triplet fermion in the loop and $m_Z/m_\Phi \to 0$, the NLO QCD correction to
$\Phi \to \gamma Z$ approaches the same value as in the
$\Phi \to \gamma \gamma$ case \cite{Spira:1991tj,Djouadi:2005gi}.  For a
generic scalar $\Phi$, the correction can be applied by the replacement
\begin{align}
  A_f^{\texttt{T1},S} & \to A_f^{\texttt{T1},S} \left(1 - \frac{\as}{\pi}  \right) ,\\
  A_f^{\texttt{T1},P} & \to A_f^{\texttt{T1},P},
\end{align}
i.e.\ the correction to the CP-odd part of the amplitude vanishes in this limit.
In \FD\ we implement this correction for $m_\Phi/\smc{m}_f < 0.8$ and $m_Z/m_\Phi < 0.75$, where
$\smc{m}_f$ is the running mass of the colored fermion running in the loop.

Finally, similarly as in the $\Phi \to \gamma \gamma$ case, we multiply the amplitude by the $(\alpha(0)/\alpha(m_\Phi))^{1/2}$ factor related to a single on-shell photon.$^{\ref{footnote:flag3}}$

\subsubsection{Special treatment of $\Phi \to g g$}

The higher\-/order corrections for the $\Phi \to gg$ decay are
implemented differently from the $\Phi \to \gamma \gamma$ case.  The
reason is that a calculation of higher\-/order corrections involves
not only virtual, but also real amplitudes.
The known higher order corrections can therefore be only
implemented at the level of the partial width, not the amplitude.

The $\Phi\to gg$ decay width, including higher order SM QCD
corrections to the top quark loop, can be written as
\begin{align}
\label{eq:hgg_master}
 \Gamma(H \to gg) ={}&
 \Gamma_\LO^\text{full} (H \to gg) \\
 &+ \Gamma_H
 \left[
 1 - \left(\frac{\as^{(6)}}{\as^{(5)}}\right)^2
 + \frac{1}{\Gamma_H^0} \left(\frac{\as^{(5)}}{\pi} \delta_{\NLO}^H
 + \left(\frac{\as^{(5)}}{\pi}\right)^2 \delta_{\NNLO}^H
 + \left(\frac{\as^{(5)}}{\pi}\right)^3 \delta_{\NNNLO}^H\right)\right], \nonumber\\
\label{eq:Agg_master}
 \Gamma(A \to gg) ={}&
 \Gamma_\LO^\text{full} (A \to gg) + \Gamma_A
 \left[
 1 - \left(\frac{\as^{(6)}}{\as^{(5)}}\right)^2
 + \frac{1}{\Gamma_A^0} \left(\frac{\as^{(5)}}{\pi} \delta_{\NLO}^A
 + \left(\frac{\as^{(5)}}{\pi}\right)^2 \delta_{\NNLO}^A\right)
 \right] ,
\end{align}
where $\Gamma_{\LO}^\full$ is the full, model specific, one-loop BSM
contribution to $\Phi \to g g$ and
\begin{align}
  \Gamma_H^0 &= \left| \frac{3}{2\tau}\left[1 + \left(1 - \frac{1}{\tau}\right)\arcsin^2(\sqrt{\tau}) \right]\right|^2 ,\\
  \Gamma_H &= \frac{m_H}{18 \pi} \left(\frac{\as^{(5)}}{\pi}\right)^2 \left|C_{Ht\bar{t}} \; \sqrt{\tau}  \right|^2 \Gamma_H^0, \\
  \Gamma_A^0 &= \left| \frac{1}{\tau} \arcsin^2(\sqrt{\tau})\right|^2, \\
  \Gamma_A &= \frac{m_A}{8 \pi} \left(\frac{\as^{(5)}}{\pi}\right)^2 \left|C_{At\bar{t}} \; \sqrt{\tau} \right|^2 \Gamma_A^0,
\end{align}
where $\tau = m_\Phi^2/(4 \smc{m}_t^2)$.  In
Eqs.~\eqref{eq:hgg_master}--\eqref{eq:Agg_master} $\as^{(n)}$
denotes the \MSbar\ strong coupling in the SM with $n$ active quark
flavors.

For the decay of a CP\-/even scalar $H$, two-, three- and
four\-/loop QCD contributions are taken into account.  For the decay
of a CP\-/odd scalar $A$, two- and three\-/loop QCD contributions are
taken into account.
The term $1 - (\as^{(6)}/\as^{(5)})^2$ converts the squared top quark contribution in $\Gamma_\LO^\text{full} (\Phi \to gg)$ from 6 to 5 flavors.

The NLO, NNLO and NNNLO QCD corrections in
Eqs.~\eqref{eq:hgg_master}--\eqref{eq:Agg_master} were obtained
analytically in
Refs.~\cite{Djouadi:1991tka,Schreck:2007um,Chetyrkin:1997iv,Baikov:2006ch,Spira:1995rr,Chetyrkin:1998mw}
by employing a heavy quark limit.  For a CP\-/even Higgs $H$ we
implement $\delta_{\NLO}^H$ at $\order{\tau^5}$
\cite{Djouadi:1991tka,Larin:1995sq,Schreck:2007um}, $\delta_{\NNLO}^H$ at
$\order{\tau^2}$ \cite{Chetyrkin:1997iv,Schreck:2007um} and
$\delta_{\NNNLO}^H$ at $\order{\tau^0}$ \cite{Baikov:2006ch}:

\newcommand{\LH}{L_H}
\newcommand{\LHH}{L_H^2}
\newcommand{\Lt}{L_t}
\newcommand{\Ltt}{L_t^2}

\begin{align}
\label{eq:hgg_nlo}
\delta_{\NLO}^H ={}&
\frac{95}{4} - \frac{7}{6} N_f + \frac{33-2 N_f}{6}\LH
\\
& + \left(\frac{5803}{540} + \frac{77}{30}\LH - \frac{14}{15}\Lt + N_f\left[ -\frac{29}{60}- \frac{7}{45}\LH \right] \right) \tau \nonumber
\\
& + \left( \frac{1029839}{189000} + \frac{16973}{12600}\LH - \frac{1543}{1575}\Lt + N_f \left[ -\frac{89533}{378000} - \frac{1543}{18900}\LH \right]\right) \tau^2 \nonumber\\
& + \left(\frac{9075763}{2976750} + \frac{1243}{1575}\LH - \frac{452}{525}\Lt + N_F \left[ -\frac{3763}{28350} - \frac{226}{4725} \LH\right]\right) \tau^3 \nonumber\\
& + \left(\frac{50854463}{27783000} + \frac{27677}{55125}\LH - \frac{442832}{606375}L_t + N_f \left[ - \frac{10426231}{127338750} - \frac{55354}{1819125}\LH \right]\right) \tau^4 \nonumber\\
& + \left(\frac{252432553361}{218513295000} + \frac{730612}{2149875}\LH + \frac{2922448}{4729725}L_t + N_f \left[ -\frac{403722799}{7449316875} - \frac{1461224}{70945875}\LH \right]\right)\tau^5, \nonumber \\
\label{eq:hgg_nnlo}
 \delta_{\NNLO}^H ={}& \frac{149533}{288} - \frac{363}{8}\zeta_2 - \frac{495}{8} \zeta_3 + \frac{3301}{16}\LH + \frac{363}{16}L_H^2 + \frac{19}{8} L_t \\
 & + N_f \left( -\frac{4157}{72} + \frac{11}{2}\zeta_2 + \frac{5}{4}\zeta_3 - \frac{95}{4}\LH -\frac{11}{4}L_H^2 + \frac{2}{3} L_t\right) \nonumber
\\
 & + N_f^2 \left( \frac{127}{108} - \frac{\zeta_2}{6} + \frac{7}{12}L_H +\frac{1}{12} L_H \right)  \nonumber\\
 & + \left(\frac{104358341}{1555200} - \frac{847}{240}\pi^2 + \frac{7560817}{69120} \zeta_3 + \left[\frac{203257}{2160} - \frac{77}{15}L_t\right]\LH + \frac{847}{80}\LHH - \frac{24751}{1080}\Lt - \frac{77}{180}\Ltt \right. \nonumber \\
 & \left. + N_f \left [ -\frac{9124273}{388800} + \frac{77}{180}\pi^2 + \frac{7}{12}\zeta_3 + \left(-\frac{67717}{6480} + \frac{14}{45}\Lt\right)\LH - \frac{77}{60}\LHH + \frac{586}{405}\Lt + \frac{7}{90}\Ltt\right] \right. \nonumber \\
 & \left. N_f^2 \left[\frac{5597}{12960} -\frac{7}{540}\pi^2 + \frac{29}{120}\LH + \frac{7}{180}\LHH\right]
 \right)\tau \nonumber\\
 &+ \left(-\frac{1279790053883}{12192768000} - \frac{186703}{100800}\pi^2 + \frac{39540255113}{232243200}\zeta_3 + \left[\frac{9158957}{189000}-\frac{16973}{3150}\Lt\right]\LH \right. \nonumber\\
 & \left. +\frac{186703}{33600}\LHH - \frac{10980293}{453600}\Lt + \frac{20059}{37800}\Ltt + N_f \left[ -\frac{64661429393}{5715360000} \right. \right. \nonumber \\
 & \left. \left. - \frac{16973}{25200}\LHH + \frac{16973}{75600}\pi^2 + \frac{1543}{5040}\zeta_3 + \left( -\frac{10306537}{1944000} + \frac{1543}{4725}\Lt \right)\LH \right. \right. \nonumber \\
 & \left. \left. +\frac{8973773}{6804000}\Lt + \frac{1543}{18900}\Ltt \right] +
 N_f^2 \left[\frac{3829289}{19440000} - \frac{1543}{226800}\pi^2 + \frac{89533}{756000}\LH + \frac{1543}{75600}\LHH \right] \right)\tau^2, \nonumber \\
 \label{eq:deltaNNNLO}
 \delta_{\NNNLO}^H ={}& 467.683620788 - \frac{8}{3}\left(\frac{19}{8} + \frac{2}{3} N_f\right) + \left[122.440972222 -  2\left(\frac{19}{8} + \frac{2}{3} N_f\right) \right]\Lt \nonumber \\
 &+ 10.9409722222 \Ltt ,
\end{align}
where $L_x\equiv\ln(\mu^2/m_x^2)$ and we set $N_f = 5$.
$\delta_{\NNNLO}^H$ has been obtained in \cite{Baikov:2006ch} for the top quark pole mass.
The terms in Eq.~\eqref{eq:deltaNNNLO} proportional to $19/8 + 2/3 N_f$ originate from the conversion to the \MSbar scheme using Eq.~\eqref{eq:mq_pole_to_mq}.
The expansions in Eqs.~\eqref{eq:hgg_nlo} and \eqref{eq:hgg_nnlo}
provide a good approximation of the full $\tau$ dependence for
$\tau \lesssim 0.7$, and are applied in \FD\ in this regime.
For the CP\-/odd Higgs $A$ we only implement the leading terms from
Refs.~\cite{Spira:1995rr,Chetyrkin:1998mw} in the large mass
expansion:
\begin{align}
  \delta_{\NLO}^A &= \frac{97}{4} - \frac{7}{6} N_f + \frac{33 - 2 N_f}{6}\ln \frac{\mu^2}{m_A^2} , \\
  \delta_{\NNLO}^A &= \frac{237311}{864} - \frac{529}{24} \zeta_2 - \frac{445}{8} \zeta_3 + 5 \ln \frac{m_A^2}{m_t^2} ,
\end{align}
where $m_t$ is the top quark pole mass and, as before, we set $N_f = 5$.
Moreover, in the expression for
$\delta_{\NNLO}^A$, we have also set $\mu = m_A$.
%

From the point of view of Higgs decays, where $\Phi \to gg$ is neither
the main discovery channel nor does it influence the total width very
significantly, the implemented corrections offer a good compromise
between accuracy and complexity.  Beyond the included QCD corrections,
further model\-/specific corrections are also known in the literature.
Two\-/loop EW corrections for the SM\-/like Higgs were found to be around
a few percent \cite{Actis:2008ts}, comparable with the four-loop QCD
corrections.  As they are small and model\-/dependent, we opt to not
include them in the current version of \FD.  The NLO QCD corrections
to squark contributions are also known by
Ref.~\cite{Muhlleitner:2006wx}.  For phenomenologically viable squark
masses the correction to the SM Higgs boson decays can be neglected
within the target precision of our program.  Although $\Phi \to gg$ is
important channel for the total width, it is not a discovery channel
and so the effect of those corrections gets significantly diluted.

In models with CP violation we include both the CP\-/even and the
CP\-/odd corrections by decomposing the coupling of the CP\-/mixture
state $\Phi$ as
$C_{\Phi t \bar{t}} = C_{H t \bar{t}} + \gamma^5 C_{A t \bar{t}}$ and
applying the corrections appropriately.

As mentioned at the end of \secref{sec:Phigamgam}, the above described implementation is an improvement over the previously available effective couplings calculation \cite{Staub:2016dxq}---it offers increased precision by including higher order terms in $\tau$ expansion for the CP-even part and generalizes handling of CP-violating models.

\subsubsection{Discussion of the Ward identity}
\label{sec:WardIdentity}

The general amplitude for a decay of a scalar $S$ into two vectors
$V_1$ and $V_2$ has the form
\begin{align}
  \amp_{S \to V_1 V_2} = \varepsilon^*_{1,\mu}
  \varepsilon^*_{2,\nu}  \Big ( & F_\eta \eta^{\mu\nu} + F_{11}
  p_1^\mu p_1^\nu + F_{12} p_1^\mu p_2^\nu
  + F_{21} p_1^\nu p_2^\mu
  + F_{22} p_2^\mu p_2^\nu
  + F_\epsilon \epsilon^{\mu\nu\alpha\beta} p_{1, \alpha} p_{2,\beta}
  \Big ) .
\end{align}
For physical vector bosons, since $\epsilon_i p_i = 0$, the terms
$F_{11}$, $F_{22}$ and $F_{12}$ do not contribute and only the three
coefficients $F_\eta$, $F_{21}$ and $F_\epsilon$ coefficients are
relevant.
For at least one massless vector in the final state, two of these,
$F_\eta$ and $F_{21}$, are related by the Ward identity
\begin{align}
\label{eq:WI}
  F_\eta = - p_1 p_2 \, F_{21} ,
\end{align}
reflecting the gauge invariance under the
$\epsilon_1^\mu\to\epsilon_1^\mu+\lambda
    p_1^\mu$ transformation.
A technical consequence of the Ward identity is that squared and
spin\-/summed matrix elements $\left|\amp_{S \to V_1 V_2}\right|^2$ in
Eqs.~\eqref{eq:SSV_square_massless},
\eqref{eq:SSV_square_massless-massive} are automatically non\-/negative.

In gauges other than the unitary gauge where loops with Goldstone
bosons have to be included, the fulfilment of the identity in Eq.~\eqref{eq:WI}
relies on the relation between $h G^+ G^-$ coupling and the external,
physical Higgs boson momentum. For example in the case of the SM, the $h G^+
G^-$ coupling is determined by $\lambda = m_h^2/v^2$, and the Ward
identity holds only if the same value for $m_h$ is used both for
evaluating the coupling and the external Higgs boson momentum
$p_h^2=m_h^2$. Otherwise the one-loop diagram with a mixed charged
Goldstone boson and $W$-boson loop contributes a term to the $F_\eta$
coefficient that violates the Ward identity.

This situation leads to a potential problem in calculations which work
in Feynman--'t Hooft gauge and
do not use the OS renormalization scheme
for the coupling constants: the external Higgs boson momentum is
always set to the squared Higgs pole mass, while internal couplings
may be evaluated as running couplings in some scheme, and the Ward
identity at some fixed order is numerically violated.\footnote{The
  violation is of course formally of higher order but can be
  numerically relevant.}
%
%
Since internally \FD\ and \FS\ work in $\MSbar/\DRbar$ renormalization
scheme and employ Feynman--'t Hooft gauge, our calculation of
$H\to\gamma V$, where $V = \gamma, Z$, suffers from this problem,
including the possibility of a negative decay width.

Numerical violation of Ward identity happens both in SM as well as in BSM models.
Our solution to this problem is as follows.  \FD\ uses the Ward identity to eliminate
$F_\eta$ in favor of $F_{21}$ before squaring the matrix element.
This ensures that the squared matrix element is non\-/negative.
Still, \FD\ calculates all form factors $F_i$ in order to test the
validity of the Ward identity. Specifically the squared matrix
elements obtained by eliminating $F_\eta$ or by eliminating $F_{21}$
in favor of $F_\eta$ are compared; if the difference is larger than
$10\%$ a warning is printed.\footnote{We regard violation of Ward identity as an indication of an existing theory uncertainty from unknown higher order contributions.}

%% file: tex/comparison.tex
\section{Comparison with existing tools}
\label{sec:comparison}

In this section we investigate the performance, generality and
accuracy of \FD and compare it with other established decay calculators.
Since higher order corrections to Higgs decays are not small, it is reasonable to expect moderate differences between results, depending on the details of the implementation.
This issue is also sometimes exacerbated by differences in the renormalization schemes in which input parameters are defined.

To discuss all the differences we apply \FD to a variety of models representing
qualitatively distinct features: the SM,  the singlet
extended SM, the type II THDM, the constrained MSSM
(CMSSM), the CP-violating MSSM, and the MRSSM. In each model we compute
Higgs-boson decays, and in the CMSSM we additionally compute squark
decays. The \FD results are compared with results from existing public
programs: \HD\ 6.53, \sHD, \THDECAY\ 1.1.4, \shit\ 1.5a, \sarah\
4.14.3/\spheno\ 4.0.4 and \SSUSY\ 4.1.9.

Among these programs, \sarah/\spheno is a generic code, like \FD, based on \sarah\
and able to treat large classes of BSM models. The other programs are
dedicated to specific models. Before discussing the detailed
comparisons we remark that the comparisons require scheme
translations because the different programs utilize different
renormalization schemes and different selections of input
parameters. Generally we adopt the following procedure.
In non\-/supersymmetric models we will start from $\MSbar$
parameters, use \FS to calculate physical masses and use them,
depending on the model, as inputs to \HD or \sHD (the exception is \THDECAY were we use directly \MSbar parameters).
In the case of the CMSSM comparison, due to the incompatibility between \HD and \FS generated SLHA output, we use \shit and its interface to \HD.\footnote{Internally, \shit\ 1.5a uses \HD\ 3.4.}
The technical details of the \sarah/\spheno setup are given in \appref{sec:spheno_setup}.

In the following comparisons the following SM parameters are used in almost
all programs:
\begin{equation}
\begin{aligned}
  m_t &= 172.76\unit{GeV}, & m_\tau &= 1.77686\unit{GeV}, & m_Z &= 91.1876\unit{GeV}, \\
  m_b^{\MSbar}(m_b^{\MSbar}) &= 4.18\unit{GeV}, & m_c^{\MSbar}(m_c^{\MSbar}) &= 1.27\unit{GeV}, &
  \alpha^{-1}(m_Z) &= 127.934, \\
  \alpha^{-1}(0) &= 137.036, & \as^{(5)}(m_Z) &= 0.1179, & V_{\abbrev{CKM}} &= \mathbb{1},\\
  \Gamma_W &= 2.085\unit{GeV}, & \Gamma_Z &= 2.4952\unit{GeV}.
\end{aligned}
\end{equation}
The exceptions are: \HD, which takes
$m_c^{\MSbar} (3\unit{GeV})$ and which we leave set to its default value of $0.986\unit{GeV}$ \cite{Djouadi:2018xqq},
and \shit which expects $m_c^{\pole}$ that we set to $1.67\unit{GeV}$ \cite{Zyla:2020zbs}.
The $m_W$ is a derived quantity in \FS, \sarah/\spheno and \SSUSY while for other codes we leave it to their respective defaults.

As various codes organize their calculations in slightly different ways, we take special care to compare like with like for a fair and uniform comparison.
To that end we set certain publicly available options as will be described through out this section and in \appref{sec:spheno_setup}.
However since the purpose is to provide a fair comparison of the results of public codes as they have been written, we do not make any changes in the source code of external programs.
We did however create custom steering files for the SSM in \sarah/\spheno and \FS and MRSSM in \sarah/\spheno.
In the case of the MRSSM this was necessary to make a meaningful comparison with the \FD\ extension of \FS, while in the case of the SSM we did it to match the \sHD.

For reproducibility, input and output files from all programs used in this comparison, as well as the model steering files mentioned above,  are attached to the \texttt{arXiv} version of this work.

\subsection{Standard Model}


\begin{table}[tb]
    \centering
    \begin{tabular}{l||cccc||c}
    \multirow{2}{*}{channel} & \multirow{2}{*}{\HD} & \sarah/\spheno & \sarah/\spheno & \multirow{2}{*}{\FD} &
    \multirow{2}{*}{LHCXSWG} 
    \\
    & & (\texttt{DECAY}) & (\texttt{DECAY1L}) & \\
    \hline
    $h \to b\bar{b}$ &
    2.384 & 2.152 & 1.832 & 2.353 & 2.381 \\

    $h \to W^+ W^-$ &
    $8.913 \cdot 10^{-1}$ & $9.301 \cdot 10^{-1}$ & --- & $8.498 \cdot 10^{-1}$ & $8.834 \cdot 10^{-1}$ \\

    $h \to \tau \bar \tau$ &
    $2.566 \cdot 10^{-1}$ & $2.499 \cdot 10^{-1}$ & $2.556 \cdot 10^{-1}$ & $2.505 \cdot 10^{-1}$ & $2.566 \cdot 10^{-1}$ \\

    $h \to c \bar c$ &
    $1.184 \cdot 10^{-1}$ & $1.015 \cdot 10^{-1}$ & $9.429 \cdot 10^{-2}$ & $1.163 \cdot 10^{-1}$ & $1.182 \cdot 10^{-1}$ \\

    $h \to Z Z$ &
    $1.087 \cdot 10^{-1}$ & $9.205 \cdot 10^{-2}$ & --- & $1.084 \cdot 10^{-1}$ & $1.084 \cdot 10^{-1}$
 \\

    \hline
    $h \to  g g$ &
    $3.347 \cdot 10^{-1}$ & $3.351 \cdot 10^{-1}$ & $1.508 \cdot 10^{-1}$ & $3.464 \cdot 10^{-1}$ & $3.354 \cdot 10^{-1}$ \\

    $h \to \gamma \gamma$ &
    $9.307 \cdot 10^{-3}$ & $1.088 \cdot 10^{-2}$ & $1.722 \cdot 10^{-2}$ & $9.146 \cdot 10^{-3}$ & $9.309 \cdot 10^{-3}$ \\

    $h \to \gamma Z$ &
    $6.298 \cdot 10^{-3}$ & --- & $< 0$ & $5.957 \cdot 10^{-3}$ & $6.320 \cdot 10^{-3}$ \\

    \hline
    total width &
    $4.111$ & $3.874$ & --- & 4.042 & $4.101$

    \end{tabular}
    \caption{Comparison of Higgs boson decay widths in the SM as
      calculated by \HD, \sarah/\spheno and \FD\ as implemented in
      \FS\ 2.6, as well as widths recommended by the LHC Higgs Cross
      Section Working Group for $m_h = 125.1\unit{GeV}$
      \cite{deFlorian:2016spz}.  All widths are given in MeV.}
    \label{tab:SM_comparison}
\end{table}

In \tabref{tab:SM_comparison} we show the comparison between different
programs for the SM Higgs boson decays, together with partial widths
recommended for this Higgs boson mass by the LHC Higgs Cross Section
Working Group (LHCXSWG) \cite{deFlorian:2016spz}.  We fix the physical
Higgs mass to $m_h=125.1\unit{GeV}$ \cite{Zyla:2020zbs} in all programs,
although this corresponds to different values of the quartic Higgs
coupling $\lambda$ in the different programs.  Dashes denote channels
which cannot be computed by the used version of \sarah/\spheno.  We
also do not report the total width from the \texttt{DECAY1L} block, as
\sarah/\spheno cannot compute some channels which are numerically important for the total width.

Generally there is very good agreement between \FD and the LHCXSWG
recommendations and \HD, which validates the results of \FD. The
largest relative deviation between \FD and \HD of below 5\% occurs for the
decay into $W^+W^-$ and $\gamma Z$.  The
relative deviations for the other channels are smaller than
$2\%$. The results
from the \sarah/\spheno \texttt{DECAY} block also agree well with the
LHCXSWG recommendations, except for the $h\to b\bar{b}$ and
$h\to \gamma\gamma$ channels, where deviations of around $10\%$ can be
observed.  As expected, the results from the \texttt{DECAY1L} block of
\sarah/\spheno\ show larger deviations due to missing Higgs\-/specific
higher\-/order corrections.  Note that several decay modes cannot be
computed by \sarah/\spheno\ in the given setup or lead to negative
outputs, as indicated in the table.%
\footnote{Except for \THDECAY, all programs discussed in this entire
  section output branching ratios and total
  Higgs decay widths rather than partial widths.  For these programs we
  define the entries of \tabref{tab:SM_comparison} and following
  tables by multiplying the total width and branching ratios; negative
  entries in the tables thus correspond to negative outputs for
  branching ratios.}

Systematic small differences exist between the various computations of
the $h\to VV$ ($V=W,Z$) decay modes.  For \HD\ and \FD\ the table
presents calculations including double off\-/shell decays of the Higgs
boson, even if $2 m_V > m_h > m_V$, as it matches more closely what is
actually reported as the $\Gamma(h \to VV)$ width in the literature
(corresponding to the default in both programs). In contrast, \spheno\
only includes single off-shell decays to gauge bosons.\footnote{As
  explained in \appref{sec:runtime_settings}, the behavior of \sarah\
  can be reproduced in \FD by setting the flag \texttt{FlexibleDecay[4] = 1}.}



\begin{figure}[tbh!]
  \centering
  \begin{minipage}{0.6\textwidth}
    \includegraphics[height=0.3\textheight]{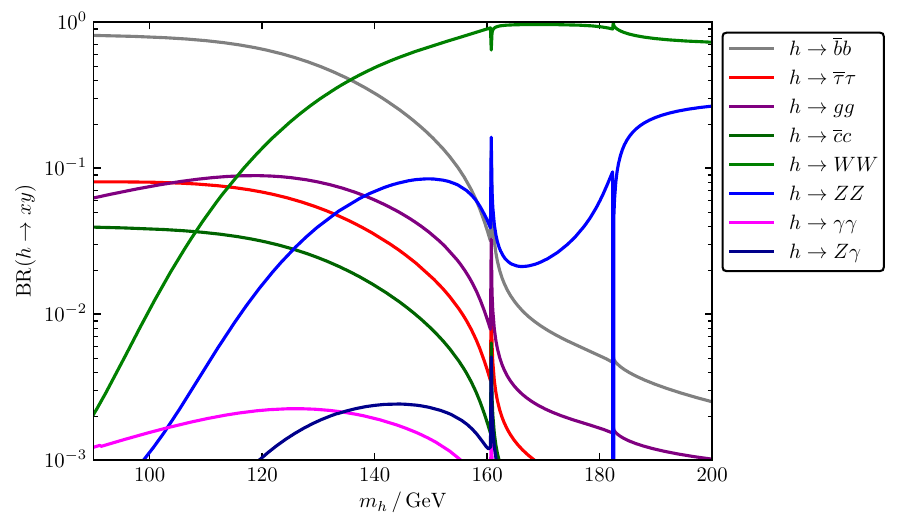}\\
    \includegraphics[height=0.3\textheight]{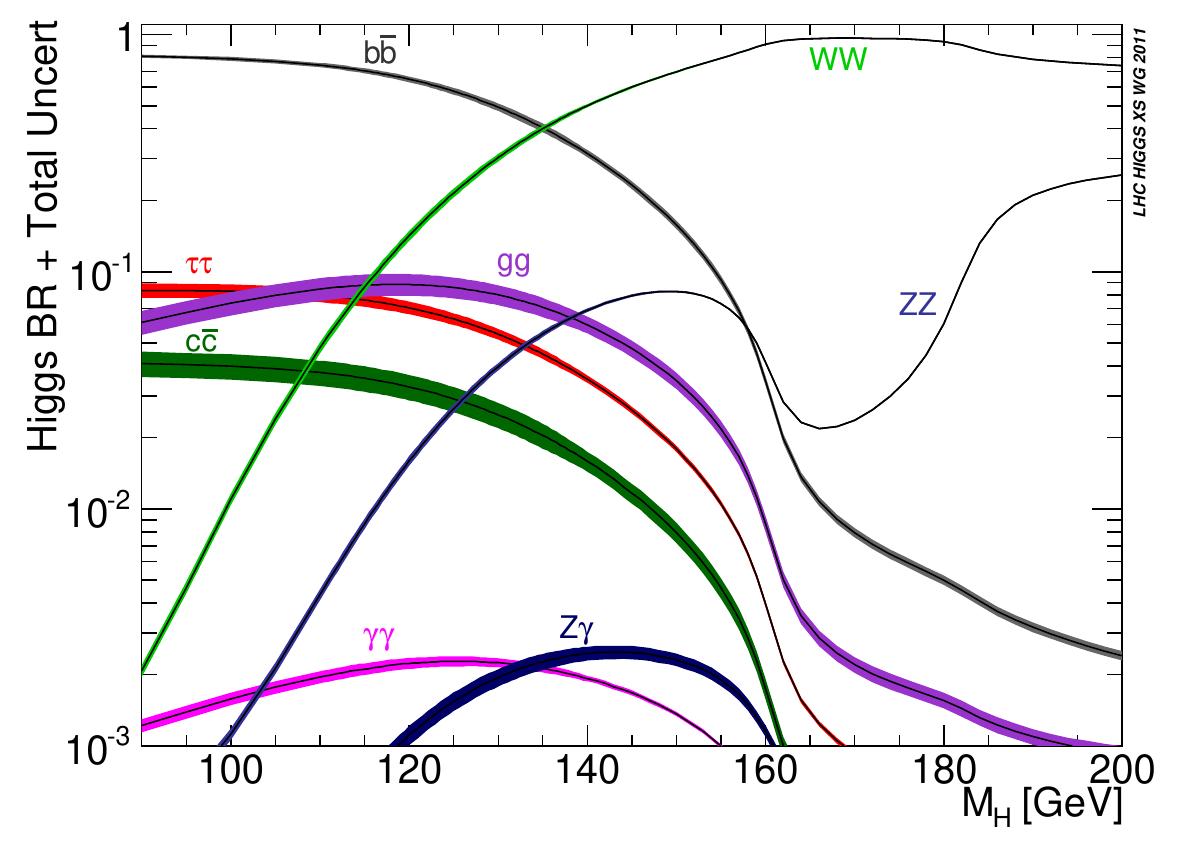}
  \end{minipage}
  \caption{SM results for Higgs branching ratios with those from the LHC Higgs Cross Section Working Group in the low mass range \cite{Denner:2011mq}.}
  \label{fig:SM_BRs_FD_vs_HXSWG_lowmass}
\end{figure}
In \figref{fig:SM_BRs_FD_vs_HXSWG_lowmass} we look at the SM Higgs
branching ratios over the Higgs mass range $m_h \in
[80,200]\unit{GeV}$ and compare to the results of the Higgs Cross
Section Working Group, which were presented in
Ref.\ \cite{Denner:2011mq}.  As can be seen in the figure our results
are in good agreement with Ref.\ \cite{Denner:2011mq} over most
masses.
However a notable difference is that our results contain noticeable
kinks at thresholds associated with twice the $W$ and twice the
$Z$ masses.
This reflects our treatment of the off-shell Higgs decays.
The Higgs XSWG uses \texttt{Prophecy4f}~\cite{Bredenstein_2006,Bredenstein_2006_2,Bredenstein_2007} which computes the $W/Z \to 4f$ process therefore not exhibiting any thresholds.

\subsection{Real singlet extension of the Standard Model}

In this section we consider the comparison to the $\mathbb{Z}_2$ real
scalar singlet extension of the SM with the potential of Eq.~(2.11) in
Ref.\ \cite{Costa:2015llh}, where the real-singlet gets a non\-/zero
VEV.
In this model, the SM\-/like mass eigenstate has reduced couplings due
to the admixture of the scalar singlet, which couples neither to gauge
bosons nor to quarks or leptons. The mixing is parametrized by an
angle $\alphah$.
We use a benchmark point inspired by the RxSM.B1 reference point of Ref.~\cite{Costa:2015llh}: $\lambda \approx 0.420$, $\lambda_S = 0.710$,
$\lambda_{HS}\approx-0.594$, and $v_s = 140.3\unit{GeV}$ at the scale $\mu = 125.1$.\footnote{This corresponds to $\lambda_S = 8.52$ and $\lambda = 0.84$ in Ref.~\cite{Costa:2015llh} because of the difference in normalization of the Higgs quartic coupling.}
In \FS, one obtains at the one-loop level $m_h = 125.1\unit{GeV}$, the
heavier Higgs mass $m_{h_2} \approx 265.3\unit{GeV}$ and a one-loop mixing
angle $\alphah \approx -0.4284$.
\begin{table}[tb]
    \centering
    \begin{tabular}{l||ccccc}
    \multirow{2}{*}{channel} & \multirow{2}{*}{\sHD} & \sarah/\spheno & \sarah/\spheno & \multirow{2}{*}{\FD}  \\
    & & (\texttt{DECAY}) & (\texttt{DECAY1L}) & \\

    \hline
    $h \to b\bar{b}$ &
    1.994 & 1.786 & 1.532 & 1.868 \\

    $h \to W^+ W^-$ &
    $7.063 \cdot 10^{-1}$ & $7.486 \cdot 10^{-1}$ & --- & $6.753 \cdot 10^{-1}$ \\

    $h \to \tau \bar \tau$ &
    $2.142 \cdot 10^{-1}$ & $2.072 \cdot 10^{-1}$ & $2.115 \cdot 10^{-1}$ & $1.987 \cdot 10^{-1}$ \\

    $h \to c \bar c$ &
    $9.762 \cdot 10^{-2}$ & $8.423 \cdot 10^{-2}$ & $7.789 \cdot 10^{-2}$ & $9.230 \cdot 10^{-2}$ \\

    $h \to Z Z$ &
    $8.786 \cdot 10^{-2}$ & $7.441 \cdot 10^{-2}$ & --- & $8.595 \cdot 10^{-2}$ \\

    \hline
    $h \to  g g$ &
    $2.636 \cdot 10^{-1}$ & $2.763 \cdot 10^{-1}$ & $1.412 \cdot 10^{-1}$ & $2.749 \cdot 10^{-1}$ \\

    $h \to \gamma \gamma$ &
    $7.830 \cdot 10^{-3}$ & $8.914 \cdot 10^{-3}$ & $1.322 \cdot 10^{-2}$ & $7.349 \cdot 10^{-3}$ \\

    $h \to \gamma Z$ &
    $5.212 \cdot 10^{-3}$ & --- & $<0$ & $4.754 \cdot 10^{-3}$ \\

    \hline
    total width & 3.378 & 3.187 & --- & 3.208

    \end{tabular}
    \caption{Comparison of Higgs boson decay widths in the real
      singlet extended SM as calculated by \FD, \sHD and
      \sarah/\spheno.  All widths are given in MeV.}
    \label{tab:SSM_comparison}
\end{table}
The results of the comparison are gathered in \tabref{tab:SSM_comparison}.
As expected, all partial widths are reduced compared to the case of
the SM in \tabref{tab:SM_comparison}, and the relative agreement
between \FD and \sHD and \sarah/\spheno is similar to the one between \FD and \HD and \sarah/\spheno in case of the
SM.
Even with the large value of $\lambda_S$, which enhances the BSM effects, the comparison is still as good as in the case of the SM implying good control over pure-BSM corrections.

\subsection{Two-Higgs Doublet Model of Type II}

The THDM is an important extension of the
SM. Due to the second Higgs doublet its Higgs and Yukawa sectors are
rich and, especially in the most general case, involve a large number
of free parameters. To be specific we focus here on the THDM of type
II, which postulates a discrete symmetry to restrict the Higgs
potential and Yukawa couplings.

In this model, the SM\-/like Higgs state arises from mixing of the two
doublets described by two mixing angles $\alphah$ and $\beta$.
Compared to the SM, its tree-level couplings to gauge bosons are
reduced by a factor $\sin(\beta-\alphah)$ and the couplings to
down-type and up-type fermions are governed by
$(-\sin\alphah/\cos\beta)$ and $(\cos\alphah/\sin\beta)$,
respectively.  In the following we consider the parameter point given
by
\begin{equation}
\begin{aligned}
\lambda_1 &= 0.05, &
\lambda_2 &= 0.13155, &
\lambda_3 &= 0.13, &
\lambda_4 &= 0.14, &
\lambda_5 &= 0.2, \\
\lambda_6 &= 0, &
\lambda_7 &= 0, &
m_{12}^2  & = 2 \cdot 10^4\unit{GeV}^2, &
\tan\beta &= 13
\end{aligned}\label{eq:THDM-PP}%
\end{equation}
in the potential notation of%
\footnote{The default type II THDM models in \FS\ and \sarah\ use
  different normalizations of $\lambda_{1,2}$ and sign convention for
  $m_{12}^2$ than \THDECAY:
  $(\lambda_{1,2})_{\text{\FS},\text{\sarah}} = \frac{1}{2}
  (\lambda_{1,2})_{\text{\THDECAY}}$,
  $-(m_{12}^2)_{\text{\sarah}} =
  (m_{12}^2)_{\text{\FS},\text{\THDECAY}}$.}
\begin{align}
\begin{split}
V ={}& m_{11}^2 \Phi_1^2 + m_{22}^2 \Phi^2 + \lambda_1 \Phi^4 + \lambda_2 \Phi^2 + \lambda_3 |\Phi_2|^2 |\Phi_1|^2 + \lambda_4 |\Phi_1 \Phi_2|^2 \\
&+ \left(-m_{12}^2 \Phi_1^\dagger \Phi_2 + \frac{\lambda_5}{2} (\Phi_2 \Phi_1)^2 + \text{h.c.} \right).
\end{split}
\end{align}
In the following we compare the calculations of \FS, \sarah/\spheno
and \THDECAY\ (with \texttt{RENSCHEM = 16}) in the \MSbar\ scheme.
The parameter point \eqref{eq:THDM-PP} corresponds to the on\-/shell
quantities
%
\begin{equation}
  \begin{aligned}
    m_h &\approx 125.1\unit{GeV}, &
    m_{h_2} &\approx 511.4\unit{GeV}, &
    m_A &= 499.1\unit{GeV}, &
    m_{H^+} &= 501.3\unit{GeV}, \\
    \alphah &\approx -7.307 \cdot 10^{-2} .
  \end{aligned}
\end{equation}
\tabref{tab:THDM_comparison} shows the Higgs decay channels in the
THDM-II as calculated by different programs.
As \THDECAY also optionally includes EW corrections, which are not present in other codes, for the sake of comparison we report results only with QCD ones.
The behavior of \FD,
\THDECAY\ and \sarah/\spheno is similar to the previous cases, and
there is very good agreement between all these programs with the
expected, slightly worse behavior of the \texttt{DECAY1L} block output
of \sarah/\spheno. The THDM\-/specific program \THDECAY shows similarly
good agreement, but there is an $\sim 5\%$ upward shift in its
$h\to b\bar{b}$ prediction compared to the other programs.
\begin{table}[ht]
    \centering
    \begin{tabular}{l||cccc}
    \multirow{2}{*}{channel} & \multirow{2}{*}{\THDECAY} & \sarah/\spheno & \sarah/\spheno & \multirow{2}{*}{\FD}\\
    & & (\texttt{DECAY}) & (\texttt{DECAY1L}) \\
    \hline

    $h \to b \bar{b}$ &
    2.237 & 2.110 & 1.759 & 2.121 \\

    $h \to W^+ W^-$ &
    $8.889 \cdot 10^{-1}$ & $8.321 \cdot 10^{-1}$ & --- & $8.504 \cdot 10^{-1}$ \\

    $h \to \tau \bar{\tau}$ &
    $2.406 \cdot 10^{-1}$ & $2.445 \cdot 10^{-1}$ & $2.483 \cdot 10^{-1}$ & $2.256 \cdot 10^{-1}$ \\

    $h \to c \bar{c}$ &
    $1.210 \cdot 10^{-1}$ & $1.014 \cdot 10^{-1}$ & $8.894 \cdot 10^{-2}$ & $1.164 \cdot 10^{-1}$ \\

    $h \to Z Z$ &
    $1.114 \cdot 10^{-1}$ & $8.124 \cdot 10^{-2}$ & --- & $1.084 \cdot 10^{-1}$ \\

    \hline
    $h \to g g$ &
    $3.262 \cdot 10^{-1}$ & $3.339 \cdot 10^{-1}$ & $1.785 \cdot 10^{-1}$ & $3.472 \cdot 10^{-1}$ \\

    $h \to \gamma \gamma$ &
    $1.005 \cdot 10^{-2}$ & $1.049 \cdot 10^{-2}$ & $1.572 \cdot 10^{-2}$ & $9.130 \cdot 10^{-3}$ \\

    $h \to \gamma Z$ &
    $6.814 \cdot 10^{-3}$ & --- & < 0 & $5.961 \cdot 10^{-3}$\\

    \hline
    total width &
    $3.944$ & 3.715 & --- & 3.786
    \end{tabular}
    \caption{Comparison of Higgs boson decay widths in the type II
      THDM as calculated by \THDECAY, \sarah/\spheno and \FD.  All
      widths are given in MeV.}
    \label{tab:THDM_comparison}
\end{table}

\subsection{Constrained Minimal Supersymmetric Standard Model \label{sec:cmssm}}

\begin{table}[ht]
    \centering
    \begin{tabular}{l||ccccc}
    \multirow{2}{*}{channel} & \multirow{2}{*}{\shit} & \multirow{2}{*}{\SSUSY} & \sarah/\spheno & \sarah/\spheno & \multirow{2}{*}{\FD} \\
    & & & (\texttt{DECAY}) & (\texttt{DECAY1L}) &\\
    \hline
    $h \to b\bar{b}$ &
    2.662 & 3.843 & 2.403 & 1.541 & 2.348 \\

    $h \to W^+ W^-$ &
    $8.342 \cdot 10^{-1}$ & $6.751 \cdot 10^{-1}$ & $5.887 \cdot 10^{-1}$ & --- & $8.141 \cdot 10^{-1}$ \\

    $h \to \tau \bar \tau$ &
    $2.595 \cdot 10^{-1}$ & $2.726 \cdot 10^{-1}$ & $2.778 \cdot 10^{-1}$ & $2.355 \cdot 10^{-1}$ & $2.499 \cdot 10^{-1}$ \\

    $h \to c \bar c$ &
    $1.183 \cdot 10^{-1}$ & $2.235 \cdot 10^{-1}$ & $1.031 \cdot 10^{-1}$ & $1.073 \cdot 10^{-1}$ & $1.160 \cdot 10^{-1}$ \\

    $h \to Z Z$ &
    $1.060 \cdot 10^{-1}$ & $7.606 \cdot 10^{-2}$ & $5.882 \cdot 10^{-2}$ & --- & $1.032 \cdot 10^{-1}$ \\

    \hline
    $h \to  g g$ &
    $2.731 \cdot 10^{-1}$ & $2.760 \cdot 10^{-1}$ & $2.993 \cdot 10^{-1}$ & $9.555 \cdot 10^{-2}$
 & $3.434 \cdot 10^{-1}$ \\

    $h \to \gamma \gamma$ &
    $9.439 \cdot 10^{-3}$ & $1.052 \cdot 10^{-2}$ & $8.580 \cdot 10^{-3}$ & $1.024 \cdot 10^{-2}$ & $9.940 \cdot 10^{-3}$ \\

    $h \to Z \gamma$ &
    $6.316 \cdot 10^{-3}$ & $6.779 \cdot 10^{-3}$ & --- & $4.303 \cdot 10^{-1}$ & $6.098 \cdot 10^{-3}$ \\

    \hline
    total width &
    4.272 & 5.386 & 3.741 & --- & 3.993
    \end{tabular}
    \caption{
    Comparison of Higgs boson decay widths in the MSSM as calculated by \FD, \HD  via \shit, \SSUSY and \sarah/\spheno (for CMSSM SPS1a slope with $m_0 = 1.4\unit{TeV}$).
    All widths in MeV.
    }
    \label{tab:CMSSM_comparison_h}
\end{table}

The MSSM is a very popular
extension of the SM. The CMSSM imposes
simple universality relationships on the fundamental MSSM input
parameters. Since this is implemented in many public programs, it is a
good choice that allows a well\-/defined comparison. The tree\-/level
Higgs sector of the MSSM is that of the type II THDM, with additional
constraints from supersymmetry.  At the loop level, non\-/type
II\-/like contributions arise via quantum corrections.

We follow the standard notation given in, e.g., Ref.\
\cite{Martin:1997ns}.
For the following comparison we use the parameter point
$m_0 = 1.4\unit{TeV}$, $m_{1/2} = 3.5\unit{TeV}$, $\tan\beta = 10$,
$\sign(\mu) = +1$, $A_0 = -1.4\unit{TeV}$.

\tabref{tab:CMSSM_comparison_h} shows a comparison of \FD\ and a
selection of well\-/known public programs that work in the \DRbar\ scheme
(for the status of further public programs we refer to the literature
\cite{Bahl:2018qog,Domingo:2019jsc}).  Note that \shit\ 1.5a uses \HD\
3.4 internally and implements QCD corrections up to the two-loop level
to $h \to gg$.

The difference in the $h \to c \bar{c}$ comes from the difference in
the running charm mass.  \shit accepts as input a pole charm mass,
which we set to $1.67\unit{GeV}$ and which is internally converted by
\shit to $m_c(m_h) = 0.69\unit{GeV}$.  Meanwhile \FS takes as input
$m_c^{\MSbar}(m_c^{\MSbar})$, which we set to $1.27\unit{GeV}$ which
then corresponds to $m_c(m_h) = 0.62\unit{GeV}$.

\subsection{Minimal R-symmetric Supersymmetric Standard Model}

The MRSSM is a non\-/minimal supersymmetric model with an unbroken
U(1) R\-/symmetry, which has received considerable attention in recent
years.  Its Higgs sector is particularly rich, containing two Higgs
doublets, one complex singlet and one complex triplet (which arise as
double superpartners of the SU(2)$\times$U(1) gauge bosons).  There
are no dedicated MRSSM public programs, but both \FS/\FD\ and
\sarah/\spheno\ are distributed with model files that generate
MRSSM\-/specific program for determining its mass spectrum and decay
widths.

For the following numerical comparison, we choose the parameter point
BMP2 from Ref.~\cite{Diessner:2014ksa}.  The comparison between the
results of the programs is shown in \tabref{tab:MRSSM_comparison} for
the SM-like Higgs boson $h$ and in \tabref{tab:MRSSM_comparison2} for
the next heavier state $H_2$.  Their masses as computed by \FS are
$124.5\unit{GeV}$ and $921.5\unit{GeV}$, respectively.\footnote{The
  corresponding masses in \sarah/\spheno are $123.8\unit{GeV}$ and
  $937.4\unit{GeV}$, respectively.}
Overall we see similar agreement between \FD\ and \sarah/\spheno
(\texttt{DECAY} block) and a similar behavior from \sarah/\spheno
(\texttt{DECAY1L} block) as in the previous cases in the case of light Higgs.

\begin{table}[h]
    \centering
    \begin{tabular}{l||ccc}
    \multirow{2}{*}{channel} & \sarah/\spheno & \sarah/\spheno & \multirow{2}{*}{\FD} \\
    & (\texttt{DECAY}) & (\texttt{DECAY1L}) & \\
    \hline
    $h \to b \bar{b}$ &
    $2.460$ & $2.079$ & $2.433$ \\

    $h \to W^+ W^-$ &
    $7.234 \cdot 10^{-1}$ & --- & $7.856 \cdot 10^{-1}$ \\

    $h \to \tau \bar \tau$ &
    $2.851 \cdot 10^{-1}$ & $2.601 \cdot 10^{-1}$ & $2.587 \cdot 10^{-1}$ \\

    $h \to c \bar{c}$ &
    $1.046 \cdot 10^{-1}$ & $1.273 \cdot 10^{-1}$ & $1.158 \cdot 10^{-1}$ \\

    $h \to Z Z$ &
    $7.686 \cdot 10^{-2}$ & --- & $9.987 \cdot 10^{-2}$ \\

    \hline

    $h \to g g$ &
    $3.186 \cdot 10^{-1}$ & $1.353 \cdot 10^{-1}$ & $3.462 \cdot 10^{-1}$ \\

    $h \to \gamma \gamma$ &
    $8.402 \cdot 10^{-3}$ & $1.007 \cdot 10^{-2}$ & $9.140 \cdot 10^{-3}$ \\

    $h \to \gamma Z$ &
    --- & $1.671 \cdot 10^{-1}$ & $5.588 \cdot 10^{-3}$ \\

    \hline
    total width &
    3.979  & --- & 4.056

    \end{tabular}
    \caption{Comparison of SM-like Higgs boson decay widths as
      calculated by \FD and \sarah/\spheno in the MRSSM for BMP2 of \cite{Diessner:2014ksa}.  All widths are given in
      MeV.}
    \label{tab:MRSSM_comparison}
\end{table}

\begin{table}[h]
    \centering
    \begin{tabular}{l||ccc}
    \multirow{2}{*}{channel} & \sarah/\spheno & \sarah/\spheno & \multirow{2}{*}{\FD} \\
    & (\texttt{DECAY}) & (\texttt{DECAY1L}) & \\
    \hline
    $H_2 \to b \bar{b}$ &
    1.274 & $7.950 \cdot 10^{-1}$ & $1.253$ \\

    $H_2 \to t \bar{t}$ &
    $4.163 \cdot 10^{-1}$ & $3.714 \cdot 10^{-1}$ & $4.178 \cdot 10^{-1}$ \\

    $H_2 \to \tau \bar{\tau}$ &
    $1.974 \cdot 10^{-1}$ & $1.770 \cdot 10^{-1}$ & $1.900 \cdot 10^{-1}$ \\

    $H_2 \to \bar{\chi}_2^0 \chi_1^0 $ &
    $1.560 \cdot 10^{-1}$ & $2.342 \cdot 10^{-4}$ & $1.557 \cdot 10^{-1}$ \\

    $H_2 \to \bar{\chi}_1^0 \chi_2^0 $ &
    $1.560 \cdot 10^{-1}$ & $1.692 \cdot 10^{-1}$ & $1.557 \cdot 10^{-1}$ \\

    $H_2 \to h h$ &
    $3.743 \cdot 10^{-3}$ & $5.624 \cdot 10^{-3}$ & $3.420 \cdot 10^{-3}$ \\

    $H_2 \to \mu \bar{\mu}$ &
    $6.829 \cdot 10^{-4}$ & $6.079 \cdot 10^{-4}$ & $6.719 \cdot 10^{-4}$ \\

    $H_2 \to W^+ W^-$ &
    $1.454 \cdot 10^{-3}$ & --- & $5.782 \cdot 10^{-4}$ \\

    $H_2 \to s \bar{s}$ &
    $4.746 \cdot 10^{-4}$ & $5.744 \cdot 10^{-3}$ & $5.756 \cdot 10^{-4}$ \\

    $H_2 \to Z Z$ &
    $7.229 \cdot 10^{-4}$ & --- & $3.014 \cdot 10^{-4}$ \\

    $H_2 \to \bar{\chi}_1^+ \chi_1^+ $ &
    $3.390 \cdot 10^{-4}$ & $3.209 \cdot 10^{-4}$ & $2.592 \cdot 10^{-4}$ \\

    $H_2 \to \bar{\rho}_1 \rho_1 $ &
    $2.747 \cdot 10^{-4}$ & $2.343 \cdot 10^{-4}$ & $1.746 \cdot 10^{-4}$ \\

    $H_2 \to \bar{\chi}_1^0 \chi_1^0 $ &
    $2.480 \cdot 10^{-7}$ & $1.149 \cdot 10^{-6}$ & $4.793 \cdot 10^{-5}$ \\

    $H_2 \to \bar{\chi}_2^0 \chi_2^0 $ &
    $1.678 \cdot 10^{-8}$ & $3.437 \cdot 10^{-7}$ & $2.880 \cdot 10^{-5}$ \\

    \hline
    $H_2 \to g g$ &
    $4.339 \cdot 10^{-4}$ & $2.499 \cdot 10^{-4}$ & $2.599 \cdot 10^{-4}$ \\

    $H_2 \to \gamma \gamma$ &
    $1.520 \cdot 10^{-6}$ & $1.847 \cdot 10^{-6}$ & $1.845 \cdot 10^{-6}$ \\

    $H_2 \to \gamma Z$ &
    --- & $7.762 \cdot 10^{-7}$ & $2.735 \cdot 10^{-7}$\\

    $H_2 \to h \gamma$ &
    --- & $3.759 \cdot 10^{-4}$ & 0 \\

    $H_2 \to A_2 \gamma$ &
    --- & $6.157 \cdot 10^{-10}$ & 0\\

    \hline
    total width &
    2.208 & --- & 2.179

    \end{tabular}
    \caption{Comparison of heavy Higgs boson decay widths in the MRSSM as
      calculated by \FD and \sarah/\spheno.  All widths are given in
      GeV.}
    \label{tab:MRSSM_comparison2}
\end{table}

In the case of heavy Higgs decays we observe much worse agreement.
The disagreement of some of the important channels can be tracked down
\begin{itemize}
 \item disagreement in $H_2 \to g g$ comes from \sarah including in the \texttt{DECAY} column higher order corrections that are not appropriate, since they were derived in the limit $m_t \gg m_H$, whereas in this scenario we actually have  $m_t \ll m_{H_2}$,
 \item disagreement between \FD and \texttt{DECAY} in $H_2 \to \gamma \gamma$ comes from the NLO QCD correction. This correction in \sarah is negative and around < 5\% while above the 2 top threshold it should be positive and $\gtrsim 20\%$ \cite{Djouadi:1993ji,Melnikov:1993tj},
 \item \sarah result for $H_2 \to h \gamma$ cannot be correct as this process is forbidden by CP-symmetry,
  \item result for $H_2 \to A_2 \gamma$ (where $A_2$ is the lightest, physical pseudoscalar Higgs boson) should be 0 as the amplitude for physical (i.e. transversal) photons that also fulfils the Ward identity is identically 0.
\end{itemize}

\subsection{Squark decays in the Minimal Supersymmetric Standard Model}

The \FD\ module is also capable for automatically generating $1\to2$
decays for BSM particles.  As an example, in
Tab.~\ref{tab:CMSSM_comparison_b1} and \ref{tab:CMSSM_comparison_b2} we present results for decays of the first and second sbottom
squarks in the CMSSM parameter point discussed in \secref{sec:cmssm}.
For a uniform comparison with other codes, we disable
higher\-/order corrections in
\texttt{SDECAY}~\cite{Muhlleitner:2003vg}.

\begin{table}[ht]
    \centering
    \begin{tabular}{l||ccccc}
    \multirow{2}{*}{channel} & \multirow{2}{*}{\shit} & \multirow{2}{*}{\SSUSY} & \sarah/\spheno & \multirow{2}{*}{\FD} \\
    & & & (\texttt{DECAY}) & \\
    \hline
    $\tilde{b}_1 \to \tilde \chi_1^- t$ &
    26.931 & 26.569 & 27.061 & 26.380 \\

    $\tilde{b}_1 \to \tilde \chi_2^- t$ &
    26.690 & 33.160 & 25.931 & 26.371 \\

    $\tilde{b}_1 \to \tilde t_1 W^-$ &
    23.434 & 23.906 & 23.903 & 23.635 \\

    $\tilde{b}_1 \to \tilde \chi^0_2 b$ &
    13.389 & 13.318 & 13.419 & 13.239 \\

    $\tilde{b}_1 \to \tilde \chi^0_1 b$ &
    $7.617 \cdot 10^{-1}$ & $7.635 \cdot 10^{-1}$ & $6.807 \cdot 10^{-1}$ & $7.650 \cdot 10^{-1}$ \\

    $\tilde{b}_1 \to \tilde \chi^0_4 b$ &
    $3.420 \cdot 10^{-1}$ & $4.308 \cdot 10^{-1}$ & $3.927 \cdot 10^{-1}$ & $3.575 \cdot 10^{-1}$ \\

    $\tilde{b}_1 \to \tilde \chi^0_3 b$ &
    $3.078 \cdot 10^{-1}$ & $4.010 \cdot 10^{-1}$ & $3.404 \cdot 10^{-1}$ & $3.311 \cdot 10^{-1}$ \\

    \hline
    total width &
    91.856 & 98.548 & 91.728 & 91.079
    \end{tabular}
    \caption{Comparison of $\tilde{b}_1$ decay widths in the MSSM as
      calculated by \shit, \SSUSY\ and \sarah/\spheno and \FD.  All
      widths are given in GeV.}
    \label{tab:CMSSM_comparison_b1}
\end{table}

\begin{table}[ht]
    \centering
    \begin{tabular}{l||ccccc}
    \multirow{2}{*}{channel} & \multirow{2}{*}{\shit} & \multirow{2}{*}{\SSUSY} & \sarah/\spheno & \multirow{2}{*}{\FD} \\
    & & & (\texttt{DECAY}) & \\
    \hline
    $\tilde{b}_2 \to \tilde{\chi}_1^0 b$&
    3.126 & 3.115 & 3.152 & 3.119 \\

    $\tilde{b}_2 \to \tilde{\chi}_2^- t$&
    $7.730 \cdot 10^{-1}$ & 1.046 & $8.021 \cdot 10^{-1}$ & $8.702 \cdot 10^{-1}$ \\

    $\tilde{b}_2 \to \tilde{\chi}_4^0 b$&
    $3.560 \cdot 10^{-1}$ & $4.657 \cdot 10^{-1}$ & $3.821 \cdot 10^{-1}$ & $3.849 \cdot 10^{-1}$ \\

    $\tilde{b}_2 \to \tilde{\chi}_3^0 b$&
    $3.487 \cdot 10^{-1}$ & $4.550 \cdot 10^{-1}$ & $3.719 \cdot 10^{-1}$ & $3.747 \cdot 10^{-1}$ \\

    $\tilde{b}_2 \to \tilde{t}_1 W^-$&
    $6.164 \cdot 10^{-2}$ & $9.891 \cdot 10^{-2}$ & $7.273 \cdot 10^{-2}$ & $1.099 \cdot 10^{-1}$ \\

    $\tilde{b}_2 \to \tilde{t}_2 W^-$&
    $3.790 \cdot 10^{-2}$ & $5.942 \cdot 10^{-2}$ & $3.489 \cdot 10^{-2}$ & $6.608 \cdot 10^{-2}$ \\

    $\tilde{b}_2 \to \tilde{\chi}_1^- t$&
    $1.671 \cdot 10^{-2}$ & $2.906 \cdot 10^{-2}$ & $1.916 \cdot 10^{-2}$ & $3.939 \cdot 10^{-2}$ \\

    $\tilde{b}_2 \to \tilde{b}_1 Z$&
    $2.104 \cdot 10^{-2}$ & $3.297 \cdot 10^{-2}$ & $1.837 \cdot 10^{-2}$ & $3.715 \cdot 10^{-2}$ \\

    $\tilde{b}_2 \to \tilde{b}_1 h_1$&
    $2.553 \cdot 10^{-2}$ & $3.319 \cdot 10^{-2}$ & $2.009 \cdot 10^{-2}$ & $2.633 \cdot 10^{-2}$ \\

    $\tilde{b}_2 \to \tilde{\chi}_2^0 b$&
    $8.435 \cdot 10^{-3}$ & $1.461 \cdot 10^{-2}$ & $9.808 \cdot 10^{-3}$ & $1.972 \cdot 10^{-2}$ \\

    \hline
    total width &
    4.775 & 5.349 & 4.883 & 5.048 \\
    \end{tabular}
    \caption{Comparison of $\tilde{b}_2$ decay widths in the MSSM as
      calculated by \shit, \SSUSY\ and \sarah/\spheno and \FD.  All
      widths are given in GeV.}
    \label{tab:CMSSM_comparison_b2}
\end{table}

For this parameter point, the pole masses as calculated by \FD are as follows: $m_{\tilde b_2} = 6.074\unit{TeV}$,  $m_{\tilde b_1} = 5.874\unit{TeV}$, $m_{\tilde t_1} = 4.937\unit{TeV}$, $m_{\tilde t_2} = 5.882\unit{TeV}$, $m_{\tilde \chi_i^0} = \{1.571, 2.889, 3.715, 3.719\}\unit{TeV}$ and $m_{\tilde \chi_i^-} = \{2.889, 3.720\}\unit{TeV}$.

Overall one sees a rather good agreement between most of the codes.
Some of the bigger differences in \tabref{tab:CMSSM_comparison_b2}, e.g.\ for $\tilde{b}_2 \to \tilde{t}_2 W^-$, originate from differences in sub-leading, off-diagonal entries in mass matrices.
The $\tilde{b}_2$ is mostly a right handed squark with only a percent level admixture of a left-handed component. Therefore all of its weak decays suffer from a large uncertainty as an exact value of this mixing will vary substantially from code to code.

%% file: tex/limitations.tex
\section{Limitations and future extensions}
\label{sec:limitations}

Although \FD is designed with generality in mind, the current version still has some limitations on the decays or models we support. These are:
\begin{itemize}
\item Decays of fermions and vector bosons are not supported.
\item Decays of a color octet into color octets are not
  supported. Other combinations, like for example
  $8\to3\otimes\bar{3}$ or $3 \to 8 \otimes3$ are supported.
\item Decays containing vertices which cannot be decomposed into a
  single product of Lorentz and color structure, e.g.\ the
  quartic\-/gluon vertex, are not supported.
\item In general only $1 \to 2$ decays are supported. The exception
  are the decays of scalar and pseudoscalar Higgs bosons to $ZZ$ and
  $W^+ W^-$ pairs, where we include single and double off\-/shell
  decays, assuming SM decays of $W^\pm$ and $Z$ bosons.
\item Models with forced fermion flavor-conservation (where leptons and quarks are not combined into their respective multiplets --- \texttt{NoFV} models in \FS\ nomenclature) are not supported.
\item It is assumed that the BSM model contains the SM as a subset.
  For example, BSM models where a SM\-/like weak mixing angle cannot
  be defined are not supported.
\end{itemize}
Future extensions removing these limitations and widening the scope of
\FD are planned. In particular adding decays for fermions and vectors
is a priority.  In the near future we will also add an interface to 
\HB\ and \HS\ to make it easier to use the output of
\FD in those programs.  The program is actively developed and carefully
tested for correctness with nightly units tests, minimising the
possibility of bugs being introduced during development.  Users who
find they need new features or nonetheless still encounter bugs can
report these to the \FS developers at
\url{https://github.com/FlexibleSUSY/FlexibleSUSY/issues}
or by contacting one of the developers directly.

%% file: tex/conclusions.tex
\section{Conclusions}
\label{sec:conclusions}

We have presented \FD, a computer program for the calculation of
scalar decays in a broad class of BSM models with a special emphasis on
precise predictions of Higgs boson decays. \FD is fully integrated
into the \FS framework.  It enables decay calculations for existing
\FS models with minimal modifications to the setup, and is switched on
by default for many models with new installations of \FS.

To achieve high precision of scalar and pseudoscalar Higgs boson
decays for the following final states, we have included:
\begin{itemize}
\item $q\bar{q}$: full mass dependent, one\-/loop QCD corrections; up
  to four\-/loop corrections in massless QCD; one\-/loop corrections
  in massless QED; massless, two\-/loop, mixed QCD--QED correction;
  leading order top-mass corrections for decays into $q \neq t$,
\item $l^+l^-$: one\-/loop QED corrections in the massless limit,
\item $VV$: single and double off-shell decays,
\item $\gamma \gamma$: two\-/loop QCD corrections to decay through
  fermion and scalar loops; use of $\alpha$ in the Thomson limit to
  minimize impact of two\-/loop QED/EW corrections,
\item $H\to gg$: up to four\-/loop QCD corrections through top-loop,
  including $m_t$ suppressed terms,
\item $A\to gg$: up to three\-/loop QCD corrections in the leading
  $m_t$ approximation,
\item $\gamma Z$: two\-/loop QCD corrections to quark loop in the
  heavy quark limit; $\alpha$ treatment as in the $\gamma \gamma$
  case.
\end{itemize}
The corrections in CP\-/violating models are implemented by combining
scalar and pseudoscalar limits.
No weak corrections are applied in the current version.

We have carried out detailed numerical tests in many models and
comparisons with other existing codes wherever possible, demonstrating
the accuracy, generality and reliability of \FD.
In the comparisons we found particularly good agreement with \HD in
models which \HD can handle. We also found mostly good agreement with
the \sarah/\spheno \texttt{DECAY} block results.
As various programs organize their calculations in different ways discrepancies between all of them are expected.
Their sources are understood and explained throughout the comparison section.

A noteworthy property of our approach is the use of a decoupling
variant of the $\MSbar/\DRbar$ renormalization scheme: pure BSM
parameters are $\MSbar/\DRbar$ renormalized, but SM-like parameters
are defined in a decoupling scheme. The separation between the two
kinds of parameters is automatically
derived from information provided in the model
files.
This has two advantages. On the one hand it leads to a
correct decoupling limit, i.e.\ the BSM results approach the SM ones
for heavy BSM masses within 10\%. On the other hand, it allows seamless
integration into the $\MSbar/\DRbar$ framework of \FS with the
possibility for connection to studies of, e.g., evolution under the
renormalization group equations or the effective scalar potential.

As illustrated by \figref{fig:MSSMEFTHiggs_decoupling} the decoupling
scheme leads to reliable predictions not only at high BSM scales,
where the SM prediction is recovered. Importantly, it also enables
accurate predictions of characteristic BSM signatures, allowing tests
of BSM scenarios against data.
In this case the prediction of the decoupling scheme is almost indistinguishable from a standard \DRbar calculation.
This allows for a smooth interpolation between regions of heavy and light BSM physics.

%% file: tex/acknowledgements.tex
\section*{Acknowledgements}

We thank Michael Spira for help regarding \HD and Malin Sjödahl for
allowing us to use \texttt{ColorMath}.
This research was supported in parts by the National Science Centre,
Poland, the HARMONIA project under contract UMO-2015/\allowbreak
18/\allowbreak M/\allowbreak ST2/\allowbreak 00518 (2016-2021).
The work of W.K. was partially supported by the German Research
Foundation (DFG) under grants number STO 876/4-1 and STO 876/2-2.
The work of D.H. was supported by the Grant Agency of the Czech
Republic (GACR) under the contract number 17-04902S.
The work of P.A. is supported by the Australian Research Council Future
Fellowship grant FT160100274.
The work of A.V.\ was supported by the German DFG Collaborative
Research Centre \textit{P\textsuperscript{3}\!H: Particle Physics
  Phenomenology after the Higgs Discovery} (CRC TRR 257).

%% file: tex/configuration.tex
\section{Configuration of \FD}
\label{sec:configuration}

\subsection{\FD\ build options}
\label{sec:create_code}

The generation of the decays in \FS\ is controlled by the
\code{FSCalculateDecays} variable in the \FS\ model file.  For a given
model \modelname{<model>} this file is located at
\path{model_files/<model>/FlexibleSUSY.m.in}.  The model file is read
when calling \code{createmodel}.  The \code{FSCalculateDecays}
variable can take on 2 values: \code{True} (calculate decays, default)
or \code{False}.
The particles for which the decays shall be calculated must be given
in the \code{FSDecayParticles} variables in the model file.  Allowed
values are: \code{Automatic} (default) or a list of particles.
Currently, \code{Automatic} expands to a list containing the neutral
scalar, pseudoscalar and charged Higgs bosons of the model.  In a
future version the meaning of \code{Automatic} might change to include
also other particles as well.

As an example, one might add decays to the \FS version of the
\modelname{THDMII} by adding the following lines to the model file
\path{model_files/THDMII/FlexibleSUSY.m.in}:
\begin{lstlisting}
FSCalculateDecays = True;
FSDecayParticles = Automatic; (* expands to {hh, Ah, Hm} *)
\end{lstlisting}
In the example the value of \code{FSDecayParticles} expands to the list
\code{\{hh, Ah, Hm\}}, where \code{hh}, \code{Ah} and \code{Hm} are
the names of the neutral scalar, pseudoscalar and charged Higgs bosons
in the model.

To compute decay width, \FS\ must be configured with at least one of
the dedicated libraries for evaluation of one-loop integrals.  Up to
\FS\ 2.4.2 only to \LT\ was supported.  In \FS\ 2.5.0 support for
\Collier\ was added.  \FS\ can be configured with both libraries at
the same time as
\begin{lstlisting}
./configure --with-loop-libraries=collier,looptools [...]
\end{lstlisting}
where the ellipsis stand for potential further configuration options.
If needed, the location of \LT\ and \Collier\ can be specified through
the following parameters:
\begin{lstlisting}
--with-looptools-libdir=    Path to search for LoopTools libraries
--with-looptools-incdir=    Path to search for LoopTools headers
--with-collier-libdir=      Path to search for COLLIER libraries
--with-collier-incdir=      Path to search for COLLIER modules
\end{lstlisting}
Note that for usage with \FS, \Collier\ has to compiled as a static
library, in position independent mode (see Ref.~\cite{FSREADME} for
detailed instructions).  See \code{./configure -h} for all available
configuration options.

\subsection{\FD\ runtime options}
\label{sec:runtime_settings}

To calculate the decays for a given parameter point with a spectrum
generator generated by \FS, the following flags must be set (see also
\tabref{tab:FS_SLHA_configuration_FS-block}):
\begin{itemize}
\item The calculation of the SM and BSM particle pole masses has to be enabled
  (\code{FlexibleSUSY[3] = 1} and \code{FlexibleSUSY[23] = 1}).
\item \LT\ or \Collier\ has to be selected as loop library
  (\code{FlexibleSUSY[31] = 1} or \code{2}).
\item The calculation of the decays has to be enabled
  (\code{FlexibleSUSY[32] = 1}).
\end{itemize}
The behaviour of decay module is controlled by flags in the \FD block as documented in \tabref{tab:FS_SLHA_configuration_FD-block}.

\begin{table}[tbh]
  \centering
  \begin{tabularx}{\textwidth}{clcX}
    \toprule
    Index & \mathematica symbol & Default & Description \\
    \midrule
    3 & \code{calculateStandardModelMasses} & $0$ & calculate SM pole masses (0 = no, 1 = yes) \\
    23 & \code{calculateBSMMasses} & $0$ & calculate BSM pole masses (0 = no, 1 = yes) \\
    31 & \code{loopLibrary} & $0$ & loop library (0 = \SSUSY, 1 = \Collier, 2 = \LT, 3 = \texttt{FFLite}) \\
    \bottomrule
  \end{tabularx}
  \caption{Entries for the \FS SLHA input block and corresponding
    \mathematica\ symbols to specify the runtime configuration options
    of \FS which are relevant for \FD.  Flags 3 and 23 are
    automatically set to 1 if decays are enabled (see flag 0 in
    \tabref{tab:FS_SLHA_configuration_FD-block}).  Flag 31 must be set to
    1 or 2 to calculate the decays.}
  \label{tab:FS_SLHA_configuration_FS-block}
\end{table}
\begin{table}[tbh]
  \centering
  \begin{tabularx}{\textwidth}{clcX}
    \toprule
    Index & \mathematica symbol & Default & Description \\
    \midrule
    0 & -- & $1$ & calculate decays (0 = no, 1 = yes) \\
    1 & \code{minBRtoPrint} & $10^{-5}$ & minimum BR to print\\
    2 & \code{maxHigherOrderCorrections} & 4 & include higher order corrections in decays (0 = \LO, 1 = \NLO, 2 = \NNLO, 3 = \NNNLO, 4 = \NNNNLO) \\
    3 & \code{alphaThomson} & 1 & use $\smc\alpha(m)$ or Thomson $\alpha(0)$ in decays to $\gamma \gamma$ and $\gamma Z$ (0 = $\smc\alpha(m)$, 1 = $\alpha(0)$) \\
    4 & \code{offShellVV} & 2 & decays into off\-/shell $VV$ pair (0 = no off\-/shell decays, 1 = single off\-/shell decays if $m_V < m_H < 2 m_H$, double off\-/shell if $m_H < m_V$, 2 = double off\-/shell decays if $m_H < 2 m_V$) \\
    \bottomrule
  \end{tabularx}
  \caption{Entries for the \code{FlexibleDecay} SLHA input block and
    corresponding \mathematica\ symbols to specify the runtime
    configuration options for \FD.}
  \label{tab:FS_SLHA_configuration_FD-block}
\end{table}

%% file: tex/matrix_elements.tex
\section{Matrix elements}
\label{app:ME}
The matrix elements required in the calculation of the two-body partial widths
are parametrized in terms of a set of standard matrix
elements and associated form-factors \cite{Denner:1991kt,Gigg:thesis}.  For a particular
two-body decay $X \to AB$, external leg wavefunctions and polarizations are
factored out to write the matrix element $\amp_{X \to AB}$ in the form
\begin{equation}
  \amp_{X \to AB} = \amp^{IJK} x_I x_J x_K ,
\end{equation}
where the indices $I$, $J$, and $K$ collectively denote the Lorentz
and Dirac indices for the initial and final states, and $x_I$, $x_J$, and $x_K$
the appropriate external leg factors.  In turn, the quantity $\amp^{IJK}$
is expressed in terms of a set of Lorentz scalar form-factors and covariant
operators,
\begin{equation}
  \amp^{IJK} = \sum_i F_i \amp_i^{IJK} ,
\end{equation}
where the operators $\amp_i^{IJK}$ are chosen to be the same for all
processes of a given type (i.e., $S \to F F$, $S \to S V$, and so on).
The standard matrix elements
\begin{equation}
  \amp_i = \amp_i^{IJK} x_I x_J x_K
\end{equation}
then depend solely on the kinematic variables and polarization factors and may
be calculated independently of the particular model at hand.  The matrix element
for a particular process may then be written generically in terms of the
model-dependent form-factors as
\begin{equation}
  \amp_{X \to A B} = \sum_i F_i \amp_i .
\end{equation}
Similarly, the unpolarized squared matrix element reads
\begin{equation}
  \overline{\sum} |\amp_{X \to A B}|^2 = \sum_{i, j} F_i F_j^*
  \overline{\sum} \amp_i \amp_j^\dagger ,
\end{equation}
where $\overline{\sum}$ denotes the summation over final helicities and
polarizations and averaging over initial helicities and polarizations; note
that, in general, additional summations over gauge indices not shown above are
also carried out to derive appropriate multiplicity factors.  To evaluate the
squared amplitude for a process in a given model, \FS computes the contributions
in that model to the form-factors $F_i$, which are then substituted into the
appropriate generic expression for the squared amplitude.  In the following,
we summarize the conventions used by \FS for the standard matrix elements
$\amp_i$ for each class of two-body decays and the resulting formulas
for the squared amplitudes.

\subsection{$S \to S S$}
The simplest case is the decay of a scalar to a pair of scalars, for which
the Lorentz structure of the matrix element is trivial.  A single form-factor
$F$ is used to parametrize the matrix element,
\begin{equation}
  \amp_{S \to S S} = F ,
\end{equation}
and the unpolarized squared amplitude is simply
\begin{equation}
  \overline{\sum} | \amp_{S \to S S} |^2 = |F|^2 .
\end{equation}

\subsection{$S \to F F$}
The decay of a scalar $S$ to a pair of fermions $F_1$, $F_2$ is described by
the form-factors $F_L$ and $F_R$, with the general matrix element taking the
form
\begin{equation}
  \amp_{S \to F F} = F_L \bar{u}(p_{F_1}, s_{F_1}) P_L
  v(p_{F_2}, s_{F_2}) + F_R \bar{u}(p_{F_1}, s_{F_1}) P_R
  v(p_{F_2}, s_{F_2}),
\end{equation}
where $P_L$, $P_R$ are standard left- and right-handed projection operators,
and $p_{F_i}$, $s_{F_i}$ denote the $4$-momentum and spin of the final state
fermion $F_i$, respectively.  The corresponding unpolarized squared amplitude
is
\begin{equation}
  \overline{\sum} | \amp_{S \to F F} |^2 =
  \left ( m_S^2 - m_{F_1}^2 - m_{F_2}^2 \right ) \left ( |F_L|^2 + |F_R|^2
  \right ) - 2 m_{F_1} m_{F_2} \left ( F_L F_R^* + F_L^* F_R \right ) ,
\end{equation}
where $m_S$, $m_{F_1}$, and $m_{F_2}$ are the masses of the decaying scalar and
the final state fermions.

\subsection{$S \to S V$}
The matrix element for the decay of a scalar $S_1$ into a scalar $S_2$ and a
massive vector $V$ is parametrized using a single form-factor $F$ according to
\begin{equation}
  \amp_{S_1 \to S_2 V} = F \varepsilon^{r_V *}_\mu(p_V) \left ( p_{S_1}^\mu
  + p_{S_2}^\mu \right ) ,
\end{equation}
where $\varepsilon^{r_V*}_\mu(p)$ is a polarization vector for the final
state vector boson, and $p_{S_1}$, $p_{S_2}$, and $p_V$ are the $4$-momenta of
the initial and final states.
The squared amplitude reads
\begin{equation}
  \overline{\sum} | \amp_{S_1 \to S_2 V} |^2 =
  \frac{|F|^2}{m_V^2} \left [ m_{S_1}^4 + \left ( m_V^2 - m_{S_2}^2 \right )^2
    - 2 m_{S_1}^2 \left ( m_{S_2}^2 + m_V^2 \right ) \right ] .
\end{equation}
with $m_{S_1}$ and $m_{S_2}$ the masses of the initial and final state scalars, respectively, and $m_V$ the mass of the vector.

For massless vector, like the photon, the amplitude that fulfils the Ward identity and where the external photon is physical (i.e. transversal) is identically 0.\footnote{Eq.~B.36 of Ref.~\cite{Gigg:thesis} gives a non-zero, and negative, result for the squared amplitude. This can only be valid if one considers the $1\to2$ process as a part of a larger amplitude, which is not the case we consider here.}

\subsection{$S \to V V$}
To also take into account the possibility of loop-induced decays, the
decomposition of the matrix element for the decay of a scalar into two vectors
$V_1$ and $V_2$ is chosen to be
\begin{align}
  \amp_{S \to V_1 V_2} &= \varepsilon^{r_{V_1} *}_\mu (p_{V_1})
  \varepsilon^{r_{V_2} *}_\nu (p_{V_2}) \Big ( F_\eta \eta^{\mu\nu} + F_{11}
  p_{V_1}^\mu p_{V_1}^\nu + F_{12} p_{V_1}^\mu p_{V_2}^\nu
  + F_{21} p_{V_1}^\nu p_{V_2}^\mu \nonumber \\
  & \quad {} + F_{22} p_{V_2}^\mu p_{V_2}^\nu
  + F_\epsilon \epsilon^{\mu\nu\alpha\beta} p_{V_1 \alpha} p_{V_2 \beta}
  \Big ) ,
\end{align}
where $p_{V_1}$ and $p_{V_2}$ are the $4$-momenta of the final state vectors,
and $\varepsilon^{r_{V_i} *}_\mu (p_{V_i})$ the corresponding polarizations.
There are three cases for the resulting squared amplitude, depending on
whether both vector bosons are massless, only one is, or both have non\-/zero
masses.  If both of the vector boson masses $m_{V_1} = m_{V_2} = 0$, then
the generic unpolarized squared amplitude is given by
\begin{align}
\label{eq:SSV_square_massless}
  \overline{\sum} | \amp_{S \to V V} |^2 &=
  4 | F_\eta|^2
  + m_S^2 \Re(F_\eta F_{21}^*)
  + \frac{1}{2} m_S^4 | F_\epsilon|^2,
  \quad ( m_{V_1} = m_{V_2} = 0 ) .
\end{align}
If $m_{V_1} = 0$ and $m_{V_2} \neq 0$, then the squared amplitude is given by
\begin{align}
\label{eq:SSV_square_massless-massive}
  \overline{\sum} | \amp_{S \to V V} |^2 &= 3 | F_\eta |^2
  + \frac{1}{4} \left ( m_S^2 - m_{V_2}^2 \right )^2 \left ( 2 | F_\epsilon |^2
  - | F_{21} |^2 \right ), \quad ( m_{V_1} = 0, \, m_{V_2} \neq 0 ) .
\end{align}
If $m_{V_2} = 0$ and $m_{V_1} \neq 0$ instead, then the squared amplitude takes
the same form as above with the replacements $m_{V_2} \to m_{V_1}$ and
$F_{11} \to F_{22}$.  Finally, if both vector bosons are massive, then the
generic squared amplitude is given by
\begin{align}
  \overline{\sum} | \amp_{S \to V V} |^2 &=
  \frac{|F_\eta|^2}{4 m_{V_1}^2 m_{V_2}^2} \left [ m_S^4 + m_{V_1}^4 + m_{V_2}^4
    + 10 m_{V_1}^2 m_{V_2}^2 - 2 m_S^2 \left ( m_{V_1}^2 + m_{V_2}^2 \right )
    \right ] \nonumber \\
  & \quad {} + \frac{|F_{21}|^2}{16 m_{V_1}^2 m_{V_2}^2} \left [ m_S^4
    + \left ( m_{V_1}^2 - m_{V_2}^2 \right )^2 - 2 m_S^2 \left ( m_{V_1}^2
    + m_{V_2}^2 \right ) \right ]^2 \nonumber \\
  & \quad {} + \frac{|F_\epsilon|^2}{2} \left [ \left ( m_S^2 - m_{V_1}^2
    - m_{V_2}^2 \right )^2 - 4 m_{V_1}^2 m_{V_2}^2 \right ] \nonumber \\
  & \quad {} + \frac{1}{8 m_{V_1}^2 m_{V_2}^2} \left ( F_\eta F_{21}^*
  + F_\eta^* F_{21} \right ) \Big [ m_S^6 - 3 m_S^4 \left ( m_{V_1}^2
    + m_{V_2}^2 \right ) \nonumber \\
  & \quad {} - \left ( m_{V_1}^2 - m_{V_2}^2 \right )^2 \left (
    m_{V_1}^2 + m_{V_2}^2 \right ) + m_S^2 \left ( 3 m_{V_1}^4
    + 2 m_{V_1}^2 m_{V_2}^2 + 3 m_{V_2}^4 \right ) \Big ] , \quad
  ( m_{V_1}, \, m_{V_2} > 0 ) .
\end{align}

The color algebra is performed using a custom version of the
\texttt{ColorMath} package \cite{Sjodahl:2012nk} distributed together
with \FS.

%% file: tex/spheno_setup.tex
\section{\spheno setup}
\label{sec:spheno_setup}

In the comparison to \spheno, we can directly use the same
$\MSbar/\DRbar$ Lagrangian parameters from \FS\ as inputs, since the
two programs have a very similar setup.  For every model we use the
default version of the model file distributed with \spheno\ and \FS.
There are small differences between the \FS\ and \spheno\ setups,
which we discuss on a model\-/by\-/model basis. The generic technical
setup for \spheno\ and \FS/\FD\ is chosen as similarly as possible, to
focus on differences in the actual decay computations:
\begin{itemize}
\item We switch on the running to the scale of a decaying particle:
  flag 14 in block \texttt{SPhenoInput} is set to 1 (default is 0); in
  \FD\ the \MSbar/\DRbar\ parameters are always run to the scale of
  the decaying particle.\footnote{Note that in \spheno\ this running
    is automatically turned off if the mass of the decaying particle
    is above the renormalization scale of the model (e.g.\ above
    $1\unit{TeV}$ in SUSY models in case SPA convention
    \cite{AguilarSaavedra:2005pw} is used).}
\item The mass of the decaying Higgs boson is determined at the
  one-loop level and without the supersymmetry\-/specific effective
  field theory methods of Refs.\
  \cite{Athron:2016fuq,Staub:2017jnp,Athron:2017fvs,Kwasnitza:2020wli}
  (flags 66 and 67 in block \texttt{SPhenoInput} are set to 0).
\item Tree-level running masses are used in internal lines in the
  computation of loop\-/induced decays (this is always the case in the
  \texttt{DECAY} block in \spheno\ and in \FD; for the
  \texttt{DECAY1L} block in \spheno\ we set flag 1118 in block
  \texttt{DECAYOPTIONS} to 0).
\end{itemize}
\sarah/\spheno\ has two separate modules for decay calculations, whose
results are provided in the output blocks \texttt{DECAY} and
\texttt{DECAY1L}, respectively. The first is tailored to the
calculation of Higgs decays~\cite{Staub:2011dp,Staub:2012pb}, contains
SM\-/like higher\-/order corrections similar to the ones in \HD and
\FD, however it does not compute the loop\-/induced $h\to\gamma Z$
partial width. The second decay computation (\texttt{DECAY1L} output
block) is more general and based on the general full one\-/loop
computation of Ref.\ \cite{Goodsell:2017pdq}, but contains no specific
higher\-/order corrections and is thus expected to be less precise in case
of Higgs boson decays.

%% file: flexibledecays.bbl
\begin{thebibliography}{100}
\expandafter\ifx\csname url\endcsname\relax
  \def\url#1{\texttt{#1}}\fi
\expandafter\ifx\csname urlprefix\endcsname\relax\def\urlprefix{URL }\fi
\expandafter\ifx\csname href\endcsname\relax
  \def\href#1#2{#2} \def\path#1{#1}\fi

\bibitem{Aad:2012tfa}
G.~Aad, et~al., {Observation of a new particle in the search for the Standard
  Model Higgs boson with the ATLAS detector at the LHC}, Phys. Lett. B716
  (2012) 1--29.
\newblock \href {http://arxiv.org/abs/1207.7214} {\path{arXiv:1207.7214}},
  \href {https://doi.org/10.1016/j.physletb.2012.08.020}
  {\path{doi:10.1016/j.physletb.2012.08.020}}.

\bibitem{Chatrchyan:2012xdj}
S.~Chatrchyan, et~al., {Observation of a new boson at a mass of 125 GeV with
  the CMS experiment at the LHC}, Phys. Lett. B716 (2012) 30--61.
\newblock \href {http://arxiv.org/abs/1207.7235} {\path{arXiv:1207.7235}},
  \href {https://doi.org/10.1016/j.physletb.2012.08.021}
  {\path{doi:10.1016/j.physletb.2012.08.021}}.

\bibitem{Sirunyan:2018koj}
A.~M. Sirunyan, et~al., {Combined measurements of Higgs boson couplings in
  proton\textendash{}proton collisions at $\sqrt{s}=13\,\text {Te}\text {V} $},
  Eur. Phys. J. C 79~(5) (2019) 421.
\newblock \href {http://arxiv.org/abs/1809.10733} {\path{arXiv:1809.10733}},
  \href {https://doi.org/10.1140/epjc/s10052-019-6909-y}
  {\path{doi:10.1140/epjc/s10052-019-6909-y}}.

\bibitem{ATLAS-CONF-2018-031}
\href{https://cds.cern.ch/record/2629412}{{Combined measurements of Higgs boson
  production and decay using up to 80 fb$^{-1}$ of proton--proton collision
  data at $\sqrt{s}=$ 13 TeV collected with the ATLAS experiment}}, Tech. Rep.
  ATLAS-CONF-2018-031, CERN, Geneva (Jul 2018).
\newline\urlprefix\url{https://cds.cern.ch/record/2629412}

\bibitem{Bernon:2014vta}
J.~Bernon, B.~Dumont, S.~Kraml, {Status of Higgs couplings after run 1 of the
  LHC}, Phys. Rev. D90 (2014) 071301.
\newblock \href {http://arxiv.org/abs/1409.1588} {\path{arXiv:1409.1588}},
  \href {https://doi.org/10.1103/PhysRevD.90.071301}
  {\path{doi:10.1103/PhysRevD.90.071301}}.

\bibitem{Djouadi:1997yw}
A.~Djouadi, J.~Kalinowski, M.~Spira, {HDECAY: A Program for Higgs boson decays
  in the standard model and its supersymmetric extension}, Comput. Phys.
  Commun. 108 (1998) 56--74.
\newblock \href {http://arxiv.org/abs/hep-ph/9704448}
  {\path{arXiv:hep-ph/9704448}}, \href
  {https://doi.org/10.1016/S0010-4655(97)00123-9}
  {\path{doi:10.1016/S0010-4655(97)00123-9}}.

\bibitem{Djouadi:2018xqq}
A.~Djouadi, J.~Kalinowski, M.~Muehlleitner, M.~Spira, {HDECAY: Twenty$_{++}$
  years after}, Comput. Phys. Commun. 238 (2019) 214--231.
\newblock \href {http://arxiv.org/abs/1801.09506} {\path{arXiv:1801.09506}},
  \href {https://doi.org/10.1016/j.cpc.2018.12.010}
  {\path{doi:10.1016/j.cpc.2018.12.010}}.

\bibitem{Krause:2018wmo}
M.~Krause, M.~Mühlleitner, M.~Spira, {2HDECAY - A program for the Calculation
  of Electroweak One-Loop Corrections to Higgs Decays in the Two-Higgs-Doublet
  Model Including State-of-the-Art QCD Corrections} (2018).
\newblock \href {http://arxiv.org/abs/1810.00768} {\path{arXiv:1810.00768}}.

\bibitem{Krause:2016oke}
M.~Krause, R.~Lorenz, M.~Muhlleitner, R.~Santos, H.~Ziesche, {Gauge-independent
  Renormalization of the 2-Higgs-Doublet Model}, JHEP 09 (2016) 143.
\newblock \href {http://arxiv.org/abs/1605.04853} {\path{arXiv:1605.04853}},
  \href {https://doi.org/10.1007/JHEP09(2016)143}
  {\path{doi:10.1007/JHEP09(2016)143}}.

\bibitem{Krause:2016xku}
M.~Krause, M.~Muhlleitner, R.~Santos, H.~Ziesche, {Higgs-to-Higgs boson decays
  in a 2HDM at next-to-leading order}, Phys. Rev. D 95~(7) (2017) 075019.
\newblock \href {http://arxiv.org/abs/1609.04185} {\path{arXiv:1609.04185}},
  \href {https://doi.org/10.1103/PhysRevD.95.075019}
  {\path{doi:10.1103/PhysRevD.95.075019}}.

\bibitem{Denner:2018opp}
A.~Denner, S.~Dittmaier, J.-N. Lang, {Renormalization of mixing angles}, JHEP
  11 (2018) 104.
\newblock \href {http://arxiv.org/abs/1808.03466} {\path{arXiv:1808.03466}},
  \href {https://doi.org/10.1007/JHEP11(2018)104}
  {\path{doi:10.1007/JHEP11(2018)104}}.

\bibitem{Hahn:2010te}
T.~Hahn, S.~Heinemeyer, W.~Hollik, H.~Rzehak, G.~Weiglein, {FeynHiggs 2.7},
  Nucl. Phys. B Proc. Suppl. 205-206 (2010) 152--157.
\newblock \href {http://arxiv.org/abs/1007.0956} {\path{arXiv:1007.0956}},
  \href {https://doi.org/10.1016/j.nuclphysbps.2010.08.035}
  {\path{doi:10.1016/j.nuclphysbps.2010.08.035}}.

\bibitem{Bahl:2018qog}
H.~Bahl, T.~Hahn, S.~Heinemeyer, W.~Hollik, S.~Pa\ss{}ehr, H.~Rzehak,
  G.~Weiglein, {Precision calculations in the MSSM Higgs-boson sector with
  FeynHiggs 2.14}, Comput. Phys. Commun. 249 (2020) 107099.
\newblock \href {http://arxiv.org/abs/1811.09073} {\path{arXiv:1811.09073}},
  \href {https://doi.org/10.1016/j.cpc.2019.107099}
  {\path{doi:10.1016/j.cpc.2019.107099}}.

\bibitem{Eriksson:2009ws}
D.~Eriksson, J.~Rathsman, O.~Stal, {2HDMC: Two-Higgs-Doublet Model Calculator
  Physics and Manual}, Comput. Phys. Commun. 181 (2010) 189--205.
\newblock \href {http://arxiv.org/abs/0902.0851} {\path{arXiv:0902.0851}},
  \href {https://doi.org/10.1016/j.cpc.2009.09.011}
  {\path{doi:10.1016/j.cpc.2009.09.011}}.

\bibitem{Kanemura:2017gbi}
S.~Kanemura, M.~Kikuchi, K.~Sakurai, K.~Yagyu, {H-COUP: a program for one-loop
  corrected Higgs boson couplings in non-minimal Higgs sectors}, Comput. Phys.
  Commun. 233 (2018) 134--144.
\newblock \href {http://arxiv.org/abs/1710.04603} {\path{arXiv:1710.04603}},
  \href {https://doi.org/10.1016/j.cpc.2018.06.012}
  {\path{doi:10.1016/j.cpc.2018.06.012}}.

\bibitem{Kanemura:2019slf}
S.~Kanemura, M.~Kikuchi, K.~Mawatari, K.~Sakurai, K.~Yagyu, {H-COUP Version 2:
  a program for one-loop corrected Higgs boson decays in non-minimal Higgs
  sectors}, Comput. Phys. Commun. 257 (2020) 107512.
\newblock \href {http://arxiv.org/abs/1910.12769} {\path{arXiv:1910.12769}},
  \href {https://doi.org/10.1016/j.cpc.2020.107512}
  {\path{doi:10.1016/j.cpc.2020.107512}}.

\bibitem{Contino:2014aaa}
R.~Contino, M.~Ghezzi, C.~Grojean, M.~Mühlleitner, M.~Spira, {eHDECAY: an
  Implementation of the Higgs Effective Lagrangian into HDECAY}, Comput. Phys.
  Commun. 185 (2014) 3412--3423.
\newblock \href {http://arxiv.org/abs/1403.3381} {\path{arXiv:1403.3381}},
  \href {https://doi.org/10.1016/j.cpc.2014.06.028}
  {\path{doi:10.1016/j.cpc.2014.06.028}}.

\bibitem{Costa:2015llh}
R.~Costa, M.~Mühlleitner, M.~O.~P. Sampaio, R.~Santos, {Singlet Extensions of
  the Standard Model at LHC Run 2: Benchmarks and Comparison with the NMSSM},
  JHEP 06 (2016) 034.
\newblock \href {http://arxiv.org/abs/1512.05355} {\path{arXiv:1512.05355}},
  \href {https://doi.org/10.1007/JHEP06(2016)034}
  {\path{doi:10.1007/JHEP06(2016)034}}.

\bibitem{Frisch:2010gw}
W.~Frisch, H.~Eberl, H.~Hlucha, {HFOLD - a program package for calculating
  two-body MSSM Higgs decays at full one-loop level}, Comput. Phys. Commun. 182
  (2011) 2219--2226.
\newblock \href {http://arxiv.org/abs/1012.5025} {\path{arXiv:1012.5025}},
  \href {https://doi.org/10.1016/j.cpc.2011.05.007}
  {\path{doi:10.1016/j.cpc.2011.05.007}}.

\bibitem{Porod:2003um}
W.~Porod, {SPheno, a program for calculating supersymmetric spectra, SUSY
  particle decays and SUSY particle production at e+ e- colliders}, Comput.
  Phys. Commun. 153 (2003) 275--315.
\newblock \href {http://arxiv.org/abs/hep-ph/0301101}
  {\path{arXiv:hep-ph/0301101}}, \href
  {https://doi.org/10.1016/S0010-4655(03)00222-4}
  {\path{doi:10.1016/S0010-4655(03)00222-4}}.

\bibitem{Porod:2011nf}
W.~Porod, F.~Staub, {SPheno 3.1: Extensions including flavour, CP-phases and
  models beyond the MSSM}, Comput. Phys. Commun. 183 (2012) 2458--2469.
\newblock \href {http://arxiv.org/abs/1104.1573} {\path{arXiv:1104.1573}},
  \href {https://doi.org/10.1016/j.cpc.2012.05.021}
  {\path{doi:10.1016/j.cpc.2012.05.021}}.

\bibitem{Allanach:2017hcf}
B.~C. Allanach, T.~Cridge, {The Calculation of Sparticle and Higgs Decays in
  the Minimal and Next-to-Minimal Supersymmetric Standard Models: SOFTSUSY4.0},
  Comput. Phys. Commun. 220 (2017) 417--502.
\newblock \href {http://arxiv.org/abs/1703.09717} {\path{arXiv:1703.09717}},
  \href {https://doi.org/10.1016/j.cpc.2017.07.021}
  {\path{doi:10.1016/j.cpc.2017.07.021}}.

\bibitem{Baglio:2013iia}
J.~Baglio, R.~Gröber, M.~Mühlleitner, D.~T. Nhung, H.~Rzehak, M.~Spira,
  J.~Streicher, K.~Walz, {NMSSMCALC: A Program Package for the Calculation of
  Loop-Corrected Higgs Boson Masses and Decay Widths in the (Complex) NMSSM},
  Comput. Phys. Commun. 185~(12) (2014) 3372--3391.
\newblock \href {http://arxiv.org/abs/1312.4788} {\path{arXiv:1312.4788}},
  \href {https://doi.org/10.1016/j.cpc.2014.08.005}
  {\path{doi:10.1016/j.cpc.2014.08.005}}.

\bibitem{Engeln:2018mbg}
I.~Engeln, M.~Mühlleitner, J.~Wittbrodt, {N2HDECAY: Higgs Boson Decays in the
  Different Phases of the N2HDM}, Comput. Phys. Commun. 234 (2019) 256--262.
\newblock \href {http://arxiv.org/abs/1805.00966} {\path{arXiv:1805.00966}},
  \href {https://doi.org/10.1016/j.cpc.2018.07.020}
  {\path{doi:10.1016/j.cpc.2018.07.020}}.

\bibitem{Athron:2014yba}
P.~Athron, J.-h. Park, D.~Stöckinger, A.~Voigt, {FlexibleSUSY -- A spectrum
  generator generator for supersymmetric models}, Comput. Phys. Commun. 190
  (2015) 139--172.
\newblock \href {http://arxiv.org/abs/1406.2319} {\path{arXiv:1406.2319}},
  \href {https://doi.org/10.1016/j.cpc.2014.12.020}
  {\path{doi:10.1016/j.cpc.2014.12.020}}.

\bibitem{Athron:2016fuq}
P.~Athron, J.-h. Park, T.~Steudtner, D.~Stöckinger, A.~Voigt, {Precise Higgs
  mass calculations in (non-)minimal supersymmetry at both high and low
  scales}, JHEP 01 (2017) 079.
\newblock \href {http://arxiv.org/abs/1609.00371} {\path{arXiv:1609.00371}},
  \href {https://doi.org/10.1007/JHEP01(2017)079}
  {\path{doi:10.1007/JHEP01(2017)079}}.

\bibitem{Athron:2017fvs}
P.~Athron, M.~Bach, D.~Harries, T.~Kwasnitza, J.-h. Park, D.~Stöckinger,
  A.~Voigt, J.~Ziebell, {FlexibleSUSY 2.0: Extensions to investigate the
  phenomenology of SUSY and non-SUSY models} (2017).
\newblock \href {http://arxiv.org/abs/1710.03760} {\path{arXiv:1710.03760}}.

\bibitem{Staub:2009bi}
F.~Staub, {From Superpotential to Model Files for FeynArts and
  CalcHep/CompHep}, Comput.Phys.Commun. 181 (2010) 1077--1086.
\newblock \href {http://arxiv.org/abs/0909.2863} {\path{arXiv:0909.2863}},
  \href {https://doi.org/10.1016/j.cpc.2010.01.011}
  {\path{doi:10.1016/j.cpc.2010.01.011}}.

\bibitem{Staub:2010jh}
F.~Staub, {Automatic Calculation of supersymmetric Renormalization Group
  Equations and Self Energies}, Comput.Phys.Commun. 182 (2011) 808--833.
\newblock \href {http://arxiv.org/abs/1002.0840} {\path{arXiv:1002.0840}},
  \href {https://doi.org/10.1016/j.cpc.2010.11.030}
  {\path{doi:10.1016/j.cpc.2010.11.030}}.

\bibitem{Staub:2012pb}
F.~Staub, {SARAH 3.2: Dirac Gauginos, UFO output, and more}, Computer Physics
  Communications 184 (2013) pp. 1792--1809.
\newblock \href {http://arxiv.org/abs/1207.0906} {\path{arXiv:1207.0906}},
  \href {https://doi.org/10.1016/j.cpc.2013.02.019}
  {\path{doi:10.1016/j.cpc.2013.02.019}}.

\bibitem{Staub:2013tta}
F.~Staub, {SARAH 4: A tool for (not only SUSY) model builders},
  Comput.Phys.Commun. 185 (2014) 1773--1790.
\newblock \href {http://arxiv.org/abs/1309.7223} {\path{arXiv:1309.7223}},
  \href {https://doi.org/10.1016/j.cpc.2014.02.018}
  {\path{doi:10.1016/j.cpc.2014.02.018}}.

\bibitem{Allanach:2001kg}
B.~Allanach, {SOFTSUSY: a program for calculating supersymmetric spectra},
  Comput.Phys.Commun. 143 (2002) 305--331.
\newblock \href {http://arxiv.org/abs/hep-ph/0104145}
  {\path{arXiv:hep-ph/0104145}}, \href
  {https://doi.org/10.1016/S0010-4655(01)00460-X}
  {\path{doi:10.1016/S0010-4655(01)00460-X}}.

\bibitem{Allanach:2013kza}
B.~Allanach, P.~Athron, L.~C. Tunstall, A.~Voigt, A.~Williams, {Next-to-Minimal
  SOFTSUSY}, Comput.Phys.Commun. 185 (2014) 2322--2339.
\newblock \href {http://arxiv.org/abs/1311.7659} {\path{arXiv:1311.7659}},
  \href {https://doi.org/10.1016/j.cpc.2014.04.015}
  {\path{doi:10.1016/j.cpc.2014.04.015}}.

\bibitem{Goodsell:2014bna}
M.~D. Goodsell, K.~Nickel, F.~Staub, {Two-Loop Higgs mass calculations in
  supersymmetric models beyond the MSSM with SARAH and SPheno}, Eur. Phys. J.
  C75~(1) (2015) 32.
\newblock \href {http://arxiv.org/abs/1411.0675} {\path{arXiv:1411.0675}},
  \href {https://doi.org/10.1140/epjc/s10052-014-3247-y}
  {\path{doi:10.1140/epjc/s10052-014-3247-y}}.

\bibitem{Goodsell:2015ira}
M.~Goodsell, K.~Nickel, F.~Staub, {Generic two-loop Higgs mass calculation from
  a diagrammatic approach}, Eur. Phys. J. C75~(6) (2015) 290.
\newblock \href {http://arxiv.org/abs/1503.03098} {\path{arXiv:1503.03098}},
  \href {https://doi.org/10.1140/epjc/s10052-015-3494-6}
  {\path{doi:10.1140/epjc/s10052-015-3494-6}}.

\bibitem{Gabelmann:2018axh}
M.~Gabelmann, M.~M\"uhlleitner, F.~Staub, {Automatised matching between two
  scalar sectors at the one-loop level}, Eur. Phys. J. C 79~(2) (2019) 163.
\newblock \href {http://arxiv.org/abs/1810.12326} {\path{arXiv:1810.12326}},
  \href {https://doi.org/10.1140/epjc/s10052-019-6570-5}
  {\path{doi:10.1140/epjc/s10052-019-6570-5}}.

\bibitem{Harlander:2017kuc}
R.~V. Harlander, J.~Klappert, A.~Voigt, {Higgs mass prediction in the MSSM at
  three-loop level in a pure $\overline{\text{DR}}$ context} (2017).
\newblock \href {http://arxiv.org/abs/1708.05720} {\path{arXiv:1708.05720}}.

\bibitem{Harlander:2018yhj}
R.~Harlander, J.~Klappert, A.~Ochoa~Franco, A.~Voigt, {The light CP-even MSSM
  Higgs mass resummed to fourth logarithmic order}, Eur. Phys. J. C 78~(10)
  (2018) 874.
\newblock \href {http://arxiv.org/abs/1807.03509} {\path{arXiv:1807.03509}},
  \href {https://doi.org/10.1140/epjc/s10052-018-6351-6}
  {\path{doi:10.1140/epjc/s10052-018-6351-6}}.

\bibitem{Harlander:2019dge}
R.~Harlander, J.~Klappert, A.~Voigt, {The light CP-even MSSM Higgs mass
  including N$^\mathbf {3}$LO+N$^\mathbf {3}$LL QCD corrections}, Eur. Phys. J.
  C 80~(3) (2020) 186.
\newblock \href {http://arxiv.org/abs/1910.03595} {\path{arXiv:1910.03595}},
  \href {https://doi.org/10.1140/epjc/s10052-020-7747-7}
  {\path{doi:10.1140/epjc/s10052-020-7747-7}}.

\bibitem{Kwasnitza:2020wli}
T.~Kwasnitza, D.~St\"ockinger, A.~Voigt, {Improved MSSM Higgs mass calculation
  using the 3-loop FlexibleEFTHiggs approach including $x_{t}$-resummation},
  JHEP 07~(07) (2020) 197.
\newblock \href {http://arxiv.org/abs/2003.04639} {\path{arXiv:2003.04639}},
  \href {https://doi.org/10.1007/JHEP07(2020)197}
  {\path{doi:10.1007/JHEP07(2020)197}}.

\bibitem{Slavich:2020zjv}
P.~Slavich, et~al., {Higgs-mass predictions in the MSSM and beyond} (12 2020).
\newblock \href {http://arxiv.org/abs/2012.15629} {\path{arXiv:2012.15629}}.

\bibitem{Staub:2016dxq}
F.~Staub, et~al., {Precision tools and models to narrow in on the 750 GeV
  diphoton resonance}, Eur. Phys. J. C76~(9) (2016) 516.
\newblock \href {http://arxiv.org/abs/1602.05581} {\path{arXiv:1602.05581}},
  \href {https://doi.org/10.1140/epjc/s10052-016-4349-5}
  {\path{doi:10.1140/epjc/s10052-016-4349-5}}.

\bibitem{Goodsell:2017pdq}
M.~D. Goodsell, S.~Liebler, F.~Staub, {Generic calculation of two-body partial
  decay widths at the full one-loop level}, Eur. Phys. J. C77~(11) (2017) 758.
\newblock \href {http://arxiv.org/abs/1703.09237} {\path{arXiv:1703.09237}},
  \href {https://doi.org/10.1140/epjc/s10052-017-5259-x}
  {\path{doi:10.1140/epjc/s10052-017-5259-x}}.

\bibitem{Martin:2005qm}
S.~P. Martin, D.~G. Robertson, {TSIL: A Program for the calculation of two-loop
  self-energy integrals}, Comput. Phys. Commun. 174 (2006) 133--151.
\newblock \href {http://arxiv.org/abs/hep-ph/0501132}
  {\path{arXiv:hep-ph/0501132}}, \href
  {https://doi.org/10.1016/j.cpc.2005.08.005}
  {\path{doi:10.1016/j.cpc.2005.08.005}}.

\bibitem{Athron:2015rva}
P.~Athron, M.~Bach, H.~G. Fargnoli, C.~Gnendiger, R.~Greifenhagen, J.-h. Park,
  S.~Pa\ss{}ehr, D.~St\"ockinger, H.~St\"ockinger-Kim, A.~Voigt, {GM2Calc:
  Precise MSSM prediction for $(g - 2)$ of the muon}, Eur. Phys. J. C 76~(2)
  (2016) 62.
\newblock \href {http://arxiv.org/abs/1510.08071} {\path{arXiv:1510.08071}},
  \href {https://doi.org/10.1140/epjc/s10052-015-3870-2}
  {\path{doi:10.1140/epjc/s10052-015-3870-2}}.

\bibitem{Harlander:2008ju}
R.~Harlander, P.~Kant, L.~Mihaila, M.~Steinhauser, {Higgs boson mass in
  supersymmetry to three loops}, Phys. Rev. Lett. 100 (2008) 191602.
\newblock \href {http://arxiv.org/abs/0803.0672} {\path{arXiv:0803.0672}},
  \href {https://doi.org/10.1103/PhysRevLett.101.039901}
  {\path{doi:10.1103/PhysRevLett.101.039901}}.

\bibitem{Kant:2010tf}
P.~Kant, R.~V. Harlander, L.~Mihaila, M.~Steinhauser, {Light MSSM Higgs boson
  mass to three-loop accuracy}, JHEP 08 (2010) 104.
\newblock \href {http://arxiv.org/abs/1005.5709} {\path{arXiv:1005.5709}},
  \href {https://doi.org/10.1007/JHEP08(2010)104}
  {\path{doi:10.1007/JHEP08(2010)104}}.

\bibitem{Mathematica}
{Wolfram Research{,} Inc.},
  \href{https://www.wolfram.com/mathematica}{Mathematica,}, {Champaign, IL,
  2021}.
\newline\urlprefix\url{https://www.wolfram.com/mathematica}

\bibitem{gsl}
M.~Galassi, J.~Davies, J.~Theiler, B.~Gough, G.~Jungman, M.~Booth, F.~Rossi,
  \href{https://www.gnu.org/software/gsl/}{{GNU Scientific Library Reference
  Manual}} (2009).
\newline\urlprefix\url{https://www.gnu.org/software/gsl/}

\bibitem{eigenweb}
G.~Guennebaud, B.~Jacob, et~al., Eigen v3, \url{http://eigen.tuxfamily.org}
  (2010).

\bibitem{BoostLibrary}
Boost, {Boost C++ Libraries}, \url{http://www.boost.org/}, last accessed
  2020-03-10 (2020).

\bibitem{Hahn:1998yk}
T.~Hahn, M.~Perez-Victoria, {Automatized one loop calculations in
  four-dimensions and D-dimensions}, Comput. Phys. Commun. 118 (1999) 153--165.
\newblock \href {http://arxiv.org/abs/hep-ph/9807565}
  {\path{arXiv:hep-ph/9807565}}, \href
  {https://doi.org/10.1016/S0010-4655(98)00173-8}
  {\path{doi:10.1016/S0010-4655(98)00173-8}}.

\bibitem{Denner:2002ii}
A.~Denner, S.~Dittmaier, {Reduction of one loop tensor five point integrals},
  Nucl. Phys. B 658 (2003) 175--202.
\newblock \href {http://arxiv.org/abs/hep-ph/0212259}
  {\path{arXiv:hep-ph/0212259}}, \href
  {https://doi.org/10.1016/S0550-3213(03)00184-6}
  {\path{doi:10.1016/S0550-3213(03)00184-6}}.

\bibitem{Denner:2005nn}
A.~Denner, S.~Dittmaier, {Reduction schemes for one-loop tensor integrals},
  Nucl. Phys. B 734 (2006) 62--115.
\newblock \href {http://arxiv.org/abs/hep-ph/0509141}
  {\path{arXiv:hep-ph/0509141}}, \href
  {https://doi.org/10.1016/j.nuclphysb.2005.11.007}
  {\path{doi:10.1016/j.nuclphysb.2005.11.007}}.

\bibitem{Denner:2010tr}
A.~Denner, S.~Dittmaier, {Scalar one-loop 4-point integrals}, Nucl. Phys. B 844
  (2011) 199--242.
\newblock \href {http://arxiv.org/abs/1005.2076} {\path{arXiv:1005.2076}},
  \href {https://doi.org/10.1016/j.nuclphysb.2010.11.002}
  {\path{doi:10.1016/j.nuclphysb.2010.11.002}}.

\bibitem{Denner:2016kdg}
A.~Denner, S.~Dittmaier, L.~Hofer, {Collier: a fortran-based Complex One-Loop
  LIbrary in Extended Regularizations}, Comput. Phys. Commun. 212 (2017)
  220--238.
\newblock \href {http://arxiv.org/abs/1604.06792} {\path{arXiv:1604.06792}},
  \href {https://doi.org/10.1016/j.cpc.2016.10.013}
  {\path{doi:10.1016/j.cpc.2016.10.013}}.

\bibitem{Skands:2003cj}
P.~Z. Skands, et~al., {SUSY Les Houches accord: Interfacing SUSY spectrum
  calculators, decay packages, and event generators}, JHEP 07 (2004) 036.
\newblock \href {http://arxiv.org/abs/hep-ph/0311123}
  {\path{arXiv:hep-ph/0311123}}, \href
  {https://doi.org/10.1088/1126-6708/2004/07/036}
  {\path{doi:10.1088/1126-6708/2004/07/036}}.

\bibitem{Allanach:2008qq}
B.~C. Allanach, et~al., {SUSY Les Houches Accord 2}, Comput. Phys. Commun. 180
  (2009) 8--25.
\newblock \href {http://arxiv.org/abs/0801.0045} {\path{arXiv:0801.0045}},
  \href {https://doi.org/10.1016/j.cpc.2008.08.004}
  {\path{doi:10.1016/j.cpc.2008.08.004}}.

\bibitem{FlexibleSUSY}
{FlexibleSUSY github repository},
  \url{https://github.com/FlexibleSUSY/FlexibleSUSY}, accessed: September 30,
  2010.

\bibitem{Collins:1978wz}
J.~C. Collins, F.~Wilczek, A.~Zee, {Low-Energy Manifestations of Heavy
  Particles: Application to the Neutral Current}, Phys. Rev. D 18 (1978) 242.
\newblock \href {https://doi.org/10.1103/PhysRevD.18.242}
  {\path{doi:10.1103/PhysRevD.18.242}}.

\bibitem{Diessner:2017ske}
P.~Diessner, W.~Kotlarski, S.~Liebschner, D.~St\"ockinger, {Squark production
  in R-symmetric SUSY with Dirac gluinos: NLO corrections}, JHEP 10 (2017) 142.
\newblock \href {http://arxiv.org/abs/1707.04557} {\path{arXiv:1707.04557}},
  \href {https://doi.org/10.3204/PUBDB-2017-10328}
  {\path{doi:10.3204/PUBDB-2017-10328}}.

\bibitem{Haber:2000kq}
H.~E. Haber, M.~J. Herrero, H.~E. Logan, S.~Penaranda, S.~Rigolin, D.~Temes,
  {SUSY QCD corrections to the MSSM h0 $b \bar{b}$ vertex in the decoupling
  limit}, Phys. Rev. D 63 (2001) 055004.
\newblock \href {http://arxiv.org/abs/hep-ph/0007006}
  {\path{arXiv:hep-ph/0007006}}, \href
  {https://doi.org/10.1103/PhysRevD.63.055004}
  {\path{doi:10.1103/PhysRevD.63.055004}}.

\bibitem{Drees:1989du}
M.~Drees, K.-i. Hikasa, {Heavy Quark Thresholds in Higgs Physics}, Phys. Rev.
  D41 (1990) 1547.
\newblock \href {https://doi.org/10.1103/PhysRevD.41.1547}
  {\path{doi:10.1103/PhysRevD.41.1547}}.

\bibitem{PhysRevD.22.715}
E.~Braaten, J.~P. Leveille,
  \href{https://link.aps.org/doi/10.1103/PhysRevD.22.715}{Higgs-boson decay and
  the running mass}, Phys. Rev. D 22 (1980) 715--721.
\newblock \href {https://doi.org/10.1103/PhysRevD.22.715}
  {\path{doi:10.1103/PhysRevD.22.715}}.
\newline\urlprefix\url{https://link.aps.org/doi/10.1103/PhysRevD.22.715}

\bibitem{Chetyrkin:1996sr}
K.~G. Chetyrkin, {Correlator of the quark scalar currents and $\Gamma_{tot} (H
  \to \text{hadrons})$ at $\mathcal{O}(\alpha_s^3)$ in pQCD}, Phys. Lett. B390
  (1997) 309--317.
\newblock \href {http://arxiv.org/abs/hep-ph/9608318}
  {\path{arXiv:hep-ph/9608318}}, \href
  {https://doi.org/10.1016/S0370-2693(96)01368-8}
  {\path{doi:10.1016/S0370-2693(96)01368-8}}.

\bibitem{Baikov:2005rw}
P.~A. Baikov, K.~G. Chetyrkin, J.~H. Kuhn, {Scalar correlator at
  $\mathcal{O}(\alpha_s^4)$, Higgs decay into $b$-quarks and bounds on the
  light quark masses}, Phys. Rev. Lett. 96 (2006) 012003.
\newblock \href {http://arxiv.org/abs/hep-ph/0511063}
  {\path{arXiv:hep-ph/0511063}}, \href
  {https://doi.org/10.1103/PhysRevLett.96.012003}
  {\path{doi:10.1103/PhysRevLett.96.012003}}.

\bibitem{Kataev:1997cq}
A.~L. Kataev, {The Order O (alpha alpha-s) and O (alpha**2) corrections to the
  decay width of the neutral Higgs boson to the anti-b b pair}, JETP Lett. 66
  (1997) 327--330.
\newblock \href {http://arxiv.org/abs/hep-ph/9708292}
  {\path{arXiv:hep-ph/9708292}}, \href {https://doi.org/10.1134/1.567516}
  {\path{doi:10.1134/1.567516}}.

\bibitem{Chetyrkin:1995pd}
K.~Chetyrkin, A.~Kwiatkowski, {Second order QCD corrections to scalar and
  pseudoscalar Higgs decays into massive bottom quarks}, Nucl. Phys. B 461
  (1996) 3--18.
\newblock \href {http://arxiv.org/abs/hep-ph/9505358}
  {\path{arXiv:hep-ph/9505358}}, \href
  {https://doi.org/10.1016/0550-3213(95)00616-8}
  {\path{doi:10.1016/0550-3213(95)00616-8}}.

\bibitem{Pocsik:1980ta}
G.~Pocsik, T.~Torma, {On the Decays of Heavy Higgs Bosons}, Z. Phys. C 6 (1980)
  1.
\newblock \href {https://doi.org/10.1007/BF01427913}
  {\path{doi:10.1007/BF01427913}}.

\bibitem{Rizzo:1980gz}
T.~G. Rizzo, {Decays of Heavy Higgs Bosons}, Phys. Rev. D 22 (1980) 722.
\newblock \href {https://doi.org/10.1103/PhysRevD.22.722}
  {\path{doi:10.1103/PhysRevD.22.722}}.

\bibitem{Keung:1984hn}
W.-Y. Keung, W.~J. Marciano, {HIGGS SCALAR DECAYS: H ---\ensuremath{>} W+- X},
  Phys. Rev. D 30 (1984) 248.
\newblock \href {https://doi.org/10.1103/PhysRevD.30.248}
  {\path{doi:10.1103/PhysRevD.30.248}}.

\bibitem{Djouadi:2005gi}
A.~Djouadi, {The Anatomy of electro-weak symmetry breaking. I: The Higgs boson
  in the standard model}, Phys. Rept. 457 (2008) 1--216.
\newblock \href {http://arxiv.org/abs/hep-ph/0503172}
  {\path{arXiv:hep-ph/0503172}}, \href
  {https://doi.org/10.1016/j.physrep.2007.10.004}
  {\path{doi:10.1016/j.physrep.2007.10.004}}.

\bibitem{Cahn:1988ru}
R.~N. Cahn, {The Higgs Boson}, Rept. Prog. Phys. 52 (1989) 389.
\newblock \href {https://doi.org/10.1088/0034-4885/52/4/001}
  {\path{doi:10.1088/0034-4885/52/4/001}}.

\bibitem{Grau:1990uu}
A.~Grau, G.~Panchieri, R.~J.~N. Phillips, {Contributions of off-shell top
  quarks to decay processes}, Phys. Lett. B 251 (1990) 293--298.
\newblock \href {https://doi.org/10.1016/0370-2693(90)90939-4}
  {\path{doi:10.1016/0370-2693(90)90939-4}}.

\bibitem{Spira:1991tj}
M.~Spira, A.~Djouadi, P.~M. Zerwas, {QCD corrections to the H Z gamma
  coupling}, Phys. Lett. B 276 (1992) 350--353.
\newblock \href {https://doi.org/10.1016/0370-2693(92)90331-W}
  {\path{doi:10.1016/0370-2693(92)90331-W}}.

\bibitem{Gehrmann:2015dua}
T.~Gehrmann, S.~Guns, D.~Kara, {The rare decay $H\to Z\gamma$ in perturbative
  QCD}, JHEP 09 (2015) 038.
\newblock \href {http://arxiv.org/abs/1505.00561} {\path{arXiv:1505.00561}},
  \href {https://doi.org/10.1007/JHEP09(2015)038}
  {\path{doi:10.1007/JHEP09(2015)038}}.

\bibitem{Bonciani:2015eua}
R.~Bonciani, V.~Del~Duca, H.~Frellesvig, J.~M. Henn, F.~Moriello, V.~A.
  Smirnov, {Next-to-leading order QCD corrections to the decay width H
  $\rightarrow$ Z$\gamma$}, JHEP 08 (2015) 108.
\newblock \href {http://arxiv.org/abs/1505.00567} {\path{arXiv:1505.00567}},
  \href {https://doi.org/10.1007/JHEP08(2015)108}
  {\path{doi:10.1007/JHEP08(2015)108}}.

\bibitem{deBlas:2019rxi}
J.~de~Blas, et~al., {Higgs Boson Studies at Future Particle Colliders}, JHEP 01
  (2020) 139.
\newblock \href {http://arxiv.org/abs/1905.03764} {\path{arXiv:1905.03764}},
  \href {https://doi.org/10.1007/JHEP01(2020)139}
  {\path{doi:10.1007/JHEP01(2020)139}}.

\bibitem{Hahn:2000kx}
T.~Hahn, {Generating Feynman diagrams and amplitudes with FeynArts 3}, Comput.
  Phys. Commun. 140 (2001) 418--431.
\newblock \href {http://arxiv.org/abs/hep-ph/0012260}
  {\path{arXiv:hep-ph/0012260}}, \href
  {https://doi.org/10.1016/S0010-4655(01)00290-9}
  {\path{doi:10.1016/S0010-4655(01)00290-9}}.

\bibitem{Djouadi:1990aj}
A.~Djouadi, M.~Spira, J.~J. van~der Bij, P.~M. Zerwas, {QCD corrections to
  gamma gamma decays of Higgs particles in the intermediate mass range}, Phys.
  Lett. B 257 (1991) 187--190.
\newblock \href {https://doi.org/10.1016/0370-2693(91)90879-U}
  {\path{doi:10.1016/0370-2693(91)90879-U}}.

\bibitem{Djouadi:1993ji}
A.~Djouadi, M.~Spira, P.~M. Zerwas, {Two photon decay widths of Higgs
  particles}, Phys. Lett. B 311 (1993) 255--260.
\newblock \href {http://arxiv.org/abs/hep-ph/9305335}
  {\path{arXiv:hep-ph/9305335}}, \href
  {https://doi.org/10.1016/0370-2693(93)90564-X}
  {\path{doi:10.1016/0370-2693(93)90564-X}}.

\bibitem{Harlander:2005rq}
R.~Harlander, P.~Kant, {Higgs production and decay: Analytic results at
  next-to-leading order QCD}, JHEP 12 (2005) 015.
\newblock \href {http://arxiv.org/abs/hep-ph/0509189}
  {\path{arXiv:hep-ph/0509189}}, \href
  {https://doi.org/10.1088/1126-6708/2005/12/015}
  {\path{doi:10.1088/1126-6708/2005/12/015}}.

\bibitem{Melnikov:1994jb}
K.~Melnikov, M.~Spira, O.~I. Yakovlev, {Threshold effects in two photon decays
  of Higgs particles}, Z. Phys. C 64 (1994) 401--406.
\newblock \href {http://arxiv.org/abs/hep-ph/9405301}
  {\path{arXiv:hep-ph/9405301}}, \href {https://doi.org/10.1007/BF01560100}
  {\path{doi:10.1007/BF01560100}}.

\bibitem{Aglietti:2006tp}
U.~Aglietti, R.~Bonciani, G.~Degrassi, A.~Vicini, {Analytic Results for Virtual
  QCD Corrections to Higgs Production and Decay}, JHEP 01 (2007) 021.
\newblock \href {http://arxiv.org/abs/hep-ph/0611266}
  {\path{arXiv:hep-ph/0611266}}, \href
  {https://doi.org/10.1088/1126-6708/2007/01/021}
  {\path{doi:10.1088/1126-6708/2007/01/021}}.

\bibitem{Maierhofer:2012vv}
P.~Maierhöfer, P.~Marquard, {Complete three-loop QCD corrections to the decay
  $H -> \gamma \gamma$}, Phys. Lett. B 721 (2013) 131--135.
\newblock \href {http://arxiv.org/abs/1212.6233} {\path{arXiv:1212.6233}},
  \href {https://doi.org/10.1016/j.physletb.2013.02.040}
  {\path{doi:10.1016/j.physletb.2013.02.040}}.

\bibitem{Degrassi:2005mc}
G.~Degrassi, F.~Maltoni, {Two-loop electroweak corrections to the Higgs-boson
  decay H ---> gamma gamma}, Nucl. Phys. B 724 (2005) 183--196.
\newblock \href {http://arxiv.org/abs/hep-ph/0504137}
  {\path{arXiv:hep-ph/0504137}}, \href
  {https://doi.org/10.1016/j.nuclphysb.2005.06.027}
  {\path{doi:10.1016/j.nuclphysb.2005.06.027}}.

\bibitem{Actis:2008ts}
S.~Actis, G.~Passarino, C.~Sturm, S.~Uccirati, {NNLO Computational Techniques:
  The Cases H ---\ensuremath{>} gamma gamma and H ---\ensuremath{>} g g}, Nucl.
  Phys. B 811 (2009) 182--273.
\newblock \href {http://arxiv.org/abs/0809.3667} {\path{arXiv:0809.3667}},
  \href {https://doi.org/10.1016/j.nuclphysb.2008.11.024}
  {\path{doi:10.1016/j.nuclphysb.2008.11.024}}.

\bibitem{Fugel:2004ug}
F.~Fugel, B.~A. Kniehl, M.~Steinhauser, {Two loop electroweak correction of
  O(G(F)M(t)**2) to the Higgs-boson decay into photons}, Nucl. Phys. B 702
  (2004) 333--345.
\newblock \href {http://arxiv.org/abs/hep-ph/0405232}
  {\path{arXiv:hep-ph/0405232}}, \href
  {https://doi.org/10.1016/j.nuclphysb.2004.09.018}
  {\path{doi:10.1016/j.nuclphysb.2004.09.018}}.

\bibitem{Djouadi:1991tka}
A.~Djouadi, M.~Spira, P.~M. Zerwas, {Production of Higgs bosons in proton
  colliders: QCD corrections}, Phys. Lett. B 264 (1991) 440--446.
\newblock \href {https://doi.org/10.1016/0370-2693(91)90375-Z}
  {\path{doi:10.1016/0370-2693(91)90375-Z}}.

\bibitem{Schreck:2007um}
M.~Schreck, M.~Steinhauser, {Higgs Decay to Gluons at NNLO}, Phys. Lett. B 655
  (2007) 148--155.
\newblock \href {http://arxiv.org/abs/0708.0916} {\path{arXiv:0708.0916}},
  \href {https://doi.org/10.1016/j.physletb.2007.08.080}
  {\path{doi:10.1016/j.physletb.2007.08.080}}.

\bibitem{Chetyrkin:1997iv}
K.~G. Chetyrkin, B.~A. Kniehl, M.~Steinhauser, {Hadronic Higgs decay to order
  alpha-s**4}, Phys. Rev. Lett. 79 (1997) 353--356.
\newblock \href {http://arxiv.org/abs/hep-ph/9705240}
  {\path{arXiv:hep-ph/9705240}}, \href
  {https://doi.org/10.1103/PhysRevLett.79.353}
  {\path{doi:10.1103/PhysRevLett.79.353}}.

\bibitem{Baikov:2006ch}
P.~Baikov, K.~Chetyrkin, {Top Quark Mediated Higgs Boson Decay into Hadrons to
  Order $\alpha_s^5$}, Phys. Rev. Lett. 97 (2006) 061803.
\newblock \href {http://arxiv.org/abs/hep-ph/0604194}
  {\path{arXiv:hep-ph/0604194}}, \href
  {https://doi.org/10.1103/PhysRevLett.97.061803}
  {\path{doi:10.1103/PhysRevLett.97.061803}}.

\bibitem{Spira:1995rr}
M.~Spira, A.~Djouadi, D.~Graudenz, P.~Zerwas, {Higgs boson production at the
  LHC}, Nucl. Phys. B 453 (1995) 17--82.
\newblock \href {http://arxiv.org/abs/hep-ph/9504378}
  {\path{arXiv:hep-ph/9504378}}, \href
  {https://doi.org/10.1016/0550-3213(95)00379-7}
  {\path{doi:10.1016/0550-3213(95)00379-7}}.

\bibitem{Chetyrkin:1998mw}
K.~G. Chetyrkin, B.~A. Kniehl, M.~Steinhauser, W.~A. Bardeen, {Effective QCD
  interactions of CP odd Higgs bosons at three loops}, Nucl. Phys. B535 (1998)
  3--18.
\newblock \href {http://arxiv.org/abs/hep-ph/9807241}
  {\path{arXiv:hep-ph/9807241}}, \href
  {https://doi.org/10.1016/S0550-3213(98)00594-X}
  {\path{doi:10.1016/S0550-3213(98)00594-X}}.

\bibitem{Larin:1995sq}
S.~A. Larin, T.~van Ritbergen, J.~A.~M. Vermaseren, {The Large top quark mass
  expansion for Higgs boson decays into bottom quarks and into gluons}, Phys.
  Lett. B 362 (1995) 134--140.
\newblock \href {http://arxiv.org/abs/hep-ph/9506465}
  {\path{arXiv:hep-ph/9506465}}, \href
  {https://doi.org/10.1016/0370-2693(95)01192-S}
  {\path{doi:10.1016/0370-2693(95)01192-S}}.

\bibitem{Muhlleitner:2006wx}
M.~Muhlleitner, M.~Spira, {Higgs Boson Production via Gluon Fusion: Squark
  Loops at NLO QCD}, Nucl. Phys. B 790 (2008) 1--27.
\newblock \href {http://arxiv.org/abs/hep-ph/0612254}
  {\path{arXiv:hep-ph/0612254}}, \href
  {https://doi.org/10.1016/j.nuclphysb.2007.08.011}
  {\path{doi:10.1016/j.nuclphysb.2007.08.011}}.

\bibitem{Zyla:2020zbs}
P.~Zyla, et~al., {Review of Particle Physics}, PTEP 2020~(8) (2020) 083C01.
\newblock \href {https://doi.org/10.1093/ptep/ptaa104}
  {\path{doi:10.1093/ptep/ptaa104}}.

\bibitem{deFlorian:2016spz}
D.~de~Florian, et~al., {Handbook of LHC Higgs Cross Sections: 4. Deciphering
  the Nature of the Higgs Sector} 2/2017 (10 2016).
\newblock \href {http://arxiv.org/abs/1610.07922} {\path{arXiv:1610.07922}},
  \href {https://doi.org/10.23731/CYRM-2017-002}
  {\path{doi:10.23731/CYRM-2017-002}}.

\bibitem{Denner:2011mq}
A.~Denner, S.~Heinemeyer, I.~Puljak, D.~Rebuzzi, M.~Spira, {Standard Model
  Higgs-Boson Branching Ratios with Uncertainties}, Eur. Phys. J. C 71 (2011)
  1753.
\newblock \href {http://arxiv.org/abs/1107.5909} {\path{arXiv:1107.5909}},
  \href {https://doi.org/10.1140/epjc/s10052-011-1753-8}
  {\path{doi:10.1140/epjc/s10052-011-1753-8}}.

\bibitem{Bredenstein_2006}
A.~Bredenstein, A.~Denner, S.~Dittmaier, M.~M. Weber,
  \href{http://dx.doi.org/10.1103/PhysRevD.74.013004}{Precise predictions for
  the higgs-boson decayh→ww/zz→4leptons}, Physical Review D 74~(1) (Jul
  2006).
\newblock \href {https://doi.org/10.1103/physrevd.74.013004}
  {\path{doi:10.1103/physrevd.74.013004}}.
\newline\urlprefix\url{http://dx.doi.org/10.1103/PhysRevD.74.013004}

\bibitem{Bredenstein_2006_2}
A.~Bredenstein, A.~Denner, S.~Dittmaier, M.~Weber,
  \href{http://dx.doi.org/10.1016/j.nuclphysbps.2006.09.104}{Precision
  calculations for the higgs decays h → zz/ww → 4 leptons}, Nuclear Physics
  B - Proceedings Supplements 160 (2006) 131–135.
\newblock \href {https://doi.org/10.1016/j.nuclphysbps.2006.09.104}
  {\path{doi:10.1016/j.nuclphysbps.2006.09.104}}.
\newline\urlprefix\url{http://dx.doi.org/10.1016/j.nuclphysbps.2006.09.104}

\bibitem{Bredenstein_2007}
A.~Bredenstein, A.~Denner, S.~Dittmaier, M.~M. Weber,
  \href{http://dx.doi.org/10.1088/1126-6708/2007/02/080}{Radiative corrections
  to the semileptonic and hadronic higgs-boson decays h→ww/zz→4 fermions},
  Journal of High Energy Physics 2007~(02) (2007) 080–080.
\newblock \href {https://doi.org/10.1088/1126-6708/2007/02/080}
  {\path{doi:10.1088/1126-6708/2007/02/080}}.
\newline\urlprefix\url{http://dx.doi.org/10.1088/1126-6708/2007/02/080}

\bibitem{Martin:1997ns}
S.~P. Martin, {A Supersymmetry primer}, Adv. Ser. Direct. High Energy Phys. 21
  (2010) 1--153.
\newblock \href {http://arxiv.org/abs/hep-ph/9709356}
  {\path{arXiv:hep-ph/9709356}}, \href
  {https://doi.org/10.1142/9789812839657_0001}
  {\path{doi:10.1142/9789812839657_0001}}.

\bibitem{Domingo:2019jsc}
F.~Domingo, S.~Heinemeyer, S.~Pa\ss{}ehr, G.~Weiglein, {Precision Predictions
  for Higgs decays in the (N)MSSM}, CERN Yellow Reports: Monographs 3 (2020)
  247--266.
\newblock \href {https://doi.org/10.23731/CYRM-2020-003.247}
  {\path{doi:10.23731/CYRM-2020-003.247}}.

\bibitem{Diessner:2014ksa}
P.~Dießner, J.~Kalinowski, W.~Kotlarski, D.~Stöckinger, {Higgs boson mass and
  electroweak observables in the MRSSM}, JHEP 12 (2014) 124.
\newblock \href {http://arxiv.org/abs/1410.4791} {\path{arXiv:1410.4791}},
  \href {https://doi.org/10.1007/JHEP12(2014)124}
  {\path{doi:10.1007/JHEP12(2014)124}}.

\bibitem{Melnikov:1993tj}
K.~Melnikov, O.~I. Yakovlev, {Higgs ---\ensuremath{>} two photon decay: QCD
  radiative correction}, Phys. Lett. B 312 (1993) 179--183.
\newblock \href {http://arxiv.org/abs/hep-ph/9302281}
  {\path{arXiv:hep-ph/9302281}}, \href
  {https://doi.org/10.1016/0370-2693(93)90507-E}
  {\path{doi:10.1016/0370-2693(93)90507-E}}.

\bibitem{Muhlleitner:2003vg}
M.~Muhlleitner, A.~Djouadi, Y.~Mambrini, {SDECAY: A Fortran code for the decays
  of the supersymmetric particles in the MSSM}, Comput. Phys. Commun. 168
  (2005) 46--70.
\newblock \href {http://arxiv.org/abs/hep-ph/0311167}
  {\path{arXiv:hep-ph/0311167}}, \href
  {https://doi.org/10.1016/j.cpc.2005.01.012}
  {\path{doi:10.1016/j.cpc.2005.01.012}}.

\bibitem{FSREADME}
{Subsection `Support for alternative loop libraries' in the FlexibleSUSY
  README}, accessed June 30, 2020.

\bibitem{Denner:1991kt}
A.~Denner, {Techniques for calculation of electroweak radiative corrections at
  the one loop level and results for W physics at LEP-200}, Fortsch. Phys. 41
  (1993) 307--420.
\newblock \href {http://arxiv.org/abs/0709.1075} {\path{arXiv:0709.1075}},
  \href {https://doi.org/10.1002/prop.2190410402}
  {\path{doi:10.1002/prop.2190410402}}.

\bibitem{Gigg:thesis}
M.~A. Gigg, \href{http://etheses.dur.ac.uk/2301/1/2301_311.pdf}{{Monte Carlo
  simulations of physics beyond the standard model}}, Ph.D. thesis, Durham
  University (2008).
\newline\urlprefix\url{http://etheses.dur.ac.uk/2301/1/2301_311.pdf}

\bibitem{Sjodahl:2012nk}
M.~Sjödahl, {ColorMath - A package for color summed calculations in SU(Nc)},
  Eur. Phys. J. C 73~(2) (2013) 2310.
\newblock \href {http://arxiv.org/abs/1211.2099} {\path{arXiv:1211.2099}},
  \href {https://doi.org/10.1140/epjc/s10052-013-2310-4}
  {\path{doi:10.1140/epjc/s10052-013-2310-4}}.

\bibitem{AguilarSaavedra:2005pw}
J.~A. Aguilar-Saavedra, et~al., {Supersymmetry parameter analysis: SPA
  convention and project}, Eur. Phys. J. C 46 (2006) 43--60.
\newblock \href {http://arxiv.org/abs/hep-ph/0511344}
  {\path{arXiv:hep-ph/0511344}}, \href
  {https://doi.org/10.1140/epjc/s2005-02460-1}
  {\path{doi:10.1140/epjc/s2005-02460-1}}.

\bibitem{Staub:2017jnp}
F.~Staub, W.~Porod, {Improved predictions for intermediate and heavy
  Supersymmetry in the MSSM and beyond}, Eur. Phys. J. C 77~(5) (2017) 338.
\newblock \href {http://arxiv.org/abs/1703.03267} {\path{arXiv:1703.03267}},
  \href {https://doi.org/10.1140/epjc/s10052-017-4893-7}
  {\path{doi:10.1140/epjc/s10052-017-4893-7}}.

\bibitem{Staub:2011dp}
F.~Staub, T.~Ohl, W.~Porod, C.~Speckner, {A Tool Box for Implementing
  Supersymmetric Models}, Comput. Phys. Commun. 183 (2012) 2165--2206.
\newblock \href {http://arxiv.org/abs/1109.5147} {\path{arXiv:1109.5147}},
  \href {https://doi.org/10.1016/j.cpc.2012.04.013}
  {\path{doi:10.1016/j.cpc.2012.04.013}}.

\end{thebibliography}


\begin{thebibliography}{0}
\bibitem{1}
P.~Athron, J.~h.~Park, D.~St\"ockinger and A.~Voigt,
Comput. Phys. Commun. \textbf{190} (2015), 139-172
doi:10.1016/j.cpc.2014.12.020
[arXiv:1406.2319 [hep-ph]].

\bibitem{2}
P.~Athron, M.~Bach, D.~Harries, T.~Kwasnitza, J.~h.~Park, D.~St\"ockinger, A.~Voigt and J.~Ziebell,
Comput. Phys. Commun. \textbf{230} (2018), 145-217
doi:10.1016/j.cpc.2018.04.016
[arXiv:1710.03760 [hep-ph]].
\end{thebibliography}
